%% file: DAFx26_tmpl_v3.tex
\def\papertitle{A Production-Oriented Framework for Evaluation of SFX Generation  }
\def\paperauthorA{Mélodie Desbos}
\def\paperauthorB{Yara Bahram}
\def\paperauthorC{Eric Granger}
\def\paperauthorD{Mohammadhadi Shateri}
\documentclass[twoside,a4paper]{article}
\usepackage{etoolbox}

\usepackage{dafx26v3}
\usepackage{float}
\usepackage{cuted}
\usepackage{placeins} 
\usepackage{amsmath,amssymb,amsfonts,amsthm}
\usepackage{siunitx}
\usepackage{natbib}
\usepackage{euscript}
\usepackage[T1]{fontenc}
\usepackage[utf8]{inputenc}
\usepackage{ifpdf}
\usepackage[english]{babel}
\usepackage{caption}
\usepackage{subfig} % or can use subcaption package
\usepackage{color}
\usepackage{booktabs}
\usepackage{lipsum}

\input glyphtounicode
\ninept

\newcounter{numauth}
\newcounter{listcnt}
\newcommand\authcnt[1]{\ifdefined#1 \stepcounter{numauth} \fi}

\newcommand\addauth[1]{
\ifdefined#1 
\stepcounter{listcnt}
\ifnum \value{listcnt}<\value{numauth}
\appto\authorslist{, #1}
\else
\appto\authorslist{~and~#1}
\fi
\fi}
\authcnt{\paperauthorB}
\authcnt{\paperauthorC}
\authcnt{\paperauthorD}
\authcnt{\paperauthorE}
\authcnt{\paperauthorF}
\authcnt{\paperauthorG}
\authcnt{\paperauthorH}
\authcnt{\paperauthorI}
\authcnt{\paperauthorJ}
\def\authorslist{\paperauthorA}
\addauth{\paperauthorB}
\addauth{\paperauthorC}
\addauth{\paperauthorD}
\addauth{\paperauthorE}
\addauth{\paperauthorF}
\addauth{\paperauthorG}
\addauth{\paperauthorH}
\addauth{\paperauthorI}
\addauth{\paperauthorJ}
\usepackage{booktabs}
\usepackage{tabularx}
\usepackage{xcolor}
\usepackage{array}
\usepackage[table]{xcolor}
\usepackage{float}

\newcommand{\outlineboxxxx}[1]{
\begingroup 
\setlength{\fboxsep}{0.1ex}% inner padding 
\setlength{\fboxrule}{1.3pt}% border thickness 
\fcolorbox[HTML]{16FF20}{FFFFFF}{\strut #1}% 
\endgroup 
}

\newcommand{\outlineboxxxxx}[1]{
\begingroup 
\setlength{\fboxsep}{0.1ex}% inner padding 
\setlength{\fboxrule}{1.3pt}% border thickness 
\fcolorbox[HTML]{6FB1FF}{FFFFFF}{\strut #1}% 
\endgroup 
}

\newcommand{\outlineboxxxxxx}[1]{
\begingroup 
\setlength{\fboxsep}{0.1ex}% inner padding 
\setlength{\fboxrule}{1.3pt}% border thickness 
\fcolorbox[HTML]{E67BFF}{FFFFFF}{\strut #1}% 
\endgroup 
}

\newcommand{\outlineboxxxxxxxx}[1]{
\begingroup 
\setlength{\fboxsep}{0.1ex}% inner padding 
\setlength{\fboxrule}{1.3pt}% border thickness 
\fcolorbox[HTML]{000000}{FFFFFF}{\strut #1}% 
\endgroup 
}
\usepackage{times}
\newif\ifpdf
\ifx\pdfoutput\relax
\else
   \ifcase\pdfoutput
      \pdffalse
   \else
      \pdftrue
   \fi
\fi

\ifpdf % compiling with pdflatex
  \usepackage[pdftex,
    pdftitle={\papertitle},
    pdfauthor={\authorslist},
    pdfsubject={Proceedings of the 29th International Conference on Digital Audio Effects (DAFx26)},
    colorlinks=false, % links are activated as color boxes instead of color text
    bookmarksnumbered, % use section numbers with bookmarks
    pdfstartview=XYZ % start with zoom=100% instead of full screen; especially useful if working with a big screen :-)
  ]{hyperref}
  \usepackage[pdftex]{graphicx}
\else % compiling with latex
  \usepackage[dvips]{epsfig,graphicx}
  \usepackage[dvips,
    pdftitle={\papertitle},
    pdfauthor={\authorslist},
    pdfsubject={Proceedings of the 29th International Conference on Digital Audio Effects (DAFx26)},
    colorlinks=false, % no color links
    bookmarksnumbered, % use section numbers with bookmarks
    pdfstartview=XYZ % start with zoom=100% instead of full screen
  ]{hyperref}
\fi
\usepackage[hypcap=true]{caption}
\title{\papertitle}

\affiliation
{\paperauthorA $^*$, \paperauthorB, \paperauthorC, \paperauthorD }
 {\href{https://www.etsmtl.ca}{LIVIA, Dept. of Systems Engineering}, ETS Montréal, Canada \\
 {\tt \href{mailto:melodie.desbos@livia.etsmtl.ca}{melodie.desbos@livia.etsmtl.ca}}
 }
\setcitestyle{numbers}
\begin{document}
% more pdf-tex settings:
\ifpdf % used graphic file format for pdflatex
  \DeclareGraphicsExtensions{.png,.jpg,.pdf}
\else  % used graphic file format for latex
  \DeclareGraphicsExtensions{.eps}
\fi

%\makeatletter
%\pdfbookmark[0]{\@pdftitle}{title}
%\makeatother

\maketitle

\input{S1_Abstract}
\input{S2_Introduction}
\input{S3_Related_Work}
\input{S4_Method}

\input{S6_Results}

%\input{S7_Discussion}
\input{S8_Conclusion}

\begingroup
\footnotesize  
\bibliographystyle{IEEEtranDAFx}
\bibliography{DAFx26_tmpl}
\endgroup

\input{S9_Appendix}

\end{document}

%% file: S1_Abstract.tex
\begin{abstract}

% current audio generation methods are mostly evaluated in broader TTA or unconditional settings rather than in reference-conditioned SFX production scenarios
% the paper proposes a production-oriented evaluation framework for reference-guided SFX variation
% the benchmark is built on a shared protocol on ESC-50 with reference-conditioned generation, combining objective, subjective, and qualitative analysis
% the main finding is that no single method satisfies all production requirements, although AudioX offers the strongest overall compromise for generating full reference-conditioned variation.
%\sloppy

\noindent Industrial sound design requires audio generation systems that not only produce realistic audio, but also preserve the perceptual identity of a reference, support controllable variation, and remain efficient for practical workflows. 
Existing evaluations are usually tied to text-to-audio (TTA), unconditional, or task-specific settings, limiting assessment for reference-guided sound effects (SFX) variation.
%Current methods are often evaluated in broad text-to-audio (TTA), unconditional, or task-specific settings, limiting assessment of their suitability for applications such as reference-guided sound effects (SFX) variation. 
To address this gap, we present a production-oriented evaluation framework for structured comparison of heterogeneous audio generation and editing methods. Our framework identifies nine production requirements and explicitly accounts for differences in model capabilities, enabling comparison under a common production objective. 
A two-stage protocol is introduced: (1) a reference-guided audio-to-audio (ATA) variation task, in which all methods are evaluated under the same ESC-50 SFX adaptation setup, and (2) capability-specific analyses of native operations such as SFX morphing, temporal and energy alignment, inpainting, and targeted editing. This framework combines objective metrics (including FAD, ImageBind-based reference alignment, and diversity across generated variants), together with a human study of perceptual identity preservation and transient diagnosis. 
%%
% Revealing complementary strengths and trade-offs, our study provides a practical overview of which baselines are best suited to different production needs. 
Our study reveals complementary strengths and trade-offs across baselines for different production needs.
%We provide a practical overview of suitable baselines for different production needs. 
Among the full-generation baselines evaluated under a shared ATA setting, AudioX provides the strongest overall trade-off between reference alignment and diversity while still supporting SFX morphing. Other baselines remain most suitable for specific editing operations. 
%Under the shared ATA setting, AudioX provides the strongest overall trade-off between reference alignment and diversity while still supporting SFX morphing, while other baselines remain better suited to specific editing operations.
Our framework establishes a structured evaluation and decision protocol for reference-guided SFX variation and provides a practical basis for designing future unified industrial audio generation pipelines. Audio demos and further details can be found on the accompanying web page\footnote{\label{fn:web}\href{https://melodiedesbos.github.io/A-PRODUCTION-ORIENTED-FRAMEWORK-FOR-EVALUATION-OF-SFX-GENERATION/}{https://melodiedesbos.github.io/sfx-eval-framework}}. 
%by providing a more comprehensive understanding of heterogeneous SoTA methods, their capabilities, and their behavior in real-world production settings. 
\end{abstract}

%The framework introduces a shared benchmark and comparative protocol designed to test each method’s ability to generate usable variants of a reference SFX under common production requirements. Each method is lightly adapted to ESC-50 and evaluated through a two-stage design. First, in a shared reference-conditioned audio-to-audio task, each model generates variants from a reference clip and its class label across 50 classes, yielding 4,000 generated samples per model. Second, complementary method-specific analyses probe native capabilities such as style transfer, temporal alignment, inpainting, and targeted editing. Objective evaluation combines FAD with ImageBind-based alignment to the reference and diversity across variants, together with a human study on perceptual identity preservation. Results show that AudioX provides the strongest overall compromise by balancing diversity and reference alignment, while AudioLDM favours diversity and stylization. T-Foley, A$^2$SB, and ThinkSound reveal complementary strengths under temporal guidance, energy control, targeted modification, and repair settings. Although a perfectly fair comparison cannot be guaranteed because of heterogeneous pretrained distributions, conditioning strategies, and signal representations, the proposed framework provides requirement-driven baseline selection, a shared evaluation benchmark, and complementary capability-specific analyses for reference-guided audio generation and editing systems.

%% file: S2_Introduction.tex
\section{Introduction}
\label{sec:intro}
% SFX production requires realistic audio, perceptual identity preservation, controllable variation, and efficiency for deployment, yet current methods are still only partially evaluated under those constraints.
% In production soundbanks, each class may contain only a few reusable references, leading to repetition and expensive manual variation design.
% Current evaluation protocols are heterogeneous, and systematic evaluation frameworks for reference-guided SFX variation remain limited.

%1-Production motivation
\noindent \textbf{Production Motivation}~\textemdash~
\noindent In SFX production, sound designers often rely on few reusable recordings, since manually designing variations is costly and time-consuming. %In SFX production, sound designers often rely on a limited number of reusable recordings for a given SFX class, since manually designed variation is time-consuming and costly. 
This low variability can lead to perceptible repetition %when the same sounds are repeatedly used 
in interactive media, motivating systems that can produce useful variations from a reference audio clip \cite{chung2024tfoley, smartdj2025, fang2026acfoley, liang2025audiomorphix, liu2025thinksound} or from standard TTA generation \cite{stableaudioopen2024, liu2023audioldm2, kreuk2022audiogen}, rather than synthesizing audio unconditionally \cite{zhu2023edmsound}. In practice, an effective sound design system must do more than produce plausible audio as it is subject to multiple constraints. %Indeed, SFX variations are tied to timing \cite{kong2025a2sb, liu2025thinksound}, energy signals \cite{chung2024tfoley}, acoustic identity \cite{smartdj2025}, or a given reference audio clip \cite{liang2025audiomorphix}. 
Unlike generic TTA generation, this setting requires jointly balancing realism, identity preservation, controllability, and usability. 
%In SFX production contexts, audio clips are typically organized into \emph{soundbanks}, where each semantic class may contain only a limited number of reusable references.

%2-gap in current evaluation 
\noindent\textbf{Gaps in Current Evaluations}~\textemdash~Recent audio generation progress has made SFX generation increasingly accessible. Diffusion, flow-matching, and multimodal pipelines now support %a wide range of generation and editing operations including 
style transfer \cite{liu2023audioldm, audiox2025} (or SFX morphing \cite{niu2024soundmorpher}), temporal conditioning \cite{chung2024tfoley}, targeted editing \cite{liu2025thinksound}, and inpainting \cite{kong2025a2sb}. However, these methods are typically evaluated only within their original task settings, such as TTA generation \cite{liu2023audioldm, stableaudioopen2024}, video-to-audio (VTA) synthesis \cite{liu2025thinksound, luo2023difffoley}, foley alignment \cite{chung2024tfoley, Chen2025CVPR}, or localized restoration \cite{kong2025a2sb}. %Therefore, their reported performance provides only limited insight into how well a model can generate controlled, plausible, and production-useful variations from a reference audio clip. More importantly, differences in task formulation, architecture, and evaluation criteria make cross-method comparison difficult.
Therefore, their reported performance offers limited insight into controlled and production-useful reference-guided variation. In addition, differences in task formulation, architecture, and evaluation criteria make methods difficult to compare and complicate the selection of the most suitable approach for specific SFX generation needs.

%\noindent Existing methods are relevant, but their evaluation is too task-specific and heterogeneous to tell us which ones are actually useful for production-oriented SFX variation from a given reference.

%what we propose 
% We introduce a production-oriented evaluation framework that makes heterogeneous reference-guided generation and editing methods comparable under a shared SFX-variation objective.
\noindent \textbf{Proposal of Evaluation Framework}~\textemdash~This work aims to address this gap by proposing a production-oriented evaluation framework for reference-guided SFX variation.
%such as diversity, high fidelity, controllability, and efficiency, while preserving the perceptual identity of the original sound event (e.g., footsteps, a door slam, laughter).
The proposed framework defines practical production requirements, selects complementary state-of-the-art baselines accordingly, and evaluates them through a two-stage protocol: (1) All methods are compared under a shared reference-conditioned ATA variation task using ESC-50, an open-source few-shot soundbank, to reflect a realistic production 
%constraints
usage \cite{piczak2015esc50}.
%together with a light \emph{fine-tuning} adaptation applied to 
%This framework enables the evaluation of each approach within a shared application and under realistic production constraints. 
(2) Since the evaluated baselines differ in conditioning, training, and editing capabilities, our goal is not to produce a single ranking but to characterize their suitability for creative SFX variation.
The proposed framework further provides a capability-specific analyses
%to avoid reducing all methods to a single narrow framework, 
to assess strengths and limitations of native editing operations such as reference-guided SFX morphing, temporal alignment, inpainting, and targeted editing. %and highlight practical strengths and limitations of each approach. 
%Overall, the resulting protocol supports a structured comparison of heterogeneous methods under a common production objective, while accounting as fairly as possible for the specific characteristics of each model and the requirements of industrial usage.  add space??
%This evaluation framework helps bridge the gap between the strengths of generative pipelines and their direct implementation in industrial sound design, by providing a more comprehensive understanding of heterogeneous SoTA methods, their capabilities, and their behavior in real-world production settings.
This design enables a structured and fair comparison of heterogeneous methods under a shared production objective, while creating a decision protocol to emphasize distinctive strengths of each approach. \\

%rebuttal
%\noindent\textcolor{blue}{Since the evaluated baselines differ in their conditioning signals, training protocols, and editing scopes, our goal is not to reduce their comparison to a unified ranking. Instead, the proposed framework provides an application-relevant evaluation and decision protocol that reveals how each method satisfies specific production requirements under a shared reference-conditioned ATA setting. Through capability-specific analyses, it highlights practical strengths and limitations of each approach.}

\sloppy
\noindent \textbf{Contributions}~\textemdash~
%\noindent Our contributions are as follows :
%\begin{itemize}
    %\item 
    \textbf{(i)} We introduce a production-oriented evaluation framework for reference-guided SFX variation grounded in practical requirements such as identity preservation, diversity, controllable variation, and production usability. 
    %\item We introduce a production-oriented evaluation framework for reference-guided SFX variation grounded in practical requirements such as fidelity, identity preservation, diversity without drift, controllability, targeted modification, and efficiency.
    %\item  
    \textbf{(ii)} A two-stage protocol is proposed that combines a shared reference-conditioned ATA variation task with capability-specific analyses. This enables a decision-oriented analysis and more structured comparison across heterogeneous baselines. 
    %\item  We propose a unified benchmark and protocol that combines a shared ATA generation task on ESC-50 \cite{piczak2015esc50} with method-specific analyses, enabling a structured and fair comparison across heterogeneous baselines while preserving their native capabilities. 
    %\item 
    \textbf{(iii)} Comparison of recent audio generation and editing methods through objective, perceptual, qualitative, and capability-specific evaluations. %and shows that current approaches exhibit clear trade-offs rather than a single dominant solution, with different methods addressing complementary production needs. Diversity-fidelity trade-offs and method-specific editing capabilities for reference-guided SFX variation.
    It reveals clear trade-offs and complementary strengths of different methods for reference-guided SFX variation. 
%\end{itemize}

%The contributions of this work are threefold:(i) a requirement-driven baseline selection for production-oriented SFX variation;(ii) a shared benchmark and evaluation protocol combining objective and subjective criteria for fairer comparison across heterogeneous pipelines; and(iii) a detailed analysis of diversity–fidelity trade-offs and method-specific editing capabilities for reference-guided SFX variation.

%\noindent Section~\ref{sec:method} presents the production-oriented evaluation framework, including the task formulation and method selection. Section~\ref{sec:experimental} details the experimental setup, and Section~\ref{sec:results} presents the analysis of the compared methods.

%\vspace{-10pt}

%% file: S3_Related_Work.tex
\section{Related Work} \label{sec:relatedwork}

%%%%%%%%
\subsection{Audio Generation for SFX}
% unconditional / generic audio generation,text-to-audio, toward audio editing and multimodal conditioning. position AudioLDM / AudioX / ThinkSound / T-Foley / A$^2$SB as part of a broader move from generation to controllable editing, differ in representation, conditioning, and editing scope.

In audio generation, diffusion-based models remain a dominant paradigm for high-fidelity synthesis and are increasingly used for unconditional or text-conditioned synthesis, while operating on waveform, spectrogram, or latent representations \cite{liu2023audioldm, chung2024tfoley, majumder2024tango2}. AudioLDM adopts a latent-diffusion formulation in a VAE-based latent space and relies on CLAP embeddings for audio-text conditioning \cite{liu2023audioldm}. T-Foley \cite{chung2024tfoley} is guided by sound-class and temporal-event information for Foley generation. 
Recently, flow-matching models have also been explored for controllable audio generation, conditioning the flow on multimodal semantic embeddings \cite{liu2025thinksound}.
Another common denoising architecture is the Diffusion Transformer (DiT), in which self-attention operates on multimodal features together with audio tokens or latent audio representations, while the Transformer serves as the denoising network \cite{audiox2025, floresgarcia2024sketch2sound, Chen2025CVPR, zhao2025uniform,shi2025samaudio, cheng2024mmaudio}. %An example is AudioX \cite{audiox2025} that combines text, video, and audio signals through multimodal fusion, enabling audio generation under multiple conditioning signals. 
Earlier SFX-oriented works also studied controllable SFX generation, through synthesis of environmental sounds from onomatopoeic words~\cite{okamoto2022onoma}, or from acoustic features with explicit control over pitch, loudness, timbre, and transients~\cite{liu2023ddspsfx}. These works motivate SFX generation as controllable synthesis, but lack inputs over recent generative strengths.

%%%%%%%%%
\subsection{Reference-Guided Audio Variation and Editing}

%Beyond unconditional or TTA synthesis, diffusion-based models have also been widely adopted for audio editing, where the goal is to modify an existing sample while preserving relevant acoustic characteristics such as timbre, temporal structure, or event identity. 
Beyond unconditional or TTA synthesis, audio editing aims to modify existing samples while preserving timbre, temporal structure, or event identity.
%These editing settings may include inpainting, localized temporal or spectral modification, style transfer (or sound morphing), and, in some cases, video-guided sound replacement. 
%A representative example is AudioMorphix \cite{liang2025audiomorphix}, a training-free spectrogram-based editing method based on latent diffusion models such as AudioLDM \cite{liu2023audioldm} and Tango2 \cite{majumder2024tango2}. 
Recent generative audio frameworks support editing driven either by explicit instructions~\cite{liu2025thinksound, loopcopilot2023}, reference signals~\cite{kong2025a2sb, chung2024tfoley}, or both \cite{yang2024uniaudio, ppae2024, smartdj2025, fang2026acfoley,audit2023, jia2024audioeditor, Chen2025CVPR}. 
In practice, these pipelines cover complementary objectives, including full reference-guided variation~ \cite{smartdj2025, jia2024audioeditor}, temporal control~\cite {chung2024tfoley}, localized modification such as inpainting \cite{kong2025a2sb} or region-specific editing \cite{liu2025thinksound}. 
\noindent These developments are particularly relevant to SFX generation, where the objective is often not only to synthesize plausible audio, but also to preserve event identity and temporal structure while supporting controllable variation \cite{chung2024tfoley, Chen2025CVPR}. %Consequently, SFX generation is often closer to constrained generation or editing than to open-ended audio synthesis. 
%\noindent Among the baselines considered in this work, these capabilities are instantiated in complementary ways: AudioLDM supports text-guided manipulations such as style transfer (\textcolor{blue}{e.g. sound morphing when applied to SFX \cite{niu2024soundmorpher}}); T-Foley emphasizes temporal-event control \cite{chung2024tfoley}; ThinkSound supports interactive object-centric refinement and targeted editing \cite{liu2025thinksound}; AudioX enables multimodal anything-to-audio generation \cite{audiox2025}; and A$^2$SB focuses on localized restoration through inpainting \cite{kong2025a2sb}. 
In addition, a few methods are explicitly designed to adapt to new domains under restricted-data settings, including retrieval-augmented approaches for zero-shot and few-shot TTA generation \cite{yang2024audiobox} and customized generation from a few reference audio samples \cite{yuan2025dreamaudio}. Yet these methods remain primarily generation-oriented and are not necessarily suited to editing or reference-guided SFX workflows.

%%%%%%%%%%%
\subsection{Evaluation Gaps for Production-oriented SFX Variation}
% Existing evaluation frameworks in audio generation mainly target unconditional or text-conditioned synthesis, and therefore provide limited insight into controlled, plausible, and production-useful variations from a reference sound.
% current evaluation is often method-specific, benchmarks are heterogeneous
% few unified evaluation scripts 

SFX generation focuses on producing short, semantically meaningful, and often temporally structured events tied to an action, scene, or reference. \cite{chung2024tfoley, Chen2025CVPR}. Compared with general TTA generation, it places stronger emphasis on event identity, temporal synchronization, and controllability. %, since the objective is often not only to generate plausible audio but also to match a specific timing, motion, or acoustic characteristic. 
\noindent Despite recent progress, such methods are rarely evaluated under a shared SFX-variation objective or a common benchmark, and their evaluation protocols remain tied to each method's original task: AudioLDM mainly on TTA generation \cite{kim2019audiocaps}, with style transfer, high resolution, and inpainting shown separately; T-Foley on temporal-event control, combining quality, diversity, and temporal-adherence metrics with MOS ratings; ThinkSound on VTA generation, object-focused generation, and editing, mainly on VGGSound \cite{chen2020vggsound} and MovieGen Audio Bench \cite{polyak2024moviegen}; AudioX on a large protocol spanning TTA \cite{xie2024audiotime, wang2025ttabench}, VTA, joint text-and-video, and instruction-driven benchmarks; and A$^2$SB on localized restoration. Thus, evaluations reflect different task definitions, datasets, and control assumptions, making a direct cross-method comparison inherently difficult. This limitation is important for production-oriented SFX variation, where the objective is not only to generate plausible audio but also to preserve event identity, support controllable variation, and remain feasible under practical constraints such as limited data and efficient inference. Such limitations motivate a shared evaluation framework suitable for practical SFX deployment (this being the focus of our work).

%% file: S4_Method.tex
\begin{figure*}[!t] 
  \centering
  \includegraphics[width=\linewidth]{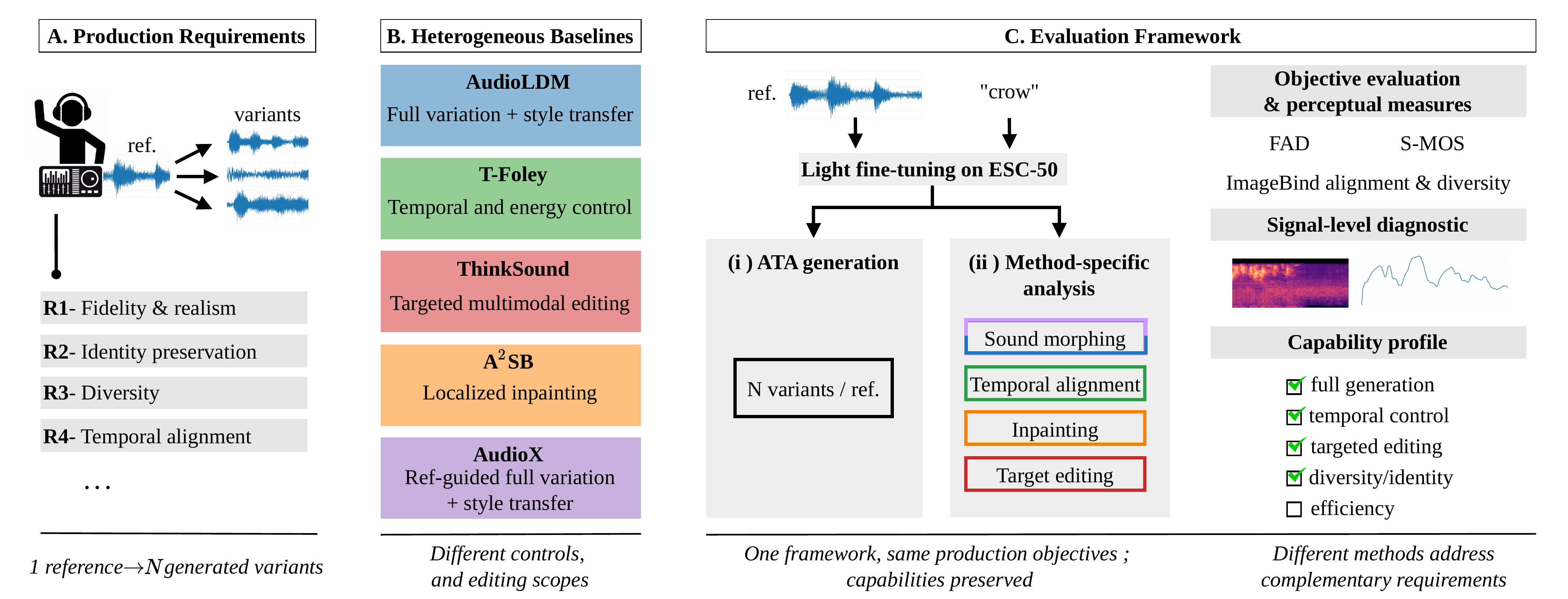}
    \caption{Overview of the proposed production-oriented evaluation framework for reference-guided SFX variation. A. In production settings, a reference sound often requires multiple diverse variations (Sec.~\ref{subsec:prod_req}); B. Existing baselines address complementary capabilities \cite{liu2023audioldm, chung2024tfoley,liu2025thinksound,kong2025a2sb,audiox2025}, including full variation generation (Sec.~\ref{subsec:heteroBase}); C. We propose an evaluation framework to assess models' abilities at generating variations of SFX in production settings (Sec.~\ref{sec:method}) and report a full capability profile for each baseline on our web page \textsuperscript{\ref{fn:web}}.}
    \vspace{-2pt}
\label{fig:motivation}
\end{figure*}

\section{Production-Oriented Evaluation Framework and Experimental Set-up}
\label{sec:method}

%Overview of the proposed production-oriented evaluation framework for reference-guided SFX variation. In production settings, a reference sound often requires multiple diverse versions under constraints such as identity preservation, controllability, or temporal alignment. Existing baselines address complementary capabilities, including full variation generation (AudioLDM \cite{liu2023audioldm}, AudioX \cite{audiox2025}), temporal-energy control (T-Foley \cite{chung2024tfoley}), targeted editing (ThinkSound \cite{liu2025thinksound}), and localized inpainting (A$^2$SB \cite{kong2025a2sb}), but are rarely compared under a shared objective. The proposed framework standardizes comparison through a common ESC-50 \cite{piczak2015esc50} reference-guided ATA setting, complemented by method-specific analyses, and evaluates each method using objective, perceptual, and qualitative criteria to characterize its relevance to different SFX production needs.

%In production settings, a reference sound often requires multiple diverse versions under constraints such as identity preservation, controllability, or temporal alignment. Existing baselines address complementary capabilities, including full variation generation (AudioLDM~\cite{liu2023audioldm}, AudioX~\cite{audiox2025}), temporal-energy control (T-Foley~\cite{chung2024tfoley}), targeted editing (ThinkSound~\cite{liu2025thinksound}), and localized inpainting (A$^2$SB~\cite{kong2025a2sb}), but are rarely compared under a shared objective.
In production settings, a reference sound often requires diverse variations under constraints such as identity preservation, controllability, and temporal alignment. Existing baselines address complementary capabilities, but are rarely compared under a shared objective.
Figure~\ref{fig:motivation} summarizes the proposed evaluation framework to standardize comparison through a common ESC-50~\cite{piczak2015esc50} reference-guided ATA setting complemented by method-specific analyses. % In this setup, each model is given a reference audio clip with minimal class-level conditioning and is required to generate multiple plausible variations. 
Each method is then evaluated using objective, perceptual, and qualitative criteria to characterize its relevance to different SFX production needs.

\input{src/tables/requirement_small}

%%%%%%%%%%%%%%%
\subsection{Production Requirements}
\label{subsec:prod_req}

\noindent %Production-ready SFX variation is not defined by high fidelity generations alone. In practical workflows, a useful method should preserve perceptual identity, support controlled variation, and remain efficient enough for iterative deployment.
Production-ready SFX variation requires more than fidelity: a useful method should preserve identity, support controlled variation, while remaining efficient for iterative deployment. In this work, the evaluation is structured around nine production requirements, each associated with an operational definition, specific evaluation signals, and typical failure modes; see Table~\ref{tab:requirement_small}. These requirements guide the baseline selection and the evaluation protocol.

\subsection{Heterogeneous Baselines}
\label{subsec:heteroBase}

We evaluate five representative baselines covering complementary SFX variation capabilities. AudioLDM \cite{liu2023audioldm} and AudioX \cite{audiox2025} for SFX morphing; T-Foley \cite{chung2024tfoley} for time-conditioned control; ThinkSound \cite{liu2025thinksound} for targeted editing; and A$^2$SB \cite{kong2025a2sb} for inpainted variations. A detailed mapping between the production requirements and each baseline can be found on our web page\textsuperscript{\ref{fn:web}}. Below, 
we provide a brief description mapping of their conditioning signals, editing scope and production strengths.
%including fidelity and realism (R1), identity preservation (R2), diversity (R3), temporal alignment (R4), energy control (R5), controllability (R6), targeted modification (R7), robustness and stability (R8), and efficiency (R9).
%.

\noindent\textbf{AudioLDM}~\cite{liu2023audioldm} %uses reference class label as restricted text input, 
can transfer the reference style to generate reference-conditioned variations. Its primary control mechanism is text conditioning, while its editing scope covers the entire generated sample. Thus, it serves as a general-purpose baseline for full-sample variation, quality and realism (R1), diversity (R3), and flexible user control through text or reference-guided editing (R6).

\noindent\textbf{T-Foley}~\cite{chung2024tfoley} uses a protocol centered on temporal-event control through sound-class and RMS envelope information for Foley generation. %, and multiple outputs are sampled from different noise seeds
This makes T-Foley particularly relevant for temporal alignment (R4) and energy control (R5), while supporting identity preservation (R2). However, it does not support targeted local edits of a specific waveform region (R7), since the reference is used only as an event condition rather than as a direct ATA edition.

%add an emphasis on this application is under-using the abilities of thinksound; as the pipeline is normally built for foundational Foley generation, interactive object-centric refinement, and targeted audio editing with high-level instructions - multimodal-to-audio. We aim to use this interactive object-centric editing but without video. (short bc detailed in part 5.3)
%Indeed, this baseline relies on a chain-of-thought reasoning process and generally benefits from more detailed instructions. The pipeline is originally designed for Foley generation, interactive object-centric refinement, and targeted audio editing with high-level instructions. 
\noindent\textbf{ThinkSound}'s~\cite{liu2025thinksound} 
%operates in a latent representation and conditions the flow on multimodal semantic embeddings. ThinkSound is evaluated on VTA generation, object-focused audio generation, and audio editing, with main comparisons conducted on VGGSound \cite{chen2020vggsound}, and outside manifold distribution evaluation on MovieGen Audio Bench\cite{polyak2024moviegen}.
control mechanism combines semantic instructions with reference conditioning, enabling targeted modifications while preserving overall plausibility. ThinkSound supports a localized editing scope, making it suitable for studying controllable semantic variation and localized content modification.
%, to generate multiple plausible variations within the same sound category. 
This allows evaluation of targeted modification (R7), controllability (R6), and semantic editing while maintaining plausible variation (R3). %More targeted control through free-form captions is possible in principle, but it is not aligned with the design choice of restricted text conditioning.

\noindent\textbf{AudioX}~\cite{audiox2025} combines text, video, and audio signals through multimodal fusion, enabling audio generation under multiple conditioning signals. AudioX spans TTA, VTA, joint text-and-video conditioning, and instruction-driven conditioning. This is particularly relevant as a recent generation model with strong identity preservation (R2), diversity without drift (R3), and controllable reference-guided generation (R6), while also providing a useful reference point for fidelity and realism (R1).

\noindent\textbf{A$^2$SB}~\cite{kong2025a2sb} masks a segment and reconstructs it from the surrounding unmasked context. Unlike the other baselines, control is provided directly through the preserved waveform context rather than text or temporal conditioning. Its editing scope is strictly local, affecting only the masked region while leaving the remainder of the signal unchanged.
%A$^2$SB is a variable-length segment of the reference audio that is masked, replaced with noise, and the missing region is sampled conditioned on the remaining unmasked context. 
This makes A$^2$SB relevant for evaluating targeted modification (R7), the preservation of an unedited context as part of identity preservation (R2), and localized variation with limited drift (R3).

\noindent Overall, the selected baselines cover four distinct editing scopes: \textbf{(i)} full-sample generation (AudioLDM, AudioX), \textbf{(ii)} temporally constrained generation (T-Foley), \textbf{(iii)} semantic reference-guided editing (ThinkSound), and \textbf{(iv)} localized waveform inpainting (A$^2$SB). This diversity enables a broader assessment of production-oriented SFX variation beyond fidelity alone.
\vspace{-5pt}

%%%%%%%%%%%%%%
\subsection{Evaluation Framework}
\label{subsec:eval_framework}
% evaluation is designed to be as fair as possible despite heterogeneous representations, pretraining data, and parameterizations; perfect fairness cannot be guaranteed, but the protocol explicitly tries to reduce representation bias.

\noindent\textbf{Task Formulation}~\textemdash~We evaluate production-oriented SFX variation through a two-stage protocol after lightweight adaptation on the few-shot soundbank ESC-50 \cite{piczak2015esc50}: (i) \textbf{ATA generation:} Given a reference audio clip and its corresponding class label, each model generates variants, which are then evaluated comprehensively. (ii)~\textbf{Method-specific analysis:} Selected methods are further analyzed in their primary audio editing setting to highlight their strengths, including SFX morphing, inpainting, temporal alignment, and targeted editing. 
%To support a faithful analysis, this section introduces a short diagnostic evaluation framework for audio editing methods applied to the generation of variants given an audio reference.
%As audio evaluation requires both objective and subjective validation, 
%The framework targets the criteria most relevant to production-oriented pipelines, namely the ability to generate high-quality and diverse variations from one or a few references while preserving event identity. The study follows a two-stage design after fine-tuning on the ESC-50 benchmark

\noindent\textbf{Dataset}~\textemdash~Experiments use ESC-50, with 2,000 five-second recordings across 50 classes. We use the official split, with folds 1-3 for training, fold 4 for validation, and fold 5 for testing, yielding 8 clips per class per fold. This low-data setting approximates practical adaptation of pretrained models to limited SFX soundbanks.
%Experiments are conducted on the ESC-50 benchmark \cite{piczak2015esc50}, which comprises 2,000 environmental recordings of 5\,s duration distributed across 50 semantic classes. 
Although relatively small in scale, the ESC-50 setting reflects practical deployment scenarios in which pretrained models are fine-tuned on limited subsets of target classes rather than trained from scratch on large-scale datasets.

\noindent\textbf{Fine-Tuning Protocol}~\textemdash~Fine-tuning is performed on top of each model's initial pretraining, with particular attention to 
%designing a lightweight adaptation that meets production requirements while 
preserving each model's original setup and framework. %Several baselines already use ESC-50 \cite{audit2023, smartdj2025, liu2023audioldm2}. 
All baselines are initialized from their released pretrained checkpoints.
%All baselines are initialized from their original pretrained checkpoints: AudioLDM from \emph{audioldm-m-full} \cite{liu2023audioldm}, T-Foley from the released pretrained model \cite{chung2024tfoley}, ThinkSound from the original checkpoints \cite{liu2025thinksound}, AudioX from the HuggingFace release \cite{audiox2025, hkustaudio_audiox_hf_2026}, and A$^2$SB from the released masking-split checkpoints \cite{kong2025a2sb}. 
%In all cases, fine-tuning is kept deliberately light to adapt each method to ESC-50 while preserving the behavior of the original model. 
Fine-tuning is deliberately lightweight to adapt each method to ESC-50 while preserving its native behavior. Encoders are kept frozen when applicable, and only task-relevant generation or conditioning layers are adapted; method-specific details are reported in our web page\textsuperscript{\ref{fn:web}}.%Appendix ~\ref{app:training_details}.
%Detailed hyperparameters and per-model training schedules are provided in the supplementary material, Section ~\ref{sec:appendix}. 

\noindent\textbf{ATA Generation Protocol}~\textemdash~For each of the 400 ESC-50 test references, we generate N = 10 variants, yielding 4,000 outputs per model. Each method receives the waveform reference and class name as minimal conditioning. %The ESC-50 test fold contains 50 classes, with 8 reference clips per class. For each reference, $N=10$ variants are generated, yielding 4\,000 outputs per model. In the shared ATA generation setting, each model receives the waveform reference clip together with the class name as minimal textual conditioning. 
ATA generation is implemented according to each baseline's native design: T-Foley uses reference-derived temporal features and RMS-envelope conditioning; ThinkSound and AudioX sample from a noisy reference latent or waveform, respectively, with class-name conditioning; and A$^2$SB is evaluated in its native inpainting regime by masking a segment of the reference and sampling multiple restorations from the same corrupted input. %In the main inpainting experiments, masked durations of 0.3\,s to 1.0\,s are used, with longer masks considered separately for robustness analysis.
Table~\ref{tab:audio_setting} reports inference time and number of generated variants per hour for each method, serving as a practical efficiency indicator (R9). Results in Tab.~\ref{tab:audio_setting} are diagnostic rather than directly comparable, since methods use different backbones, batch sizes, and sampling steps.

\label{subsec:method_spe_protocol}
\noindent\textbf{Method-specific analysis Protocol}~\textemdash Because the selected baselines differ in their conditioning signals and editing scope, we complement the shared ATA framework with a capability analysis on each method. These analyses are diagnostic and are not intended for direct cross-method comparison. For A$^2$SB, we evaluate localized inpainting by comparing short masks of 0.3-1.0,s with longer masks of 0.3-2.0,s, using metrics computed exclusively on the inpainted regions. For T-Foley, we assess the effect of its restricted pretraining manifold (only seven classes, e.g., \emph{coughing}, \emph{dog}, \emph{footsteps}, \emph{keyboard\_typing}, and \emph{rain}) by separating ESC-50 classes that overlap with its native training domain from the remaining unseen classes. For AudioLDM and AudioX, we evaluate SFX morphing using reference-to-target transformations under different noise levels $\sigma$, which control the strength of the transformation. For ThinkSound, we evaluate object-centric editing on masked regions using attenuation, enhancement, and reverberation instructions. These native analyses preserve each model's intended control mechanism while complementing the requirement-level interpretation of the shared framework.

%%%%%%%%%%%%%%%
\subsection{Evaluation Metrics and Listening Study}
Objective evaluation is performed on cropped 4\,s clips for all
methods. Distributional audio quality and realism (R1) are measured
with Fr\'echet Audio Distance (FAD) using the AudioLDM-eval
implementation~\cite{liu2023audioldm_eval}, computed per class and pooled
across classes.

\noindent \textbf{Reference alignment and diversity}~\textemdash~We use ImageBind audio
embeddings~\cite{imagebind2023} for identity preservation (R2) and
diversity (R3), since the shared audio-to-audio (ATA) protocol is
reference-conditioned rather than text-driven: an audio-text metric
such as CLAP would mainly reflect agreement with the class label
rather than preservation of the specific reference event. Let
$\hat{z}^{\text{ref}}_r$ and $\hat{z}^{\text{gen}}_{r,v}$ be the
$\ell_2$-normalised embeddings of reference $r$ and its variant $v$.
Alignment is the reference-variant cosine similarity
%$A_{r,v}=\langle \hat{z}^{\text{ref}}_r, \hat{z}^{\text{gen}}_{r,v}\rangle$,
$A_{r,v}=\mathrm{cos\_sim}(\hat{z}^{ref}_r,\hat{z}^{gen}_{r,v})= \frac{\langle \hat{z}^{ref}_r,
\hat{z}^{gen}_{r,v} \rangle}
{\|\hat{z}^{ref}_r\|_2 \, \|\hat{z}^{gen}_{r,v}\|_2}$
and diversity $D_r$ is the mean pairwise cosine distance
$D_r=\frac{2}{V_r(V_r-1)}
\sum_{1\leq i<j \leq V_r}\left(
d_{\cos}
\left(
\hat{z}^{\mathrm{gen}}_{r,i},
\hat{z}^{\mathrm{gen}}_{r,j}
\right)
\right)$, where $d_{\cos}(\cdot, \cdot)=1 - \mathrm{cos\_sim}\left( \cdot,\cdot \right)$ and $V_r{=}10$ denotes the variants of a reference,
following~\cite{xing2024seeing}. High diversity must be read jointly
with FAD and alignment, as a large spread may also indicate semantic
drift.

\noindent \textbf{Transient diagnostics}~\textemdash~Temporal behaviour (R4--R5) is assessed
from a normalised onset-strength envelope $\tilde{e}_t$, built by
summing positive log-mel increments over mel bins and rescaling to
$[0,1]$, so the comparison reflects transient shape rather than
loudness. Peaks are detected with a 40\,ms minimum spacing and
prominence $\rho{=}0.10$, yielding three measures: (i)~\emph{FWHM ratio},
the generated-to-reference ratio of median peak full-width at
half-maximum ($\to 1$ ideal; ${>}1$ smeared, ${<}1$ sharpened);
(ii)~\emph{pre-onset~$\Delta$}, the difference in normalised
pre-transient energy ($40$\,ms before vs.\ $80$\,ms after the peak;
$\to 0$ ideal, positive values indicate pre-echo); and (iii)~\emph{onset
error}, the median timing gap between matched reference and generated
onsets (tolerance $\delta_{\max}{=}150$\,ms). All detection parameters
are fixed across methods. 

\noindent \textbf{Listening study}~\textemdash~To assess perceptual
identity (R1--R2), the Similarity Mean Opinion Score (S-MOS) is rated by
15 participants on a 1--5 identity scale over reference--variation
pairs drawn from the 4{,}000 outputs per method. Participants used
headphones in a quiet environment, could replay each pair freely, and
rated 100 randomized trials per method with the prompt: \emph{rate the
identity fidelity (1--5) as the similarity to the reference
event/source, excluding loudness.} Scores are reported with 95\% CI.

\input{src/tables/computation_inference}
%Objective evaluation is performed on cropped 4\,s clips for all methods. Distributional audio-quality and realism (R1) are assessed with FAD using the AudioLDM-eval implementation \cite{liu2023audioldm_eval}, computed per class and then pooled across classes. Identity preservation (R2) and diversity (R3) are evaluated using ImageBind audio embeddings~\cite{imagebind2023}: cosine similarity between each generated sample and its corresponding reference measures reference-level alignment in the ImageBind embedding space, while the mean pairwise cosine distance among generated variants from the same reference measures diversity. %\textcolor{blue}{ImageBind is relevant as the shared ATA setting targets reference-to-variation, with direct audio-to-audio semantic comparison possible and letting the space for further conditioning on other modalities (e.g. video for temporal alignment) possible
%text-required metrics such as CLAP \noindent A listening study is further conducted to assess perceptual identity preservation (R1 and R2). For this, S-MOS is measured on reference-variation pairs sampled from the 4,000 outputs per method and rated by 15 participants on a 1–5 identity-preservation scale. See further details on the listening study and chosen metrics in Appendix~\ref{app:eval_details}.

%% file: src/tables/requirement_small.tex
\begin{table*}[!t]
\centering
\scriptsize
\setlength{\tabcolsep}{3pt}
\renewcommand{\arraystretch}{1.08}

\caption{Production requirements for reference-guided SFX variation, with evaluation signals and typical failure modes. %Additional details in Appendix Table~\ref{tab:req_sfx_variation}.
}
\label{tab:requirement_small}
\vspace{-5pt}
\begin{tabularx}{\textwidth}{@{}p{0.16\textwidth}p{0.36\textwidth}X@{}}
\toprule
\textbf{Requirement} & \textbf{Assessment} & \textbf{Typical failure modes} \\
\midrule

\textbf{R1} ~Fidelity and realism
& FAD$\downarrow$, MOS$\uparrow$, spectrogram inspection, transient diagnosis
& Transient smearing, noise, phasiness, synthetic texture, unnatural reverberation \\
\cmidrule(l{2pt}r{2pt}){1-3}

\textbf{R2} ~Identity preservation
& S-MOS$\uparrow$, ImageBind alignment$\uparrow$
& Semantic drift, generic textures, unrelated events \\
\cmidrule(l{2pt}r{2pt}){1-3}

\textbf{R3} ~Diversity without drift
& ImageBind diversity$\uparrow$, pairwise variation analysis
& Near-duplicate outputs, limited variation, identity drift \\
\cmidrule(l{2pt}r{2pt}){1-3}

\textbf{R4} ~Temporal alignment
& Energy-curve comparison, onset timing error, FWHM ratio
& Shifted onsets, stretched/compressed events, incorrect ordering \\
\cmidrule(l{2pt}r{2pt}){1-3}

\textbf{R5} ~Energy control
& Energy-curve \& envelope comparison, pre-onset $\Delta$
& Loudness drift, distorted dynamics, poor envelope matching \\
\cmidrule(l{2pt}r{2pt}){1-3}

\textbf{R6} ~Controllability
& Comparison across control settings
& Weak control response, uncontrolled drift, over-transformation \\
\cmidrule(l{2pt}r{2pt}){1-3}

\textbf{R7} ~Targeted modification
& Local qualitative and editing analysis
& Global rewriting, boundary discontinuities, unintended off-target changes \\
\cmidrule(l{2pt}r{2pt}){1-3}

\textbf{R8} Robustness and stability
& Cross-condition comparison, failure analysis
& Unstable behavior, out-of-domain failure, inconsistent quality \\
\cmidrule(l{2pt}r{2pt}){1-3}

\textbf{R9} ~Efficiency
& Inference time and compute-cost comparison
& Excessive latency, slow sampling, impractical compute cost \\
\bottomrule
\end{tabularx}
\vspace{-5pt}
\end{table*}

%% file: src/tables/computation_inference.tex
\newcolumntype{L}[1]{>{\raggedright\arraybackslash}p{#1}}
\newcolumntype{C}[1]{>{\centering\arraybackslash}p{#1}}

\begin{table}[!t]
\vspace{-10pt}
\centering
\scriptsize
\setlength{\tabcolsep}{2.7pt}
\renewcommand{\arraystretch}{1.08}

\caption{Practical inference cost under each method's setting for generating $N=10$ variants per reference over 50 ESC-50 classes with 8 references per class (4\,000 generations per model) \cite{piczak2015esc50}.  %Inference: total runtime. Values are deployment-oriented and not a strict efficiency comparison.
}
\label{tab:audio_setting}
\vspace{-5pt}
\begin{tabularx}{\columnwidth}{
L{1.50cm}
L{1.25cm}
L{1.10cm}
C{0.55cm}
C{0.55cm}
C{0.55cm}
C{1.8cm}}
\toprule
\textbf{Method} &
\textbf{Backbone} &
\textbf{Hardware} &
\textbf{Steps} &
\textbf{Batch} &
\textbf{Inference} &
\textbf{Variants/ h} \\
\midrule

\textbf{ThinkSound} (NeurIPS'25)
& MMDiT
& A100 40GB
& 24
& 2
& 0.66\,h & $\sim$6061\\

\textbf{T-Foley} (ICASSP'24)
& Wave.\ Diff.\ + FiLM
& RTX A6000
& 250
& 16
& 28.8\,h & $\sim$139\\

\textbf{A$^2$SB} (arXiv'25)
& Attn.\ UNet
& A100 40GB
& 300
& 1
& 33\,h & $\sim$121\\

\textbf{AudioX} (ICLR'26)
& DiT
& A100 40GB
& 250
& 1
& 10\,h & 400\\

\textbf{AudioLDM} (ICML'23)
& UNet (LDM)
& RTX A6000
& 200
& 1
& 0.75\,h & $\sim$5333\\
\bottomrule
\end{tabularx}
\vspace{-5pt}
\end{table}

%% file: S6_Results.tex
\newcommand{\AudioX}{\textcolor[HTML]{9366bc}{AudioX}}
\newcommand{\AudioLDM}{\textcolor[HTML]{1e76b3}{AudioLDM}}
\newcommand{\ThinkSound}{\textcolor[HTML]{d52627}{ThinkSound}}
\newcommand{\TFoley}{\textcolor[HTML]{2b9f2b}{T-Foley}}
\newcommand{\ASB}{\textcolor{orange}{A$^2$SB}}

\section{Experimental Results}
\label{sec:results}

\input{src/tables/quantitative_comparison}

\noindent This section presents results of the proposed production-oriented evaluation framework for reference-guided SFX variation. We report the shared ATA evaluation, Pareto-inspired trade-offs, and qualitative analyses, followed by method-specific results on native capabilities not fully captured by the shared framework.
%First, the evaluation of all baselines under the shared reference-conditioned ATA generation task provides a comprehensive overview of their ability to generate SFX variants. Then, method-specific analyses are conducted to highlight native capabilities not fully captured by the shared benchmark, including style transfer, temporal control, inpainting, and targeted editing.

\subsection{ATA Generation} Table~\ref{tab:audio_quantitative_overall} reports the shared reference-conditioned ATA comparison along with transient-level diagnostics, further discussed in Sec.~\ref{subsection:qualitative_results}. We evaluate each method's ability to generate variants from a reference SFX across reference preservation, quality, diversity, and perceptual identity under common production constraints.

\noindent\textbf{\AudioX}~ emerges as the strongest full-generation baseline, providing the most balanced trade-off across audio quality, identity preservation, and human evaluation. It remains consistently competitive across the main criteria, with a low FAD of 9.34, the strongest alignment score of 0.59, and the highest S-MOS among full-generation methods (3.37) [3.08, 3.66]. Its diversity score is more restrained than AudioLDM (0.27), suggesting that \AudioX~ favors reference consistency over large variation, but without collapsing to over-fitting the reference. %Its lower diversity (0.27) suggests a stronger preference for reference consistency than large variation.

%\noindent In contrast, \AudioLDM~ reaches the highest diversity among the full-generation baselines (0.43), but this gain is associated with weaker alignment (0.39), lower perceptual identity preservation, S-MOS (2.22 [1.97, 2.48]), and poor distributional quality, FAD (20.09). This suggests that increased diversity may come at the cost of identity drift and reduced perceptual quality. \ThinkSound~presents an intermediate profile, with diversity (0.32) and alignment (0.51), but still modest perceptual identity preservation in the human study, with S-MOS (2.57 [2.25, 2.89]). Its controlled variation therefore does not fully translate into stronger perceptual fidelity in this ATA setting.

\noindent In contrast, \AudioLDM~ reaches the highest diversity among full-generation baselines (0.43), but with weaker alignment (0.39), lower S-MOS (2.22 [1.97, 2.48]), and higher FAD (20.09), suggesting identity drift. \ThinkSound~ presents an intermediate profile, with diversity (0.32), alignment (0.51), and S-MOS (2.57 [2.25, 2.89]), indicating that controlled variation does not fully translate into stronger perceptual fidelity in this ATA setting.

\noindent Although \TFoley~ explicitly conditions on temporal structure and energy, it performs poorly in the shared ATA setting, with the highest full-generation FAD (24.53), the lowest S-MOS (1.89 [1.62, 2.16]), and weak alignment (0.22), despite diversity comparable to \ThinkSound~ (0.32). This suggests that temporal-event conditioning alone is insufficient to preserve the broader perceptual identity of the reference.
\ASB~ achieves the strongest S-MOS and FAD values, but should be interpreted separately because its variations are restricted to short inpainted regions of 0.3-1.0\,s while most of the reference remains preserved. On inpainted regions, it obtains FAD (4.43), S-MOS (4.81 [4.75, 4.87]), diversity (0.28), and alignment (0.49), confirming its strength for localized repair rather than full ATA generation. Among the full-generation baselines evaluated under the shared ATA setting, results indicate that \AudioX~ provides the strongest production-oriented compromise when identity preservation and perceptual fidelity are prioritized, while \AudioLDM~ and \ThinkSound~ expose different diversity-identity trade-offs that are further analyzed in Sec.~\ref{subsec:method_specificities}.

%%%%%%%%%%%%%%%%%
\subsection{Pareto-inspired Analysis in Production Settings}

\begin{figure}[H] 
  \centering
  \includegraphics[width=\linewidth]{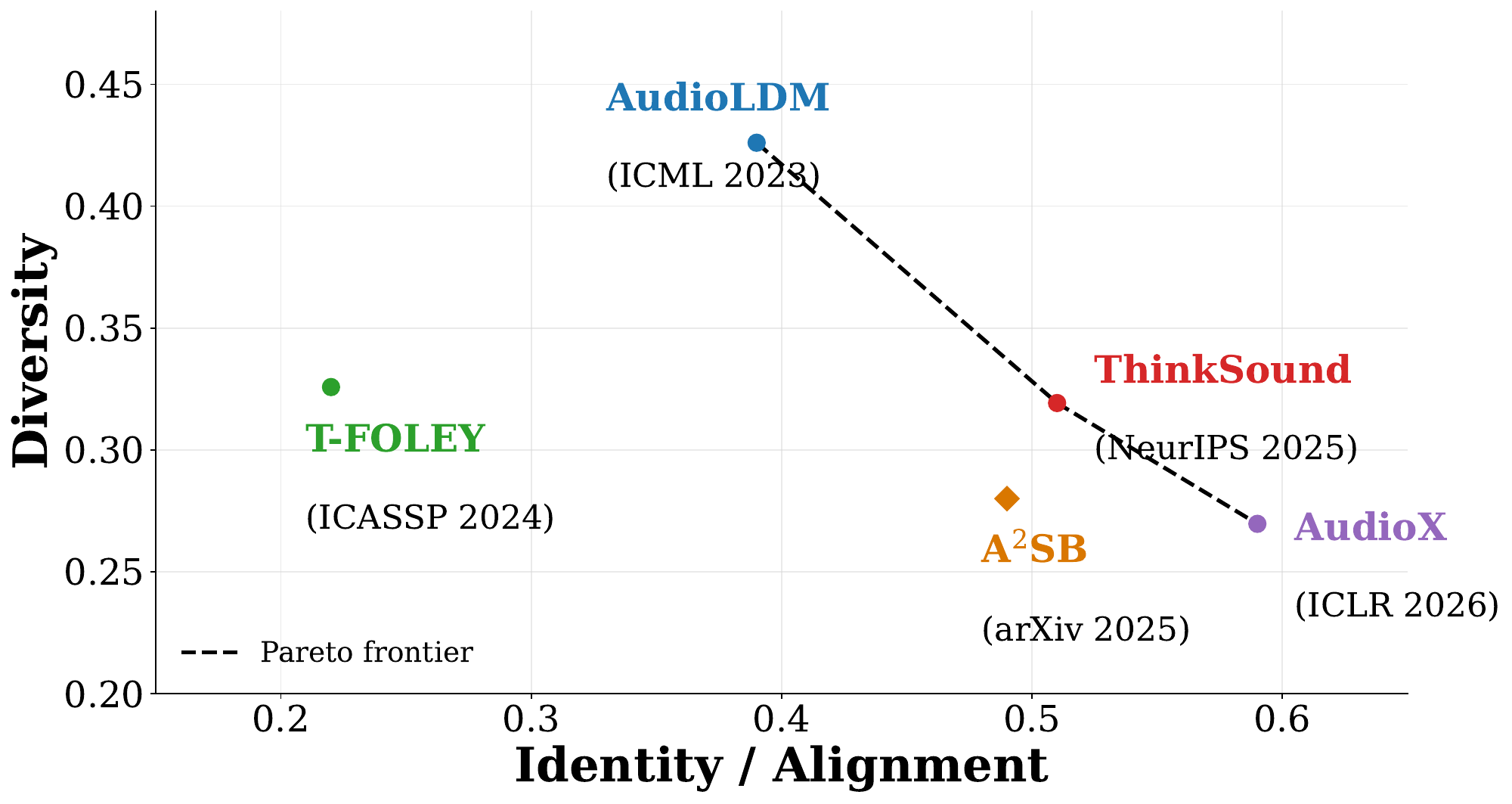}
  \vspace{-15pt}
  \caption{\textbf{Diversity–identity alignment} (R3-R2) trade-off across reference-guided SFX variation methods. Each point represents a model positioned according to its ability to preserve reference identity and generate diverse variations. Higher diversity and alignment are better.
  %Results highlight a clear trade-off, where methods achieving stronger identity alignment tend to exhibit reduced diversity, while more diverse models often compromise alignment fidelity.
  }
  \vspace{-10pt}
\label{fig:diversity-identity}
\end{figure}

\noindent Figure~\ref{fig:diversity-identity} summarizes the diversity-alignment trade-off %revealed by the proposed framework. 
\AudioLDM~ provides the highest diversity (R3: 0.43), but weaker identity alignment (R2: 0.39), while \AudioX~ reaches the strongest alignment (R2: 0.59) with lower diversity (R3: 0.27). \ThinkSound~ lies between these two regimes, with R3: 0.32 and R2: 0.51, for a moderate balance between variation and reference consistency. \TFoley~ does not occupy a favorable Pareto position, %in this shared ATA setting,
since it has diversity comparable to \ThinkSound~ (R3: 0.32), but much weaker alignment (R2: 0.22). The position of \ASB~ should again be interpreted separately, since its generation is restricted to a short inpainted segment rather than full-clip generation. Overall, the analysis confirms that higher diversity alone is insufficient when accompanied by weaker identity preservation \footnote{Further Pareto-inspired analyses are provided in our web page\textsuperscript{\ref{fn:web}}}.

%Further analyses are provided in Appendix Figure~\ref{fig:pareto_analsysis}.

%\input{src/figures/qualitative_inline}
\begin{figure*}[t!] 
  \centering
  \includegraphics[width=\linewidth]{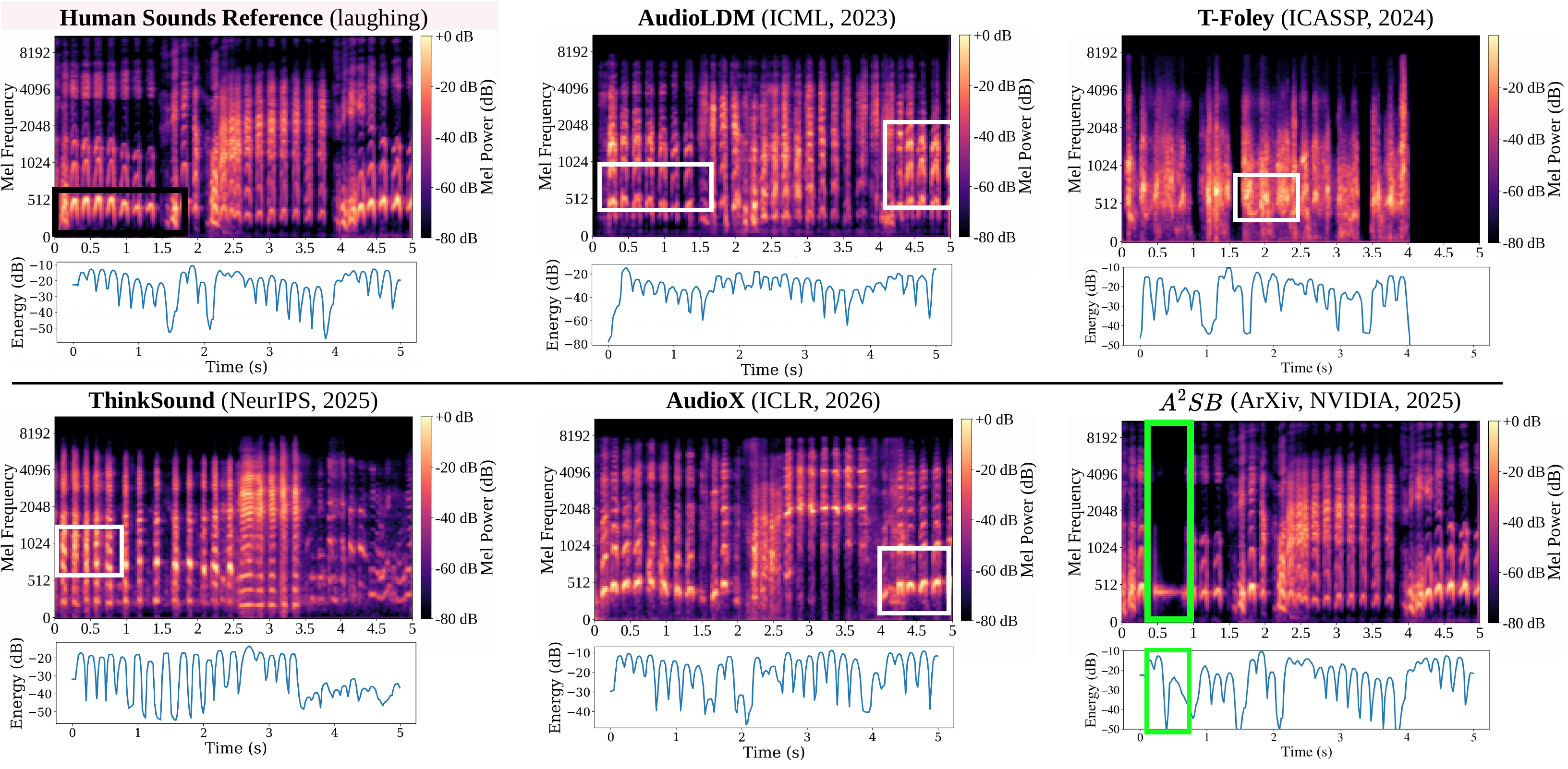}
  \vspace{-15pt}
  \caption{Qualitative comparison for the ATA variation task on the human-sound example \emph{laughing}. For \textcolor{orange}{\textbf{A$^2$SB}} inpainting, the \outlineboxxxx{masked and regenerated sections} between 0.3\,s and 1\,s are outlined. \outlineboxxxxxxxx{Similar texture to reference} is outlined in white. Top: mel-spectrogram. Bottom: energy curve. }
\label{fig:audio_laughing_melspec_energy}
\vspace{-10pt}
\end{figure*}

%%%%%%%%%%%%%%%
\subsection{Further Qualitative Analysis}
\label{subsection:qualitative_results}

Figure~\ref{fig:audio_laughing_melspec_energy} provides a visual analysis of the spectro-temporal structure and frequency behavior of the baseline generations. While the quantitative results capture overall trends in quality, alignment, and diversity, the spectrogram and energy-curve visualizations reveal additional differences in local texture preservation, temporal organization, and failure modes across methods.  In terms of spectral texture, \AudioLDM~ shows the best-preserved fine local structure, as highlighted by the white boxes, whereas \TFoley~ generations exhibit noisier textures, attenuated high-frequency content, and a reduced spectral range. \ThinkSound~ and \AudioX~ remain closer to the reference at the event level, although \ThinkSound~ shows more local spectral smoothing. The transient diagnostics support this observation: \ThinkSound~ has an FWHM ratio close to 1 but a moderate onset error (53.01\,ms), while \AudioX~ achieves the lowest onset error among full-generation methods (43.52\,ms), supporting stronger temporal organization.

\noindent These differences also appear in the temporal-energy curves. \TFoley~ sometimes follows the coarse reference energy profile, which is expected from its explicit energy conditioning, but this does not translate into better onset-level alignment, since it has the largest onset error among the full-generation baselines (63.53\,ms). This suggests that envelope following does not necessarily preserve sound texture. \ASB~ preserves the global energy evolution well because most of the reference remains unchanged, while local transient diagnostics on the inpainted region remain close to the reference, with a FWHM ratio of 0.992 and a pre-onset difference of 0.005. However, its inpainted regions may still introduce localized energy changes. Overall, the qualitative analysis complements the quantitative comparison by revealing method-specific trade-offs in local texture preservation, temporal organization, and energy behavior that are not fully captured by scalar metrics alone.

%\textcolor{blue}{Table~\ref{tab:audio_quantitative_overall} also provides a transient-level diagnostic analysis of the generated variations. Among full-length variation methods, AudioX achieves the most balanced transient preservation, with an attack-time difference close to zero, an FWHM ratio near one, and no increase in pre-onset energy. This indicates that AudioX preserves sharp onsets without introducing clear transient smearing or pre-echo artifacts. ThinkSound also produces sharp transients, as reflected by its negative attack-time difference and reduced pre-onset energy, but its higher onset error suggests weaker temporal alignment with the reference. AudioLDM exhibits mild transient softening, with a larger positive attack delay, while T-Foley shows the strongest signs of transient degradation, including onset broadening, increased pre-onset energy, and the largest onset timing error among full-length methods. A$^2$SB obtains the lowest onset error and near-neutral transient descriptors, but this is expected from its masked inpainting setting and should therefore be interpreted separately from full-length variation methods.}

%%%%%%%%%%%%%%%%%%%%%
\subsection{Method-specific Analysis}
\label{subsec:method_specificities}
Table~\ref{tab:native_diagnostics} reports compact capability-specific diagnostics under each method's native setting defined in Sec~\ref{subsec:method_spe_protocol}. These results complement the shared ATA benchmark by showing where each method is most suitable within its intended editing scope. Below, the key observations from each method are examined in greater detail.

\input{src/tables/capability_specififc_comparision}

\noindent\textcolor{orange}{\textbf{A$^2$SB}} \textbf{on inpainting}~\textemdash~Table~\ref{tab:native_diagnostics} shows that \ASB~ remains strong for local repair, but degrades as mask duration increases. Specifically, FAD rises from $4.74$ to $7.57$ and alignment stabilizes from $0.35$ to $0.31$, while diversity decreases from $0.39$ to $0.11$.  This suggests that longer masked regions are harder to reconstruct and may reduce both consistency and variation, likely because less local context remains available to guide the reconstruction. The transient diagnostic %in Appendix Table~\ref{tab:native_diagnostics_transient-diagnos}
in our web page\textsuperscript{\ref{fn:web}} further shows that both mask regimes remain temporally stable at the onset level, but a longer mask has a lower FWHM ratio ($0.89$), suggesting narrow or sharp reconstructed transients.

%Table~\ref{tab:inpainting_mask} evaluates the robustness of the inpainting pipeline under increasing masked durations, considering two regimes: short masked segments between 0.3 and 1.0 seconds, and longer segments between 0.3 and 2.0 seconds within the reference clip. Under the shorter masking regime (0.3 - 1.0\,s), the method achieves strong reconstruction fidelity, with a low FAD of 1.54 and a high ImageBind alignment of 0.959, suggesting that A$^2$SB can regenerate local audio structures while remaining consistent with the surrounding reference context. However, when the mask duration increases to 0.3 - 2.0\,s, reconstruction quality degrades noticeably: FAD rises from 1.54 to 3.92, while alignment drops from 0.959 to 0.916. This suggests the method becomes less reliable as larger portions of the signal are synthesized and less contextual information remains available to guide reconstruction. %At the same time, diversity increases slightly with longer masked segments, as the larger missing region gives the generative process more freedom and allows multiple plausible reconstructions of the masked audio segment.\\
%\input{src/tables/a2sb_inpainting}

\noindent\textcolor[HTML]{2b9f2b}{\textbf{T-Foley}} \textbf{on temporal-energy control and restricted pretraining manifold}~\textemdash~The noisy spectral texture visible in Figure~\ref{fig:audio_laughing_melspec_energy}, supports the quantitative results: although temporal-event conditioning can impose coarse structure, it does not guarantee identity preservation, and resulting variations show degraded fidelity and semantic consistency. This behavior is explained with \TFoley’s restricted pretraining manifold; Table~\ref{tab:native_diagnostics} shows substantially better within-manifold performance, with FAD improving from $25.75$ to $13.51$, S-MOS from $1.76\pm0.62$ to $3.05\pm0.78$, and alignment from $0.21$ to $0.34$, while diversity remains stable ($0.32\rightarrow0.34$). Together, these results suggest that \TFoley’s weaker ATA performance is driven more by fidelity, identity preservation, and domain-transfer limitations than by onset timing alone.

\begin{figure}[!t] 
\vspace{-10pt}
  \centering
  \includegraphics[width=\linewidth]{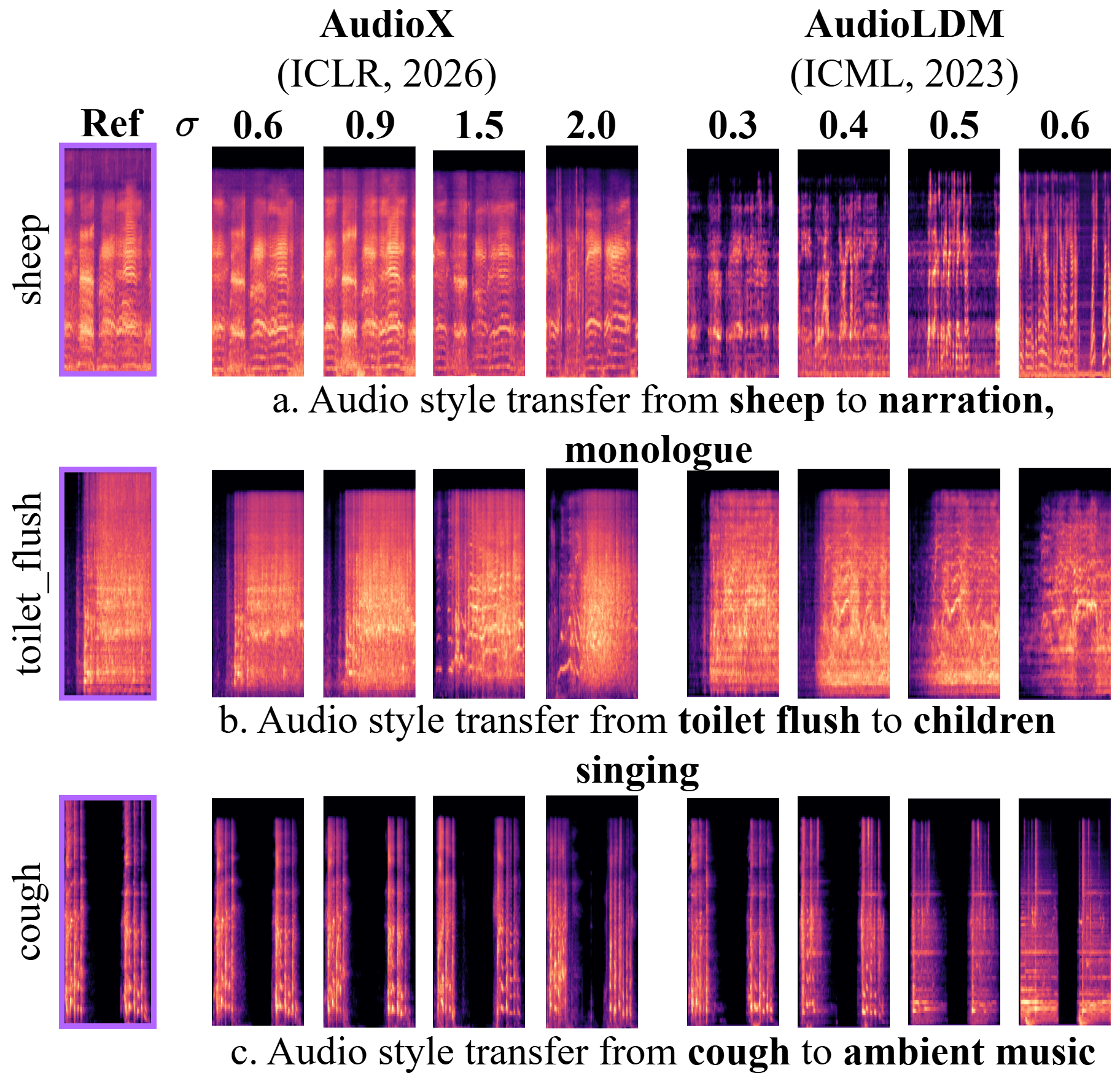}
 \vspace{-10pt}
  \caption{Ablations on \textbf{SFX Morphing task} for \textcolor[HTML]{1e76b3}{\textbf{AudioLDM}} ~\cite{liu2023audioldm} and \textcolor[HTML]{9366bc}{\textbf{AudioX}} \cite{audiox2025} on ESC-50 ~\cite{piczak2015esc50}. From left to right: the reference audio (e.g., sheep, toilet\_flush, cough) and four generated samples conditioned on the target text prompt with different initialization noise levels \textbf{$\sigma$} (transfer strength).}
\label{fig:styletransfer-melspec}
\vspace{-10pt}
\end{figure}

\noindent\textcolor[HTML]{1e76b3}{\textbf{AudioLDM}} and \textcolor[HTML]{9366bc}{\textbf{AudioX}} on \textbf{SFX Morphing Task}~\textemdash~Following the AudioLDM audio style-transfer setup \cite{liu2023audioldm} adapted to ESC-50, three examples are evaluated: toilet flush to children singing, sheep to narration/monologue, and coughing to ambient music. For smaller $\sigma$ values, the outputs remain closer to the reference, whereas larger $\sigma$ values increase adherence to the target text condition. Figure~\ref{fig:styletransfer-melspec} shows the reference audio together with generated samples obtained at different initialization noise levels $\sigma$, which control the transfer strength. At low $\sigma$, \AudioX~ preserves the source timbre and spectral structure more closely, consistent with its higher alignment ($0.77$ vs.\ $0.40$) and lower diversity ($0.10$ vs.\ $0.34$). As $\sigma$ increases, both methods trade alignment for diversity, with sharper results for AudioLDM (alignment $0.40\rightarrow0.15$; while diversity increases $0.34\rightarrow0.56$). \AudioX~ changes more progressively (alignment $0.77\rightarrow0.64$; diversity $0.10\rightarrow0.26$). The transient diagnostic in our web page\textsuperscript{\ref{fn:web}} follows the same tendency, with onset error increasing at higher $\sigma$ for both methods. This yields a clear trade-off: \AudioX~ provides smoother and more identity-preserving transformations, whereas \AudioLDM~ produces stronger transformations at the cost of weaker preservation of the original audio identity.

% For smaller \textbf{$\sigma$} values (left), the generated samples remain closer to the reference, whereas larger \textbf{$\sigma$} values produce outputs that are more strongly aligned with the text condition. Each method uses its own noise scale, consistent with its pipeline.

\noindent \textbf{\textcolor[HTML]{d52627}{ThinkSound} on object-centric editing}~\textemdash~A strength of this baseline is its ability to refine or modify sounds through user-specified regions and semantic instructions. Since the shared ATA protocol only provides restricted class-level conditioning, we further evaluate it in its native object-centric editing setting. Table~\ref{tab:native_diagnostics} shows that the edited regions remain close to the reference, with alignment ($0.64$--$0.67$) and moderate diversity ($0.19$--$0.22$).

\begin{figure}[H] 
  \centering
      \includegraphics[width=0.95\linewidth]{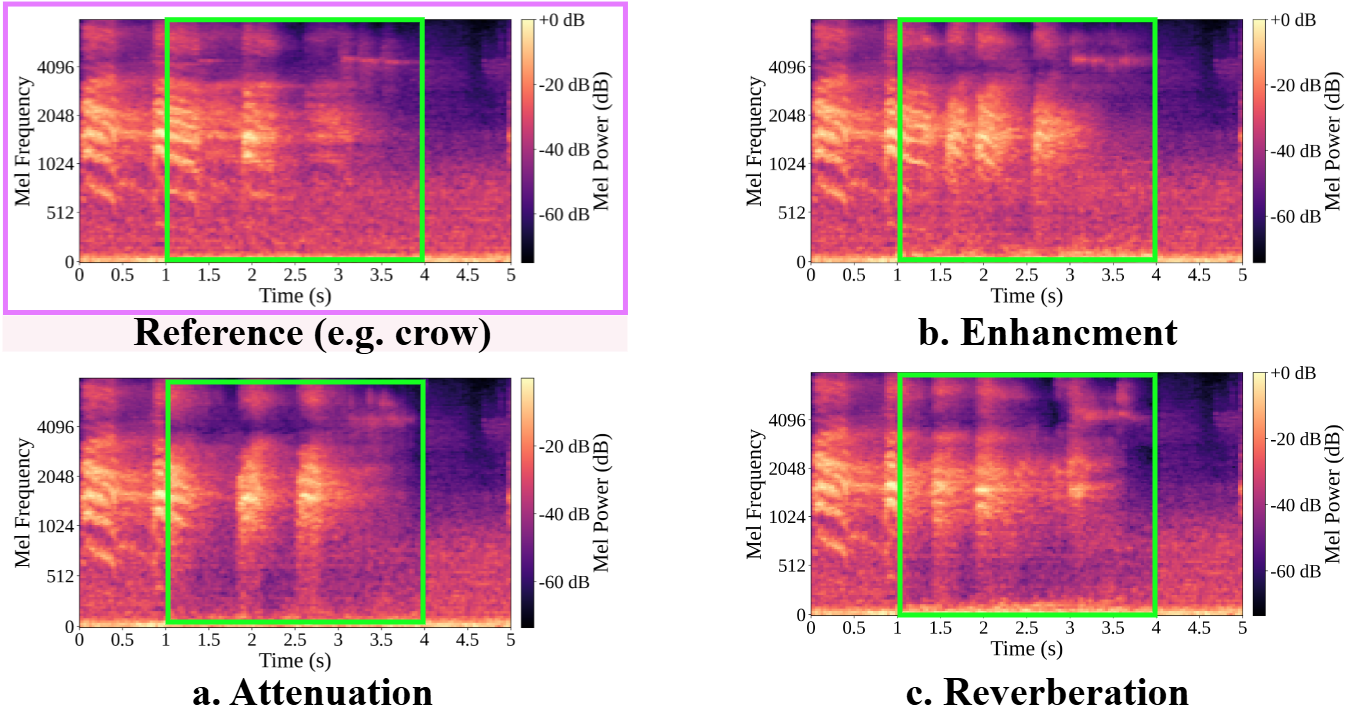}
 % \vspace{-10pt}
  \caption{Ablation of \textcolor[HTML]{d52627}{\textbf{ThinkSound}} on object-centric local editing over masked regions of the reference (e.g. \emph{crow} class from ESC-50 \cite{piczak2015esc50}). Three instructions are evaluated: a. \textcolor[HTML]{7f0000}{Attenuation}, \emph{"reduce the \{class\_name\} volume to sound far away and muffled, in the target region only."}; b. \textcolor[HTML]{d7301f}{Enhancement}, \emph{"make \{class\_name\} sound in the target region stronger and sharper."}; and c. \textcolor[HTML]{fc9272}{Reverberation}, \emph{"make \{class\_name\} event more distant and reverberant in the target region only."}.\outlineboxxxx{Masked and edited sections} [1\,s; 3\,s] are outlined. }
\label{fig:energy_compar_TS}
\vspace{-10pt}
\end{figure}

\noindent Figure~\ref{fig:energy_compar_TS} visually supports these results through light but directionally consistent fluctuations. Attenuation slightly reduces the target transient regions, enhancement increases transient strength, and reverberation introduces more diffuse repeated patterns. However, the editing scope remains limited in this setting. Since the prompts are deliberately simple and the evaluated references often contain a single dominant sound event, the model has limited semantic or acoustic structure to selectively modify. Further, the transient diagnostic in our web page\textsuperscript{\ref{fn:web}} shows reasonable local timing preservation, with onset errors between 38.3 and 43.8\,ms, although FWHM ratios above 1.1 suggest some transient broadening. As a result, \ThinkSound~ produces plausible local changes, but the edits remain relatively subtle and do not always correspond to strong, fine-grained acoustic transformations.  

%A strength of this baseline is its ability to refine or add sounds for user-specified objects through clicks or regions in the video \cite{liu2025thinksound}. In the main framework, ThinkSound may not display its best results. This is not necessarily because of limited editing capacity, but because the shared protocol provides only restricted conditioning. In this ablation, the object-centric setting is evaluated with richer semantic guidance than a simple class label, to assess whether ThinkSound can perform localized and controllable SFX editing while preserving the surrounding audio context and temporal structure. Figure~\ref{fig:energy_compar_TS} visualizes its ability to attenuate, enhance, or reverberate a target region of a given reference. As visible in the overlaid energy curves, the method largely reproduces the reference energy outside the targeted region (between 1\,s and 4\,s). Enhancement yields the clearest increase in energy, while reverberation produces a smoother and more temporally diffused profile. In contrast, attenuation remains weak, as its energy trajectory often stays above the reference. Overall, ThinkSound can modulate the reconstructed region in a prompt-dependent way, but fine-grained edit controllability remains limited in this ATA editing setting.
%Can ThinkSound modify one sound event while preserving the rest of the scene? targeted modification, controllability, temporal alignment, and context preservation

%% file: src/tables/quantitative_comparison.tex
\begin{table*}[!t]
\centering
\scriptsize
\setlength{\tabcolsep}{2pt}
\renewcommand{\arraystretch}{1.08}
\caption{Overall quantitative and transient-level diagnosis comparison of baselines for ATA on SFX variation.
FAD, S-MOS, diversity, and ImageBind-audio alignment summarize global generation quality, perceptual quality, variation diversity, and reference consistency. Transient diagnostics measure onset-local behavior using log-mel onset-strength envelopes: FWHM Ratio measures onset broadening or sharpening, Pre-onset $\Delta$ measures extra energy before the transient, and Onset Error evaluates temporal alignment. S-MOS is reported as the mean with 95\% confidence intervals over 15 raters.
%\textcolor{orange}{A$^2$SB} is an inpainting method; results are reported here only for 0.3\,s to 1\,s changes from the 5-second reference. 
\textbf{Best} and \underline{second best} on full generation methods. }
\label{tab:audio_quantitative_overall}
\vspace{-8pt}
\begin{tabularx}{\textwidth}{@{}
>{\raggedright\arraybackslash}p{0.20\textwidth}
C{0.090\textwidth}
C{0.120\textwidth}
C{0.075\textwidth}
C{0.075\textwidth}
%C{0.130\textwidth}
C{0.130\textwidth}
C{0.130\textwidth}
C{0.130\textwidth}
@{}}
\toprule

& \multicolumn{4}{c}{Overall Quantitative Results} &  \multicolumn{3}{c}{Transient Level Diagnostics} \\ 
\cmidrule(lr){2-5} \cmidrule(lr){6-8}
\textbf{Methods}
& \textbf{FAD}$\downarrow$
& \textbf{S-MOS}$\uparrow$
& \textbf{Div.}$\uparrow$
& \textbf{Align.}$\uparrow$
%& \textbf{Attack $\Delta$ ms} $\rightarrow 0$
& \textbf{FWHM} \textbf{Ratio} $\rightarrow 1$
& \textbf{Pre-onset} \textbf{$\Delta$}$\rightarrow0$
& \textbf{Onset Err.} \textbf{ms}$\downarrow$ \\[2pt]
\midrule

\textcolor[HTML]{9366bc}{\textbf{AudioX}}
{\scriptsize (ICLR, 2026)}
& \textbf{9.34}
& \textbf{3.37} [3.08, 3.66]
& 0.27 %0.67 
& \textbf{0.59}
%& \textbf{-0.26}
& \underline{0.985}
& \textbf{-0.004}
& \textbf{43.52} \\[2pt]

\textcolor[HTML]{1e76b3}{\textbf{AudioLDM}}
{\scriptsize (ICML, 2023)}
& 20.09
& 2.22 [1.97, 2.48]
& \textbf{0.43} %\textbf{0.85}
& 0.39
%& 10.19
& 0.969
& \underline{-0.005}
& 60.72 \\[2pt]

\textcolor[HTML]{d52627}{\textbf{ThinkSound}}
{\scriptsize (NeurIPS, 2025)}
& \underline{16.51}
& \underline{2.57} [2.25, 2.89]
& \underline{0.32} %0.73
& \underline{0.51}
%& \underline{-5.52}
& 0.978
& -0.011
& \underline{53.01} \\[2pt]

\textcolor[HTML]{2b9f2b}{\textbf{T-Foley}}
{\scriptsize (ICASSP, 2024)}
& 24.53
& 1.89 [1.62, 2.16]
& \underline{0.32}%\underline{0.77} 
& 0.22
%& 6.83
& \textbf{1.011}
& 0.008
& 63.53 \\[2pt]

\midrule

\textbf{\textcolor{orange}{A$^2$SB}$^*$}
{\scriptsize (ArXiv, NVIDIA, 2025)}
& 4.43
& 4.81 [4.75, 4.87]
& 0.28 %0.69 
& 0.49
%& 8.01
& 0.992
& 0.005
& 54.79 \\

\bottomrule
\multicolumn{8}{@{}p{\dimexpr\textwidth-2\tabcolsep\relax}@{}}{\scriptsize
$^*$ All reported objective metrics are computed on cropped 4\,s clips for all methods, except for \textcolor{orange}{A$^2$SB} where only inpainted regions are evaluated. }
\end{tabularx}
\vspace{-8pt}
\end{table*}

%% file: src/tables/capability_specififc_comparision.tex
\begin{table}[H]
\centering
\scriptsize
\setlength{\tabcolsep}{2pt}
\renewcommand{\arraystretch}{1.03}

\caption{Capability-specific diagnostics under each method’s native setting. Results are diagnostic and not intended for direct cross-method comparison.}
\label{tab:native_diagnostics}
\vspace{-5pt}

\begin{tabularx}{\columnwidth}{@{}>{\raggedright\arraybackslash}Xcccc@{}}
\toprule
\textbf{Methods} & \textbf{FAD}$\downarrow$ & \textbf{S-MOS}$\uparrow$ & \textbf{Div.}$\uparrow$ & \textbf{Align.}$\uparrow$  \\
\midrule

\multicolumn{5}{@{}p{\dimexpr\columnwidth-2\tabcolsep\relax}@{}}{\scriptsize \textbf{\textcolor{orange}{A$^2$SB} Inpainting} (evaluated on 3 classes)} \\
\midrule 
Mask $0.3$--$1.0$\,s &4.74 & -- & 0.39 & 0.35  \\
Mask $0.3$--$2.0$\,s & 7.57 & -- &0.11 & 0.31 \\

\midrule
\multicolumn{5}{@{}p{\dimexpr\columnwidth-2\tabcolsep\relax}@{}}{\scriptsize \textbf{SFX morphing} (evaluated on 3 classes)} \\
\midrule 
\textcolor[HTML]{9366bc}{\textbf{AudioX}} ~~~~~~ $\downarrow\sigma$ / $\uparrow\sigma$ & 4.57 / 6.64 & -- & 0.10 / 0.26 & 0.77 / 0.64 \\
\textcolor[HTML]{1e76b3}{\textbf{AudioLDM}} ~$\downarrow\sigma$ / $\uparrow\sigma$ & 18.97 / 24.09 & -- & 0.34 / 0.56 & 0.40 / 0.15\\

\midrule

\multicolumn{5}{@{}p{\dimexpr\columnwidth-2\tabcolsep\relax}@{}}{\scriptsize \textbf{\textcolor[HTML]{2b9f2b}{T-Foley} Restricted pretraining manifold}} \\
\midrule 
Seen classes (5 classes) & 13.51 & 3.05$\pm$0.78 & 0.34 & 0.34  \\
Unseen classes (45 classes) & 25.75 & 1.76$\pm$0.62 & 0.32 & 0.21 \\

\midrule
\multicolumn{5}{@{}p{\dimexpr\columnwidth-2\tabcolsep\relax}@{}}{\scriptsize \textcolor[HTML]{d52627}{\textbf{ThinkSound}} \textbf{Object-centric region editing} (evaluated on 5 classes)} \\
\midrule 
\textcolor[HTML]{7f0000}{Attenuation} Mask $1.0$--$4.0$\,s & 9.06 & -- & 0.19 & 0.67 \\
\textcolor[HTML]{d7301f}{Enhancement} Mask $1.0$--$4.0$\,s & 9.30 & -- & 0.21 & 0.64 \\
\textcolor[HTML]{fc9272}{Reverberation} Mask $1.0$--$4.0$\,s & 8.97 & -- & 0.22 & 0.64\\
%\textcolor[HTML]{7f0000}{Attenuation} & 3.29 & -- & 0.41 & 0.80 \\
%\textcolor[HTML]{d7301f}{Enhancement} & 3.33 & -- & 0.45 & 0.80 \\
%\textcolor[HTML]{fc9272}{Reverberation} & 3.12 & -- & 0.45 & 0.79 \\

\bottomrule
\end{tabularx}
\vspace{5pt}
\end{table}

%% file: S8_Conclusion.tex
\section{Discussion and Conclusion} \label{sec:conclusion}

%The proposed evaluation framework shows that production-ready SFX variation cannot be reduced to a single dominant solution. Among full-generation baselines evaluated under the shared ATA setup, \textcolor[HTML]{9366bc}{AudioX} provides the strongest overall compromise for reference-guided SFX variation. In contrast, \textcolor[HTML]{1e76b3}{AudioLDM} favors stronger diversity and stylization, \textcolor[HTML]{2b9f2b}{T-Foley} remains specialized in temporal guidance, \textcolor{orange}{A$^2$SB} is most relevant for localized inpainting and repair, and \textcolor[HTML]{d52627}{ThinkSound} shows more targeted editing potential when richer semantic guidance is provided. These results emphasize that evaluating production-ready SFX variation requires more than generic audio-quality benchmarks. A suitable pipeline must jointly balance realism, identity preservation, controllable variation, and practical usability. By providing a requirement-driven baseline selection, a shared benchmark, and complementary capability-specific analyses, this work offers a structured basis for evaluation of reference-guided audio generation and editing systems. More broadly, it points to a clear next step for model design: unified full-reference variation with explicit control, localized editing, and efficient deployment within a single framework.

Production-ready SFX variation requires an evaluation protocol able to compare heterogeneous generation and editing methods under shared production requirements. By combining a common reference-guided ATA task with capability-specific analyses, our framework makes these trade-offs explicit rather than reducing them to a single aggregate score. Under the shared reference-guided ATA generation setting, the baselines reveal complementary strengths across production needs. Among the full-generation methods, \AudioX~ provides the strongest overall compromise between fidelity, identity preservation, diversity, and reference alignment. In contrast, \AudioLDM~ remains more suitable for stronger stylization and higher variation, \TFoley~ for explicit temporal and energy control, \ASB~ for localized inpainting and repair, and \ThinkSound~ for targeted editing when richer semantic guidance is available. Thus, from a production perspective, no single baseline dominates: AudioX offers the strongest overall compromise among full-generation methods, while the most suitable choice ultimately depends on workflow priorities.

\noindent Heterogeneous baselines indicate that production-ready SFX variation can't be assessed through generic audio-quality benchmarks alone. Our framework addresses this gap through requirement-driven baseline selection, a shared ATA framework, and complementary method-specific analyses, enabling a more structured comparison while preserving native baseline strengths. Several limitations remain, including heterogeneous pretrained setups and the restricted scope of ESC-50 relative to real production soundbanks. A suitable pipeline must jointly balance realism, identity preservation, controllable variation, and practicality. More broadly, this work points to future steps for model design: unified full-reference variation with explicit control, localized editing, and efficient deployment within one framework.\\

\vspace{-5pt}
%\section{Acknowledgments}
\noindent \textbf{Acknowledgments:} 
%\vspace{-5pt}
This work was supported by the Natural Sciences and Engineering Research Council of Canada, with additional computational resources provided by the Digital Research Alliance of Canada. We thank Dr. Hugo Seuté for his valuable support, the La Forge R\&D department and the Alice sound team at Ubisoft for contributing to the problem formulation and insights in defining the production requirements.

%% file: S9_Appendix.tex
\clearpage
\section{Appendix}
\label{sec:appendix}
All material described in this appendix is available on the accompanying web page\textsuperscript{\ref{fn:web}}.

\subsection{Capability profile of evaluated baselines}

\input{src/tables/capabilty_profil_compressed}

\subsection{Additional details on Evaluated Framework}
\label{app:eval_details}

\input{src/tables/method_requirement_map}

\input{src/tables/requirement_table}

\subsubsection{Objective Evaluation} 
\label{sec:obj_eval}
\noindent Following prior audio-generation evaluations, \textbf{ImageBind} \cite{imagebind2023} is used for alignment \cite{audiox2025,liu2025thinksound, cheng2024mmaudio} and diversity. This choice is motivated by the shared ATA generation protocol, which is centered on reference-conditioned audio variation rather than text-prompt adherence. Since all methods receive only restricted class-level text, an audio-text metric such as CLAP \cite{clap} may mainly reflect agreement with the class label rather than preservation of the specific reference event. In contrast, ImageBind allows direct audio-to-audio comparison in a common multimodal embedding space.

\vspace{3mm}
\noindent\textbf{ImageBind Alignment and Diversity}~\textemdash~We compute audio embeddings using ImageBind \cite{imagebind2023} audio encoder; $z^{ref}_r$ and $z^{gen}_{r,v}$ are the embeddings of reference $r$ and its generated variants $v$.
First, embeddings are $l_2$-normalised.

\vspace{3mm}
\noindent\textbf{Alignment $A_{r,v}$} is computed as the cosine similarity \cite{audiox2025,liu2025thinksound, cheng2024mmaudio} between each variant $v$ and its reference $r$:
\begin{equation}
A_{r,v}=
 \mathrm{cos\_sim}(\hat{z}^{ref}_r,
\hat{z}^{gen}_{r,v})
=
\frac{\langle \hat{z}^{ref}_r,
\hat{z}^{gen}_{r,v} \rangle}
{\|\hat{z}^{ref}_r\|_2 \, \|\hat{z}^{gen}_{r,v}\|_2}.
\end{equation}

Although ImageBind is mainly used for alignment, we follow the Seeing and Hearing \cite{xing2024seeing} method to estimate diversity through the ImageBind-space distance.

\vspace{3mm}
\noindent\textbf{Diversity $D_r$} is computed for each reference as the average pairwise semantic distance between normalised embeddings of its generated variants:
\begin{equation}
\begin{split}
D_r 
&= \frac{2}{V_r(V_r-1)}
\sum_{1\leq i<j \leq V_r}\left(
d_{\cos}
\left(
\hat{z}^{\mathrm{gen}}_{r,i},
\hat{z}^{\mathrm{gen}}_{r,j}
\right)
\right) \\
&=\frac{2}{V_r(V_r-1)}
\sum_{1\leq i<j \leq V_r}
\left(
1 - \mathrm{cos\_sim}
\left(
\hat{z}^{\mathrm{gen}}_{r,i},
\hat{z}^{\mathrm{gen}}_{r,j}
\right)
\right) \\
&= \frac{2}{V_r(V_r-1)}
\sum_{1\leq i<j \leq V_r}
\left(
1 -
\frac{
\left\langle
\hat{z}^{\mathrm{gen}}_{r,i},
\hat{z}^{\mathrm{gen}}_{r,j}
\right\rangle
}{
\|\hat{z}^{\mathrm{gen}}_{r,i}\|_2
\,
\|\hat{z}^{\mathrm{gen}}_{r,j}\|_2
}
\right).
\end{split}
\end{equation} with the number of variants per reference $V_r=10$ and $\|\hat{z}^{\mathrm{gen}}_{r,i}\|_2\,\|\hat{z}^{\mathrm{gen}}_{r,j}\|_2 = 1$ since embeddings are already $l_2$-normalized. To compute diversity, we follow prior work \cite{xing2024seeing}, which measures semantic distance in ImageBind space as $d(x,y)=1-cos\_sim(x,y)$. We extend the same distance to measure pairwise cosine distance among generated variants. This allows quantifying their spread among variations in the ImageBind embedding space. This metric should be interpreted jointly with FAD and reference-alignment scores. While diversity captures the spread among generated variants, high diversity alone may also reflect semantic drift. Combining these metrics provides a more balanced view of generation quality, identity preservation, and diversity among generated variations.
\vspace{3mm}

To sum up, ImageBind alignment measures reference-variant similarity, while ImageBind diversity measures variation among generations from the same reference. These metrics capture semantic alignment and embedding-level diversity, whereas temporal synchronization and transient preservation are evaluated separately using signal-level onset diagnostics.

\vspace{3mm}
\noindent \textbf{Transient-level Diagnosis} is computed through onset-level descriptors from the log-mel spectrogram of each reference and generated clip. 
Let $\ell_{m,t}$ denote the log-mel spectrogram, where $m$ is the mel-frequency bin and $t$ is the time frame. 
We first build a normalized onset-strength envelope in order to focus on transient shape rather than absolute loudness. We keep only positive temporal increases in the log-mel spectrogram:
\begin{equation}
D_{m,t} = \max(\ell_{m,t+1} - \ell_{m,t}, 0),
\end{equation}
and obtain the raw onset-strength envelope by summing over mel bins:
\begin{equation}
e_t = \sum_m D_{m,t}.
\end{equation}

The envelope is then normalized to $[0,1]$:
\begin{equation}
\tilde{e}_t =
\frac{e_t - \min_t e_t}
{\max_t(e_t - \min_t e_t) + \epsilon}.
\end{equation}

This normalization makes the comparison more sensitive to onset shape and timing than to global energy differences. Transient peaks $p$ are detected from $\tilde{e}_t$ using a minimum peak distance of $40$ ms and a peak prominence threshold $\rho=0.10$.
From these peaks, we report three diagnostic metrics.

\vspace{3mm}
\noindent\textbf{Full Width at Half Maximum (FWHM) ratio}~\textemdash~For each detected peak $p$, the full width at half maximum is defined as:
\begin{equation}
\mathrm{FWHM}(p) = \omega_p \Delta_{\mathrm{ms}},
\end{equation}
where $\omega_p$ is the peak width in frames and $\Delta_{\mathrm{ms}}$ is the frame duration in milliseconds. 
For each clip, we summarize $\mathrm{FWHM}(p)$ by taking the median over all detected peaks. 
The generated-reference FWHM ratio is then:
\begin{equation}
R_{\mathrm{FWHM}} =
\frac{\mathrm{FWHM}_{\mathrm{gen}}}
{\mathrm{FWHM}_{\mathrm{ref}} + \epsilon}.
\end{equation}
A value close to $1$ indicates better transient-width preservation. 
Values above $1$ suggest broader or more smeared transients, while values below $1$ indicate sharper or narrower transients than the reference.

\vspace{3mm}
\noindent\textbf{Pre-onset $\Delta$}~\textemdash~For each peak $p$, we compare the normalized onset energy before and after the transient:
\begin{equation}
R_{\mathrm{pre}}(p) =
\frac{
\sum_{t=\max(0,p-q_{\mathrm{pre}})}^{p-1} \tilde{e}_t
}{
\sum_{t=p}^{\min(T,p+q_{\mathrm{post}})-1} \tilde{e}_t + \epsilon
},
\end{equation}
where $q_{\mathrm{pre}}$ and $q_{\mathrm{post}}$ correspond to $40$ ms and $80$ ms windows, respectively. 
For each clip, $R_{\mathrm{pre}}(p)$ is summarized by the median over detected peaks. 
The reported pre-onset difference is:
\begin{equation}
\Delta_{\mathrm{pre}} =
R_{\mathrm{pre,gen}} - R_{\mathrm{pre,ref}}.
\end{equation}

Values close to $0$ indicate similar pre-onset behavior. 
Positive values indicate extra energy before the transient, which may correspond to pre-echo or early energy leakage.

\vspace{3mm}
\noindent\textbf{Onset error}~\textemdash~Let $T_{\mathrm{ref}}=\{\tau_i^{\mathrm{ref}}\}$ and $T_{\mathrm{gen}}=\{\tau_j^{\mathrm{gen}}\}$ be the detected onset times of the reference and generated clip. 
For each reference onset, we match the nearest unused generated onset:
\begin{equation}
j^*(i) =
\arg\min_{j \in U}
\left|
\tau_j^{\mathrm{gen}} - \tau_i^{\mathrm{ref}}
\right|,
\end{equation}
where $U$ is the set of unmatched generated onsets. 
A match is accepted only if the distance is below the tolerance threshold $\delta_{\max}=150$ ms. 
For accepted matches, the onset error is;
\begin{equation}
E_{\mathrm{onset}} =
1000 \cdot
\mathrm{median}_i
\left|
\tau_{j^*(i)}^{\mathrm{gen}} - \tau_i^{\mathrm{ref}}
\right|.
\end{equation}

Lower values indicate better temporal alignment between reference and generated events.

\vspace{3mm}
All transient-detection parameters are fixed across methods to ensure a consistent diagnostic protocol. 
We set the minimum peak distance to $40\,\mathrm{ms}$ to avoid counting multiple fluctuations from the same transient as separate onsets, and use a peak prominence threshold of $\rho=0.10$ to suppress weak noise peaks in the normalized onset envelope. 
For the pre-onset diagnostic, we compare a $40\,\mathrm{ms}$ window before the peak with an $80\,\mathrm{ms}$ post-onset window, capturing short energy leakage relative to the local event energy. 
For onset matching, we use a tolerance of $\delta_{\max}=150\,\mathrm{ms}$, which allows moderate timing deviations while avoiding matches between unrelated events. 
These values are fixed analysis parameters and are not tuned per method.

\subsubsection{Subjective Evaluation} 
\label{subsec:subjeval}
Further details on the subjective evaluation are provided as follows. All participants were instructed to perform the listening tests with headphones in a quiet environment. Each reference and variation pair could be replayed as many times as needed by the participant. Trials are anonymized and randomized, and scores are reported with 95\% confidence intervals. For each method, the evaluation trials consisted of 100 test sets randomly sampled from a total of 4000 reference-variation test pairs. Each test set used the same evaluation prompt: \emph{"For each trial: listen to the Reference, then the Candidate. Rate the identity fidelity (1–-5), defined as the similarity to the reference event/source (excluding loudness)."}

%clean paragraph

%add details ont ransient evaluation and formulation

\subsection{Additional Training Details}
\label{app:training_details}
\noindent All baselines are initialized from their original pretrained checkpoints: AudioLDM from \emph{audioldm-m-full} \cite{liu2023audioldm}, T-Foley from the released pretrained model \cite{chung2024tfoley}, ThinkSound from the original checkpoints \cite{liu2025thinksound}, AudioX from the HuggingFace release \cite{audiox2025}, and A$^2$SB from the released masking-split checkpoints \cite{kong2025a2sb}.
 For latent diffusion pipelines, e.g., AudioLDM and ThinkSound, the text and audio encoders are frozen, and only the diffusion backbone and conditioning projection layers are adapted. For T-Foley, semantic conditioning is kept fixed, and only the class embeddings, MLP embeddings, and FiLM conditioning layers are updated. For AudioX, fine-tuning follows the original \emph{stable-audio-tools} pipeline, with ESC-50 clips zero-padded to the model's fixed 11\,s window and conditioning restricted to the class name and the noised reference clip. For A$^2$SB, fine-tuning follows the original inpainting setup.%, with two runs matched to the pretrained masking splits and additional masked-duration experiments used to probe robustness and editing range.
 
 The ESC-50 test fold contains 50 classes, with 8 reference clips per class. For each reference, $N=10$ variants are generated, yielding 4\,000 outputs per model. In the shared ATA generation setting, each model receives the waveform reference clip together with the class name as minimal textual conditioning.
\noindent\textbf{Latent diffusion pipelines}~\textemdash~ For AudioLDM and ThinkSound, the text and audio encoders are frozen to avoid overly sharp adaptation in the few-shot setting. Only the diffusion backbone (UNet or MMDiT) and the conditioning projection layers that map conditioning embeddings into the backbone channels are fine-tuned. For ThinkSound, conditioning features are pre-extracted into a latent-directory dataset, whereas for AudioLDM the conditioning layers are explicitly adapted. ESC-50 is converted into a latent-directory dataset before feature extraction and training. ThinkSound is fine-tuned for 10 epochs of 150 steps, while AudioLDM is fine-tuned for 200 training steps. The number of updates is intentionally kept low to reduce overfitting and distribution drift.

\noindent\textbf{Waveform diffusion pipeline}~\textemdash~ For T-Foley, fine-tuning is reduced from the original 500 epochs to 25 epochs with 250 steps per epoch, for a total of 6,250 training steps. T-Foley operates at 22\,kHz. Since fine-tuning starts from pretrained checkpoints, the original sampling rate is preserved to avoid distribution mismatch.

\noindent\textbf{AudioX adaptation}~\textemdash~ AudioX is fine-tuned for 20 epochs with 200 steps per epoch, for a total of 4,000 training steps, using the original \emph{stable-audio-tools} training pipeline \cite{audiox2025}. Because AudioX operates on a fixed 11\,s window, each 5\,s ESC-50 clip is zero-padded to 11\,s and trained with a padding-mask loss, so that the diffusion objective is computed only on the real unpadded region. During training, conditioning uses only the class name of the clip, while the optional audio and video prompt modalities are provided as empty inputs to satisfy the model interface.

\noindent\textbf{A$^2$SB adaptation and masking setup}~\textemdash~ For A$^2$SB, two fine-tuning runs are performed to match the pretrained masking splits and support evaluation in the 0.3-1.0\,s masking regime. One run is initialized from the 0.0-0.5\,s checkpoint and the other from the 0.5-1.0\,s checkpoint. To further probe robustness and extend editing ability, additional fine-tuning is performed under increasing masked durations. Two masking regimes are considered: short masked segments between 0.3 and 1.0\,s, and longer masked segments between 0.3 and 2.0\,s within the reference audio.
For fine-tuning on ESC-50, waveforms are converted into STFT features following the original pipeline setup, with \texttt{n\_fft}=2048, \texttt{hop\_length}=512, and a sampling rate of 44.1\,kHz. An inpainting corruption is then applied by masking a random noisy time segment, and the model is trained to reconstruct a clean STFT from the corrupted reference.

\noindent\textbf{Sampling-rate considerations}~\textemdash~ ThinkSound operates at 44\,kHz, T-Foley at 22\,kHz, and A$^2$SB at 44.1\,kHz in its original representation. Since all methods are adapted from pretrained checkpoints, their native sampling configurations are preserved during fine-tuning rather than forcing a common 16\,kHz training setup, which could introduce distribution mismatch. Resampling to 16\,kHz is applied only at evaluation time when required for fair comparison or listening tests.

\begin{figure}[H] 
  \centering
  \includegraphics[width=\linewidth]{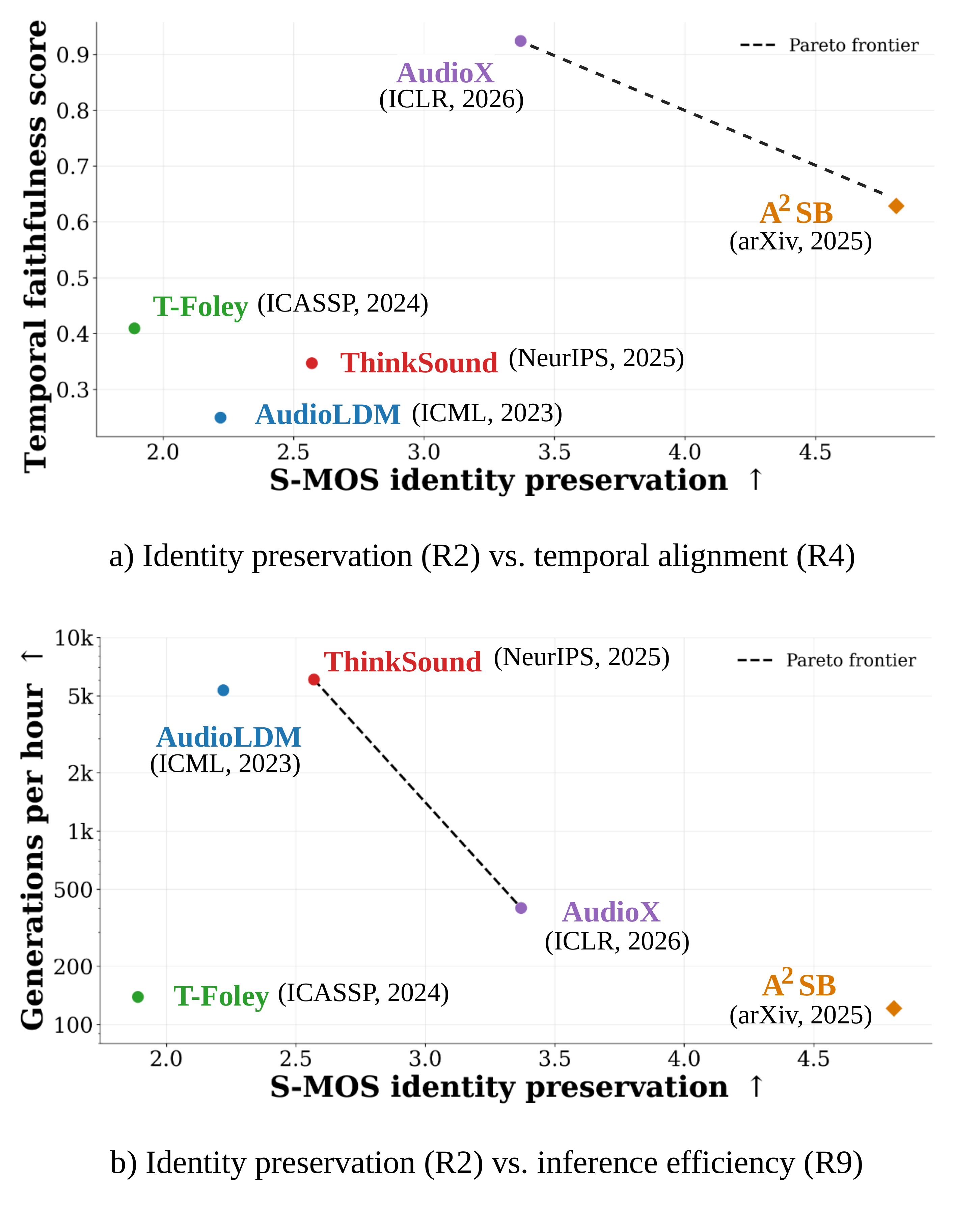}
  \caption{\textbf{Pareto-Inspired Analysis} across reference-guided SFX variation methods. We claim for \emph{pareto-inspired} as for \textbf{b)} methods may not be direct equals in inference efficiency due to different backbones, batch size and audio representations. For all evaluated values; higher is better.
  }
  \vspace{-10pt}
\label{fig:pareto_analsysis}
\end{figure}

\subsection{Additional Results and Evaluation}
\label{subsec:add_res_eval}
\noindent Table~\ref{tab:req_sfx_variation} further details the production requirements, their associated evaluation signals and diagnostics, and where they are addressed in the paper. To complement this mapping, Table~\ref{tab:req_sfx_variation_methods} provides a compact summary of each method's capability profile, highlighting how well each method fulfills the requirements for SFX variation generation. 

\noindent Figures~\ref{fig:Audio_melspec_a2a} and~\ref{fig:Audio_energycurve_a2a} provide additional qualitative evidence for the shared ATA variation task. They complement the main quantitative comparison in Figure~\ref{fig:audio_laughing_melspec_energy}, by visualizing how each baseline preserves or alters spectro-temporal texture and energy behavior across further representative classes, such as \emph{crow}, \emph{birds\_chirping}, \emph{handsaw}.

\noindent In Figure~\ref{fig:pareto_analsysis}, the Pareto-inspired analysis shows that different methods occupy distinct production trade-offs. For identity preservation vs. temporal alignment, AudioX and A$^2$SB define the frontier: AudioX provides the strongest temporal faithfulness, while A$^2$SB achieves the highest identity preservation. However, this observation should be taken with caution, as the listening test S-MOS considers the full clip, while A$^2$SB is an inpainting method and preserve unchanged regions of the reference. For identity preservation vs. efficiency, the frontier includes ThinkSound and AudioX, moving from high-throughput generation to higher identity preservation.

\input{src/tables/transient_eval_capacityspe}

\input{src/figures/melspec_overall}
\input{src/figures/energy_curve_overall}

%% file: src/tables/capabilty_profil_compressed.tex
\definecolor{capgood}{HTML}{2b9f2b}
\definecolor{capmid}{HTML}{d99a00}
\definecolor{capbad}{HTML}{d52627}
% cell helpers: 50% tint of each verdict color on white
\newcommand{\cg}{\cellcolor{capgood!50}}
\newcommand{\cma}{\cellcolor{capmid!50}}
\newcommand{\cb}{\cellcolor{capbad!50}}

\begin{table}[H]
\centering
\footnotesize
\setlength{\tabcolsep}{4pt}
\renewcommand{\arraystretch}{1.15}
\captionof{table}{Capability profile on production requirements for
reference-guided SFX variation across representative audio generation
and editing methods. Color encodes each method's suitability per
requirement: \textcolor{capgood}{native strength},
\textcolor{capmid}{supported / limited},
\textcolor{capbad}{weak / not supported}. Further details on webpage.}
\label{tab:req_sfx_variation_methods_small}

\begin{tabular}{@{}>{}p{0.45\columnwidth}*{5}{c}@{}}
\toprule
& \rotatebox{90}{\textbf{AudioX}}
& \rotatebox{90}{\textbf{AudioLDM}}
& \rotatebox{90}{\textbf{ThinkSnd.}}
& \rotatebox{90}{\textbf{T-Foley}}
& \rotatebox{90}{\textbf{A$^2$SB}} \\
\midrule
\textbf{R1} Fidelity \& realism     & \cg  & \cb  & \cma & \cb  & \cg  \\
\textbf{R2} Identity preservation   & \cg  & \cb  & \cma & \cb  & \cg  \\
\textbf{R3} Diversity without drift & \cma & \cg  & \cma & \cb  & \cma \\
\textbf{R4} Temporal alignment      & \cg  & \cb  & \cma & \cma & \cma \\
\textbf{R5} Energy control          & \cma & \cb  & \cma & \cg  & \cma \\
\textbf{R6} Controllability         & \cg  & \cma & \cma & \cb  & \cg  \\
\textbf{R7} Targeted modification   & \cb  & \cb  & \cg  & \cb  & \cg  \\
\textbf{R8} Robustness \& stability & \cg  & \cb  & \cma & \cb  & \cb  \\
\textbf{R9} Efficiency              & \cma & \cg  & \cg  & \cb  & \cb  \\
\bottomrule
\end{tabular}
\end{table}

%% file: src/tables/method_requirement_map.tex
\newcolumntype{R}{>{\raggedright\arraybackslash}p{0.16\textwidth}}
\newcolumntype{M}{>{\centering\arraybackslash}X}
\newcommand{\good}[1]{\textbf{\textcolor[HTML]{2b9f2b}{#1}}}
\newcommand{\midd}[1]{\textbf{\textcolor[HTML]{d99a00}{#1}}}
\newcommand{\bad}[1]{\textbf{\textcolor[HTML]{d52627}{#1}}}

\begin{table*}[t!]
\centering
\scriptsize
\setlength{\tabcolsep}{3pt}
\renewcommand{\arraystretch}{1.15}

\captionof{table}{Capability profile on production requirements for reference-guided SFX variation across representative audio generation and editing methods. Scores summarize the main evidence from the shared ATA benchmark and capability-specific analyses; local editing and inpainting results are diagnostic and not directly comparable to full-generation settings.}
\label{tab:req_sfx_variation_methods}

\begin{tabularx}{0.95\textwidth}{@{}R M M M M M@{}}
\toprule
\textbf{Requirement} &
\shortstack{{\textbf{AudioX}} \\ {\scriptsize (ICLR, 2026)}}&
\shortstack{{\textbf{AudioLDM}} \\ {\scriptsize (ICML, 2023)}}&
\shortstack{{\textbf{ThinkSound}} \\ {\scriptsize (NeurIPS, 2025)}}&
\shortstack{{\textbf{TFoley}} \\ {\scriptsize (ICASSP, 2024)}}  &
\shortstack{\textbf{{A$^2$SB}} \\ {\scriptsize (ArXiv, NVIDIA, 2025)}}   \\
\midrule

\textbf{R1} Fidelity \& realism
& \good{Strong full generation}; low FAD (9.34) and best full-generation S-MOS (3.37)
& \bad{Limited}; high FAD (20.09) despite plausible local texture
& \midd{Moderate}; FAD (16.51), stronger in local editing FAD (8.97--9.30)
& \bad{Weak}; highest full-generation FAD (24.53) and noisy spectra
& \good{Strong local}; FAD (4.43) on inpainted regions, not full generation \\

\cmidrule(l{2pt}r{2pt}){1-6}
\textbf{R2} Identity preservation
& \good{Strong}; best full-generation alignment (0.59) and S-MOS (3.37)
& \bad{Limited}; lower alignment (0.39) and S-MOS (2.22)
& \midd{Moderate}; alignment (0.51) in ATA, higher local alignment (0.64--0.67)
& \bad{Weak}; lowest alignment (0.22) and S-MOS (1.89)
& \good{Strong local}; S-MOS (4.81) and preserved context, but localized setting \\

\cmidrule(l{2pt}r{2pt}){1-6}
\textbf{R3} Diversity without drift
& \good{Balanced}; diversity (0.27) with strongest alignment (0.59)
& \midd{High diversity / drift}; best diversity (0.43) but weaker alignment (0.39)
& \midd{Moderate}; diversity (0.32) with alignment (0.51); local edits remain subtle
& \bad{Drift-prone}; diversity (0.32) but weak alignment (0.22)
& \midd{Localized}; diversity (0.28) in ATA, mask-size sensitive in native analysis \\

\cmidrule(l{2pt}r{2pt}){1-6}
\textbf{R4} Temporal alignment
& \good{Strongest full generation}; lowest onset error (43.52\,ms), FWHM (0.985)
& \bad{Limited}; onset error (60.72\,ms), FWHM (0.969)
& \midd{Moderate}; onset error (53.01\,ms); native edits (38.3--43.8\,ms)
& \midd{Mixed}; explicit temporal conditioning but onset error (63.53\,ms)
& \midd{Moderate local}; onset error (54.79\,ms), FWHM (0.992) \\

\cmidrule(l{2pt}r{2pt}){1-6}
\textbf{R5} Energy control
& \midd{Implicit}; preserves temporal organization but no explicit energy control
& \bad{Limited}; no explicit energy condition
& \midd{Prompt-based}; attenuation/enhancement/reverb produce light local fluctuations
& \good{Native strength}; RMS-envelope conditioning for temporal-energy control
& \midd{Context-preserving}; unmasked energy mostly preserved, not explicit control \\

\cmidrule(l{2pt}r{2pt}){1-6}
\textbf{R6} Controllability
& \good{Strong}; smooth SFX morphing with $\sigma$ control, Align. (0.77 $\rightarrow$ 0.64)
& \midd{Strong but less stable}; $\sigma$ increases variation, Div. (0.34 $\rightarrow$ 0.56), Align. (0.40 $\rightarrow$ 0.15)
& \midd{Local but limited}; semantic edits are directionally consistent but subtle under simple instructions
& \bad{Limited}; class and RMS controls, no local semantic editing
& \good{Native mask control}; editable region/duration, but longer masks degrade FAD (4.74 $\rightarrow$ 7.57) \\

\cmidrule(l{2pt}r{2pt}){1-6}
\textbf{R7} Targeted modification
& \bad{Limited}; full variation / morphing rather than local target editing
& \bad{Limited}; global style or morphing, no localized edit
& \good{Native strength}; object-centric edits with FAD (8.97--9.30), but limited fine-grained acoustic control
& \bad{Not supported}; temporal generation, no waveform-region edit
& \good{Native strength}; masked inpainting on 0.3--1.0\,s regions \\

\cmidrule(l{2pt}r{2pt}){1-6}
\textbf{R8} Robustness \& stability
& \good{Good}; stable morphing trade-off, FAD (4.57 $\rightarrow$ 6.64)
& \bad{Sensitive}; stronger $\sigma$ causes identity drop, Align. (0.40 $\rightarrow$ 0.15)
& \midd{Stable but subtle}; local edits keep Align. (0.64--0.67), but effect size depends on instruction and sound complexity
& \bad{Domain-sensitive}; seen vs.\ unseen FAD (13.51 $\rightarrow$ 25.75)
& \bad{Mask-sensitive}; longer masks reduce performance, Div. (0.39 $\rightarrow$ 0.11) \\

\cmidrule(l{2pt}r{2pt}){1-6}
\textbf{R9} Efficiency
& \midd{Moderate}; 10\,h for 4k generations, $\sim$400 variants/h
& \good{Fast}; 0.75\,h, $\sim$5333 variants/h
& \good{Fastest}; 0.66\,h, $\sim$6061 variants/h
& \bad{Slow}; 28.8\,h, $\sim$139 variants/h
& \bad{Slowest}; 33\,h, $\sim$121 variants/h \\

\bottomrule
\end{tabularx}
\end{table*}

%% file: src/tables/requirement_table.tex
\newcolumntype{Y}{>{\raggedright\arraybackslash}X}

\begin{table*}[t!]
\centering
\scriptsize
\setlength{\tabcolsep}{3pt}
\renewcommand{\arraystretch}{1.15}

\captionof{table}{Production requirements for reference-guided SFX variation.
Each requirement is paired with its operational definition, evaluation signals, reported evidence, and typical limitations or failure modes.}
\label{tab:req_sfx_variation}

\begin{tabularx}{\textwidth}{@{}p{0.035\textwidth} p{0.12\textwidth} Y Y p{0.15\textwidth} Y@{}}
\toprule
\textbf{ID} &
\textbf{Requirement} &
\textbf{Operational definition} &
\textbf{Evaluation signals / diagnostics} &
\textbf{Reported in} &
\textbf{Limitations and typical failure modes} \\
\midrule

R1 & Fidelity \& realism
& Preserve transient and spectral details while remaining perceptually plausible and free of obvious artifacts as a real-world SFX.
& FAD$\downarrow$ (plausibility-quality), MOS$\uparrow$, S-MOS$\uparrow$, spectrogram inspection \cite{liu2023audioldm, audiox2025}
& Table~\ref{tab:audio_quantitative_overall}, Fig.~\ref{fig:Audio_melspec_a2a} and Fig.~\ref{fig:audio_laughing_melspec_energy}
& FAD is distributional and should not be interpreted as direct perceptual fidelity. Typical failures include transient smearing, noise, phasiness, synthetic texture, and unnatural reverberation. \\

\cmidrule(l{2pt}r{2pt}){1-6}
R2 & Identity preservation
& Preserve the reference event class and perceptual identity \cite{fang2026acfoley, liang2025audiomorphix}.
& S-MOS$\uparrow$, ImageBind audio-reference alignment$\uparrow$ \cite{imagebind2023}
& Table~\ref{tab:audio_quantitative_overall}
& Embedding similarity captures semantic alignment but not fine acoustic details. Typical failures include semantic drift, generic textures, and unrelated events. \\

\cmidrule(l{2pt}r{2pt}){1-6}
R3 & Diversity without drift
& Generate plausible variations in texture, environment, or rendering while preserving event identity \cite{fang2026acfoley,kong2025a2sb,liang2025audiomorphix}.
& ImageBind diversity$\uparrow$ \cite{imagebind2023}, pairwise ImageBind distance, pairwise variation analysis
& Fig.~\ref{fig:diversity-identity}
& High diversity may also reflect semantic drift. Typical failures include near-duplicate outputs, limited variation, and identity drift. \\

\cmidrule(l{2pt}r{2pt}){1-6}
R4 & Temporal alignment
& Follow target temporal structure or explicit timing cues, including onset pattern, event order, and envelope shape \cite{chung2024tfoley, liu2025thinksound}.
& Energy-curve comparison \cite{chung2024tfoley}, envelope correlation, onset error, DTW on energy curves, temporal analysis \cite{liu2025thinksound}
& Fig.~\ref{fig:Audio_melspec_a2a}, Fig.~\ref{fig:audio_laughing_melspec_energy} and Fig.~\ref{fig:Audio_energycurve_a2a},  additional diagnostics in Sec.~\ref{subsection:qualitative_results} and in Tab.~\ref{tab:native_diagnostics_transient-diagnos}
& Temporal alignment is not directly measured by ImageBind similarity. Typical failures include shifted onsets, stretched or compressed events, and incorrect ordering. \\

\cmidrule(l{2pt}r{2pt}){1-6}
R5 & Energy control
& Match, preserve, or controllably modify a target loudness profile or energy trajectory \cite{chung2024tfoley}.
& Energy-curve analysis, RMS envelope comparison, energy-curve correlation, envelope comparison
& Fig.~\ref{fig:Audio_melspec_a2a}, Fig.~\ref{fig:audio_laughing_melspec_energy} and Fig.~\ref{fig:Audio_energycurve_a2a}, additional diagnostics in Sec.~\ref{subsection:qualitative_results} and in Tab.~\ref{tab:native_diagnostics_transient-diagnos}
& Energy curves provide only a coarse signal-level proxy. Typical failures include loudness drift, distorted dynamics, and poor envelope matching. \\

\cmidrule(l{2pt}r{2pt}){1-6}
R6 & Controllability
& Modulate the strength or nature of the transformation through explicit user or model controls \cite{audiox2025, liu2025thinksound, smartdj2025}.
& Control-based comparison across settings, noise-level $\sigma$ sweep, edit-strength analysis
& Fig.~\ref{fig:styletransfer-melspec}, Fig.~\ref{fig:energy_compar_TS}
& Control variables are method-specific and not always directly comparable. Typical failures include weak control response, uncontrolled drift, and over-transformation. \\

\cmidrule(l{2pt}r{2pt}){1-6}
R7 & Targeted modification
& Modify specific segments or attributes while preserving the rest of the signal \cite{liang2025audiomorphix, liu2025thinksound, kong2025a2sb}.
& Local qualitative analysis, editing evaluation, masked-region metrics, boundary analysis, local spectral distance
& A$^{2}$SB / ThinkSound diagnostics
& Localized editing is not directly comparable to full-clip generation. Typical failures include global rewriting, boundary discontinuities, and unintended off-target changes. \\

\cmidrule(l{2pt}r{2pt}){1-6}
R8 & Robustness \& stability
& Maintain consistent behavior across challenging references, controls, mask sizes, domain shifts, or generation conditions \cite{kong2025a2sb,liu2025thinksound}.
& Cross-condition comparison, failure analysis, mask-duration analysis, in-/out-manifold comparison
& Table~\ref{tab:native_diagnostics}
& Diagnostic only; does not cover all real production conditions. Typical failures include unstable behavior, out-of-domain failure, and inconsistent quality. \\

\cmidrule(l{2pt}r{2pt}){1-6}
R9 & Efficiency
& Remain practical for iterative production use in runtime, sampling cost, and deployment constraints \cite{liu2023audioldm, kong2025a2sb}.
& Inference-time comparison, runtime, sampling steps, batch size, hardware, sample rate, cost discussion
& Table~\ref{tab:audio_setting}
& Values are deployment-oriented and not fully normalized across implementations. Typical failures include excessive latency, slow sampling, and impractical compute cost. \\

\bottomrule
\end{tabularx}
\end{table*}

%% file: src/tables/transient_eval_capacityspe.tex
\begin{table}[t]
\centering
\scriptsize
\setlength{\tabcolsep}{2pt}
\renewcommand{\arraystretch}{1.03}

\caption{Capability-specific transient diagnostics under each method’s native setting. Results are diagnostic and not intended for direct cross-method comparison.}
\label{tab:native_diagnostics_transient-diagnos}
\vspace{-10pt}

\begin{tabularx}{\columnwidth}{@{}
>{\raggedright\arraybackslash}p{0.41\columnwidth}
C{0.160\columnwidth}
C{0.170\columnwidth}
C{0.170\columnwidth}
C{0.170\columnwidth}
@{}}

\toprule
\textbf{Methods} 
& \textbf{FWHM} \textbf{Ratio} $\rightarrow 1$
& \textbf{Pre-onset} \textbf{$\Delta$}$\downarrow$
& \textbf{Onset Err.} \textbf{ms}$\downarrow$ \\[2pt]

\midrule

\multicolumn{4}{@{}p{\dimexpr\columnwidth-2\tabcolsep\relax}@{}}{\scriptsize \textbf{\textcolor{orange}{A$^2$SB} Inpainting} (evaluated on 3 classes)} \\
\midrule 
Mask $0.3$--$1.0$\,s & 0.97 & -0.02 & 55.07  \\
Mask $0.3$--$2.0$\,s  & 0.89 & 0.01 & 51.99 \\

\midrule
\multicolumn{4}{@{}p{\dimexpr\columnwidth-2\tabcolsep\relax}@{}}{\scriptsize \textbf{SFX morphing} (evaluated on 3 classes)} \\
\midrule 
\textcolor[HTML]{9366bc}{\textbf{AudioX}} ~~~~~~ $\downarrow\sigma$ / $\uparrow\sigma$   & 0.95 / 0.87 &  0.007 / -0.02  &  58.61 / 67.78  \\
\textcolor[HTML]{1e76b3}{\textbf{AudioLDM}} ~$\downarrow\sigma$ / $\uparrow\sigma$  &  0.95 / 1.03  &  -0.02 / -0.03  &  64.82 / 86.78 \\

\midrule

\multicolumn{4}{@{}p{\dimexpr\columnwidth-2\tabcolsep\relax}@{}}{\scriptsize \textbf{\textcolor[HTML]{2b9f2b}{T-Foley} Restricted pretraining manifold}} \\
\midrule 
In-manifold (5 classes)& 0.97 & 0.02 & 65.28  \\
Out-manifold (45 classes)  & 1.01 & 0.007 & 63.33 \\

\midrule
\multicolumn{4}{@{}p{\dimexpr\columnwidth-2\tabcolsep\relax}@{}}{\scriptsize \textcolor[HTML]{d52627}{\textbf{ThinkSound}} \textbf{Object-centric region editing} (evaluated on 5 classes)} \\
\midrule 
\textcolor[HTML]{7f0000}{Attenuation} Mask $1.0$--$4.0$\,s & 1.13 & 0.013 & 39 \\
\textcolor[HTML]{d7301f}{Enhancement} Mask $1.0$--$4.0$\,s & 1.15 & 0.01& 38.3  \\
\textcolor[HTML]{fc9272}{Reverberation}Mask $1.0$--$4.0$\,s  & 1.12 & 0.01 & 43.8\\

\bottomrule
\end{tabularx}
\vspace{-10pt}
\end{table}

%% file: src/figures/melspec_overall.tex
\begin{figure*}[t]
\centering
\footnotesize
\setlength{\tabcolsep}{2pt}
\renewcommand{\arraystretch}{1.03}

% ---- widths ----
\newlength{\labelw}
\setlength{\labelw}{0.16\textwidth}
\newlength{\imgw}
\setlength{\imgw}{0.265\textwidth}

% ---- Reference ----
\newcommand{\refcrow}{\includegraphics[width=\imgw,height=\imgw,keepaspectratio]{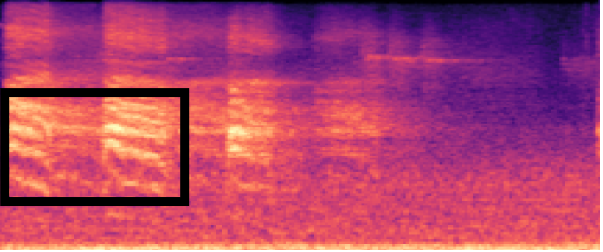}}
\newcommand{\reflaugh}{\includegraphics[width=\imgw,height=\imgw,keepaspectratio]{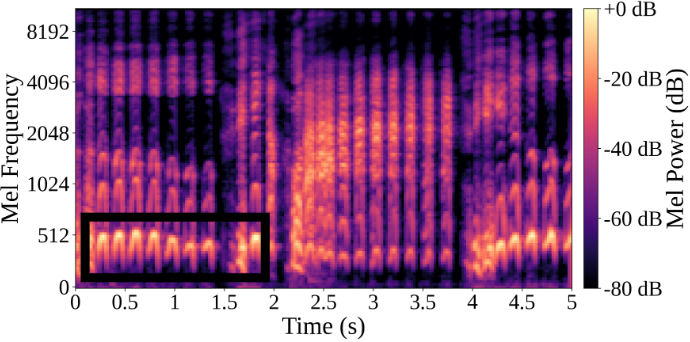}}
\newcommand{\refhandsaw}{\includegraphics[width=\imgw,height=\imgw,keepaspectratio]{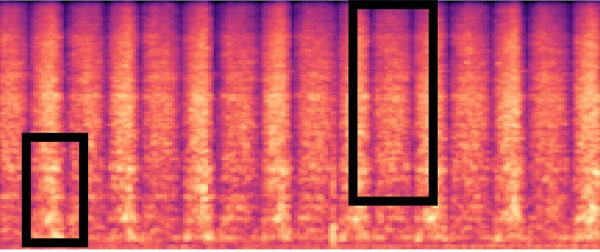}}
\newcommand{\refbirds}{\includegraphics[width=\imgw,height=\imgw,keepaspectratio]{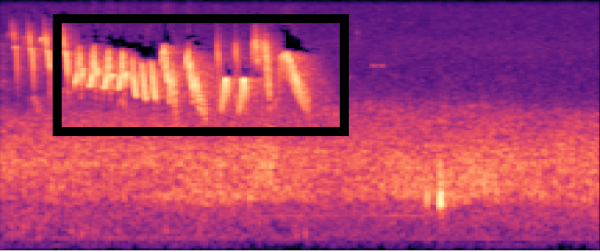}}
% ---- ThinkSound ----
\newcommand{\TScrow}{\includegraphics[width=\imgw,height=\imgw,keepaspectratio]{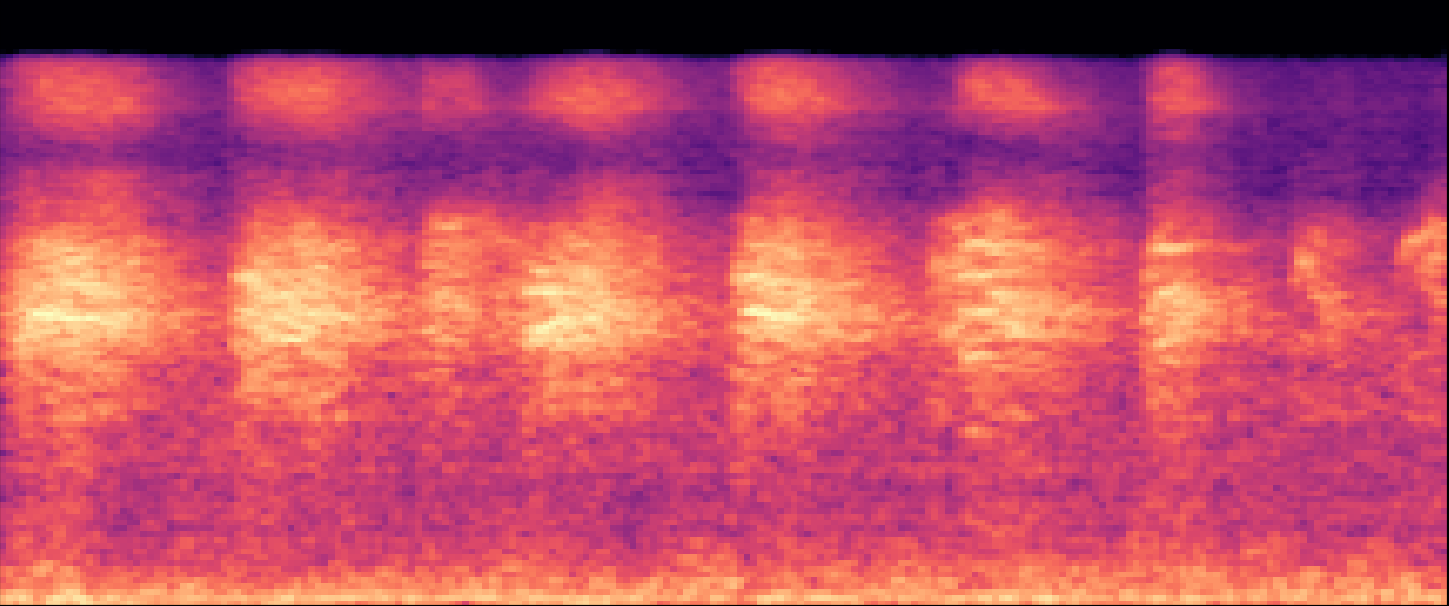}}
\newcommand{\TSlaugh}{\includegraphics[width=\imgw,height=\imgw,keepaspectratio]{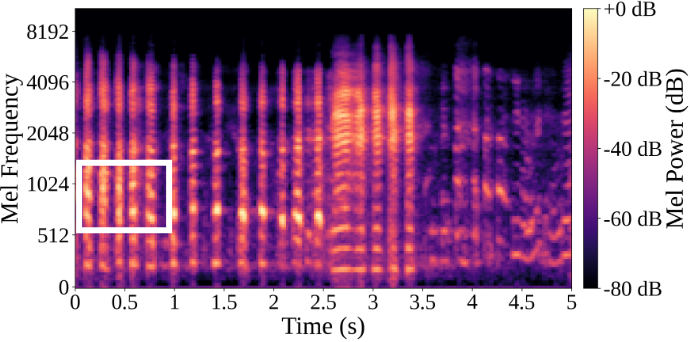}}
\newcommand{\TShandsaw}{\includegraphics[width=\imgw,height=\imgw,keepaspectratio]{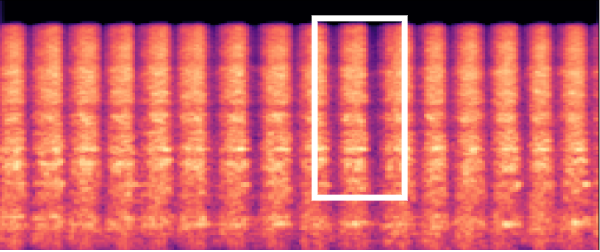}}
\newcommand{\TSbirds}{\includegraphics[width=\imgw,height=\imgw,keepaspectratio]{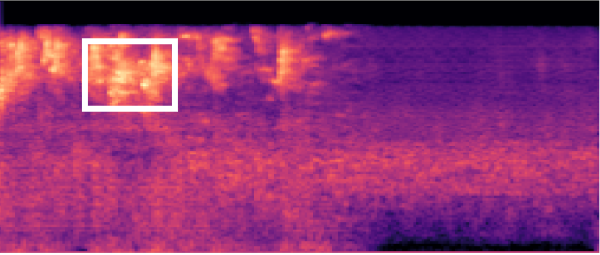}}

% ---- AudioLDM ----
\newcommand{\ALcrow}{\includegraphics[width=\imgw,height=\imgw,keepaspectratio]{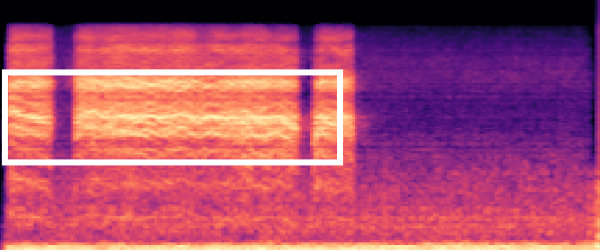}}
\newcommand{\ALlaugh}{\includegraphics[width=\imgw,height=\imgw,keepaspectratio]{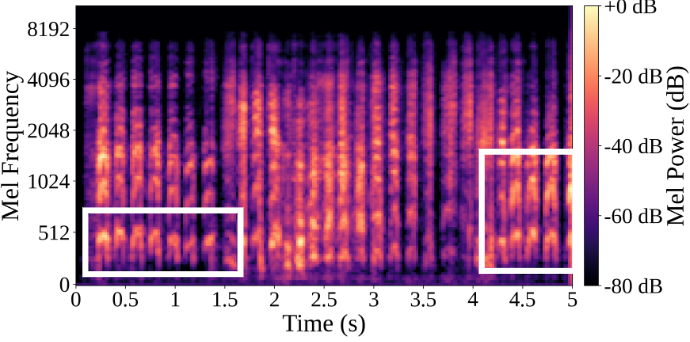}}
\newcommand{\ALhandsaw}{\includegraphics[width=\imgw,height=\imgw,keepaspectratio]{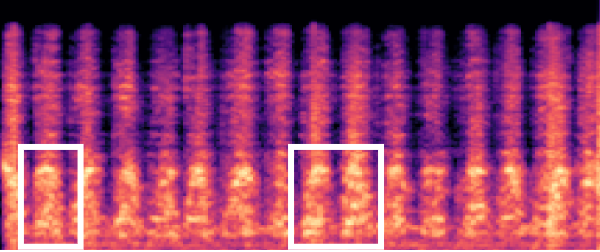}}
\newcommand{\ALbirds}{\includegraphics[width=\imgw,height=\imgw,keepaspectratio]{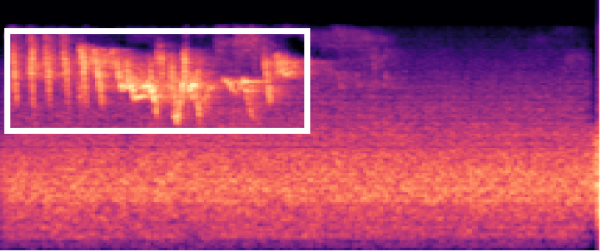}}

% ---- T-Foley ----
\newcommand{\TFcrow}{\includegraphics[width=\imgw,height=\imgw,keepaspectratio]{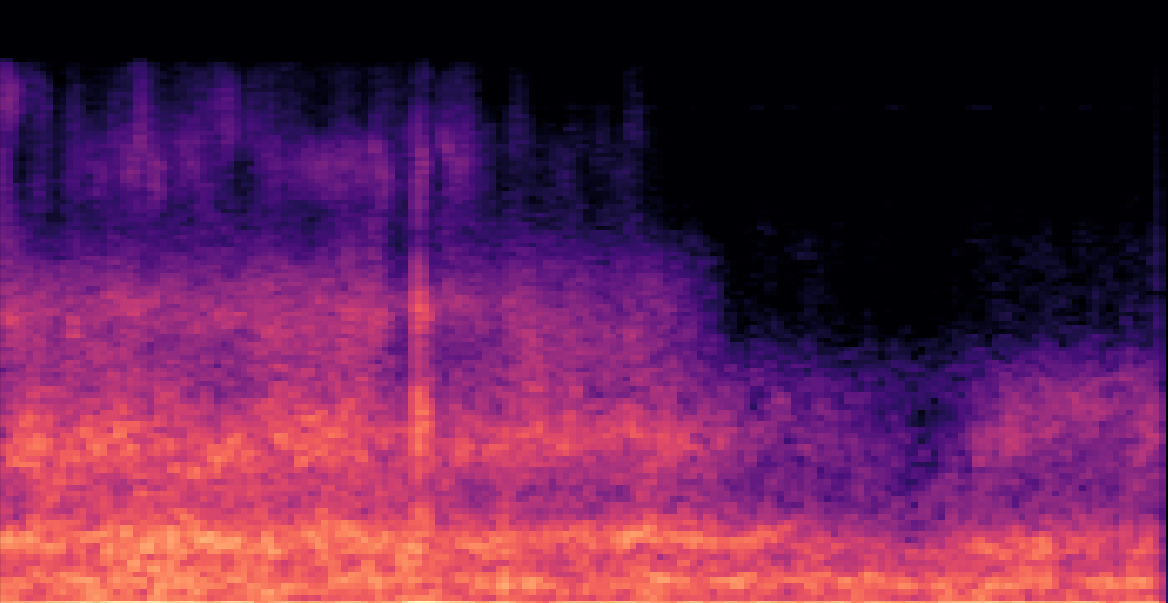}}
\newcommand{\TFlaugh}{\includegraphics[width=\imgw,height=\imgw,keepaspectratio]{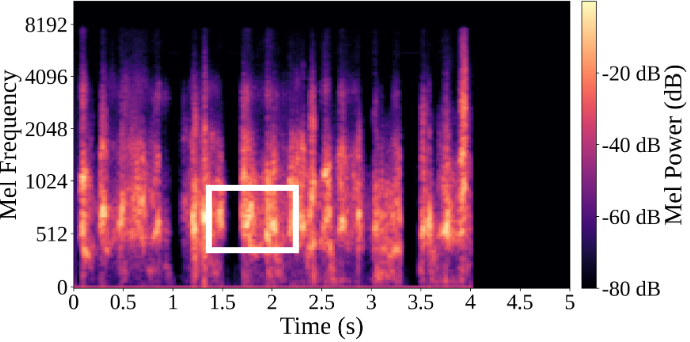}}
\newcommand{\TFhandsaw}{\includegraphics[width=\imgw,height=\imgw,keepaspectratio]{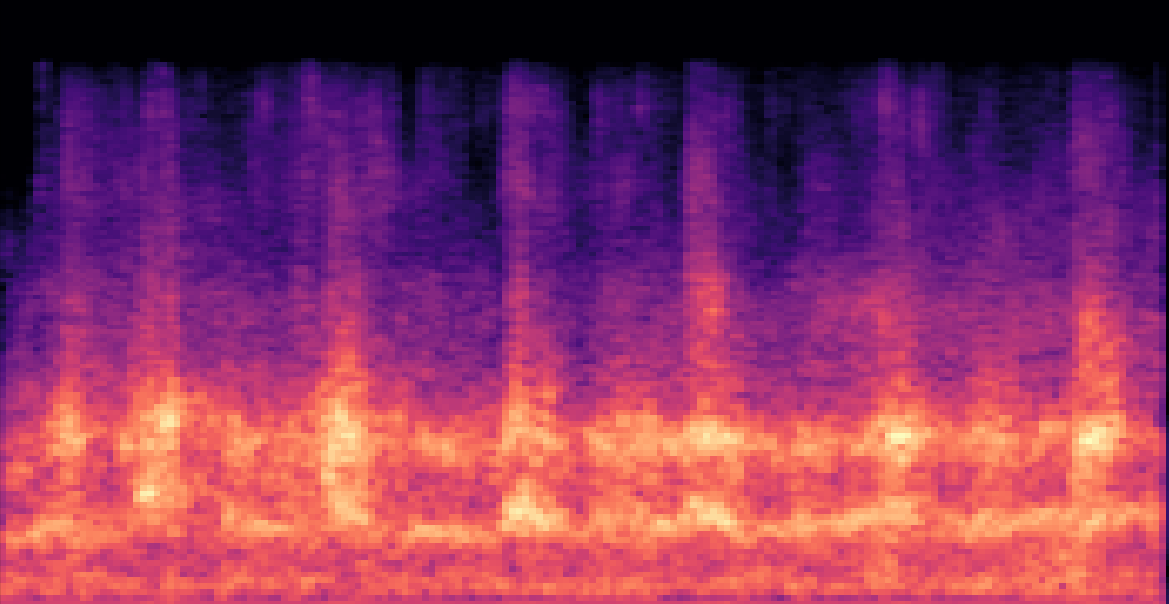}}
\newcommand{\TFbirds}{\includegraphics[width=\imgw,height=\imgw,keepaspectratio]{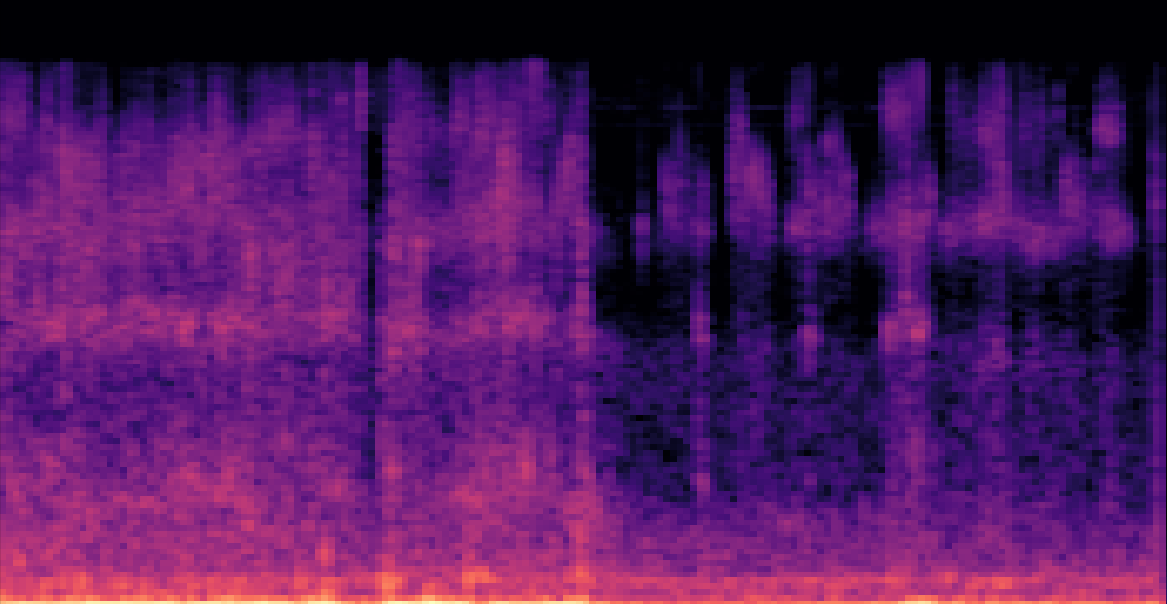}}

% ---- AudioX ----
\newcommand{\AXcrow}{\includegraphics[width=\imgw,height=\imgw,keepaspectratio]{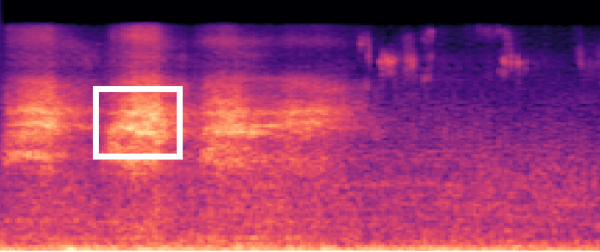}}
\newcommand{\AXlaugh}{\includegraphics[width=\imgw,height=\imgw,keepaspectratio]{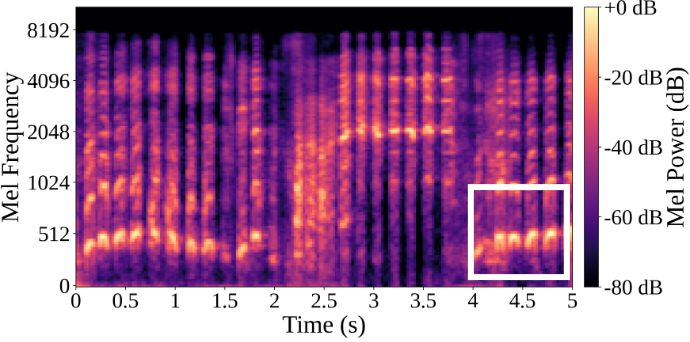}}
\newcommand{\AXhandsaw}{\includegraphics[width=\imgw,height=\imgw,keepaspectratio]{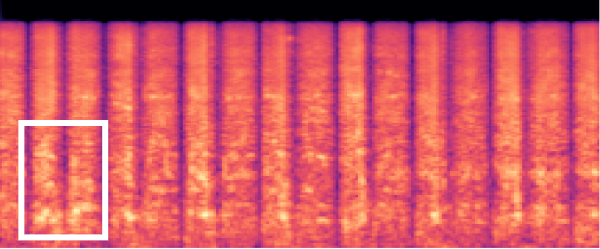}}
\newcommand{\AXbirds}{\includegraphics[width=\imgw,height=\imgw,keepaspectratio]{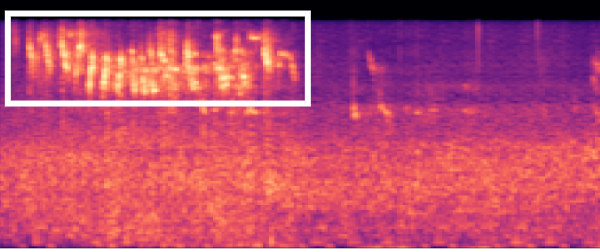}}

% ---- A2SB ----
\newcommand{\ASBcrow}{\includegraphics[width=\imgw,height=\imgw,keepaspectratio]{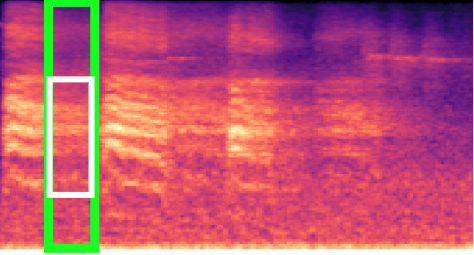}}
\newcommand{\ASBlaugh}{\includegraphics[width=\imgw,height=\imgw,keepaspectratio]{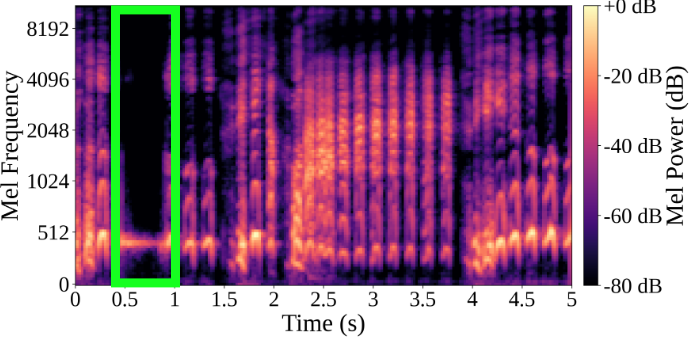}}
\newcommand{\ASBhandsaw}{\includegraphics[width=\imgw,height=\imgw,keepaspectratio]{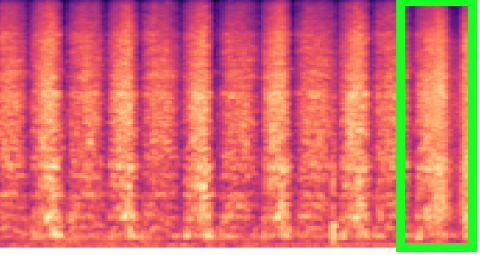}}
\newcommand{\ASBbirds}{\includegraphics[width=\imgw,height=\imgw,keepaspectratio]{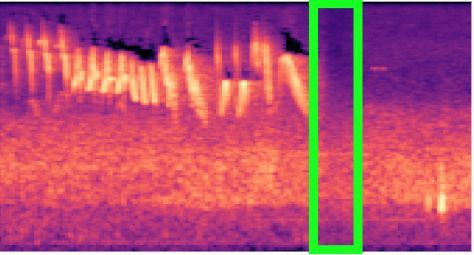}}

\begingroup
%\newcolumntype{C}[1]{>{\raggedright\arraybackslash}p{#1}}

\setlength{\arrayrulewidth}{0.7pt}
\arrayrulecolor{black}

\begin{tabular}{@{}>{\raggedright\arraybackslash}m{\labelw}ccc@{}}
&
\shortstack[c]{{\footnotesize\bfseries Animals}\\{\footnotesize (crow)}} &
\shortstack[c]{\footnotesize\bfseries Natural Sounds\\\footnotesize (chirping birds)} &
\shortstack[c]{{\footnotesize\bfseries Exterior Sounds}\\{\footnotesize (handsaw)}} \\[2pt]

\cellcolor{purple!5}\textbf{Reference}
& \cellcolor{purple!5}\outlineboxxxxxx{\refcrow}
& \cellcolor{purple!5}\outlineboxxxxxx{\refbirds}
& \cellcolor{purple!5}\outlineboxxxxxx{\refhandsaw} \\[2pt]

\arrayrulecolor{black!90}\cmidrule(lr){1-4}

\shortstack[l]{\bfseries\textit{AudioLDM}\\[-1pt](ICML, 2023)}
& \ALcrow & \ALbirds & \ALhandsaw \\
\arrayrulecolor{black!20}\cmidrule(lr){1-4}

\shortstack[l]{\bfseries\textit{T-FOLEY*}\\[-1pt](ICASSP, 2024)}
& \TFcrow & \TFbirds & \TFhandsaw \\
\arrayrulecolor{black!20}\cmidrule(lr){1-4}

\shortstack[l]{\bfseries\textit{ThinkSound}\\[-1pt](NeurIPS, 2025)}
& \TScrow & \TSbirds & \TShandsaw \\
\arrayrulecolor{black!20}\cmidrule(lr){1-4}

\shortstack[l]{\bfseries\textit{A$^2$SB*}\\[-1pt](ArXiv, 2025)}
& \ASBcrow & \ASBbirds & \ASBhandsaw \\
\arrayrulecolor{black!20}\cmidrule(lr){1-4}

\shortstack[l]{\bfseries\textit{AudioX}\\[-1pt](ICLR, 2026)}
& \AXcrow & \AXbirds & \AXhandsaw \\
\arrayrulecolor{black!90}\cmidrule(lr){1-4}

\multicolumn{4}{p{\textwidth}}{\footnotesize
$^*$ Generated clip duration is 4 seconds, following the model restrictions.}
\end{tabular}
\endgroup

\caption{Additional qualitative comparison of ATA variations using mel-spectrograms on three representative ESC-50 classes: \emph{crow}, \emph{chirping birds}, and \emph{hand\_saw}. Each row corresponds to one baseline and each column to one reference class. The figure highlights differences in local texture preservation, temporal organization, and variation behavior across methods. For \textcolor{orange}{A$^2$SB} inpainting, the \outlineboxxxx{ masked and regenerated regions} are outlined, since only a localized segment of the reference is modified. \outlineboxxxxxxxx{Black boxes} indicate prominent texture regions in the reference, while white boxes highlight similar texture patterns preserved in the generated outputs.}
\label{fig:Audio_melspec_a2a}
\end{figure*}

%% file: src/figures/energy_curve_overall.tex
\begin{figure*}[!t]
\centering
\footnotesize
\setlength{\tabcolsep}{1pt}
\renewcommand{\arraystretch}{1.02}

% ---- widths tuned for two-column DAFx layout ----
\newlength{\labelww}
\setlength{\labelww}{0.15\textwidth}
\newlength{\imgww}
\setlength{\imgww}{0.265\textwidth}

% ---- helper for vertically centered method labels ----
\newcommand{\methodcell}[2]{%
  \parbox[c]{\labelww}{\raggedright\bfseries\textit{#1}\\[-1pt](#2)}}

% ---- Reference ----
\newcommand{\refcrow}{\includegraphics[width=\imgww,height=\imgww,keepaspectratio]{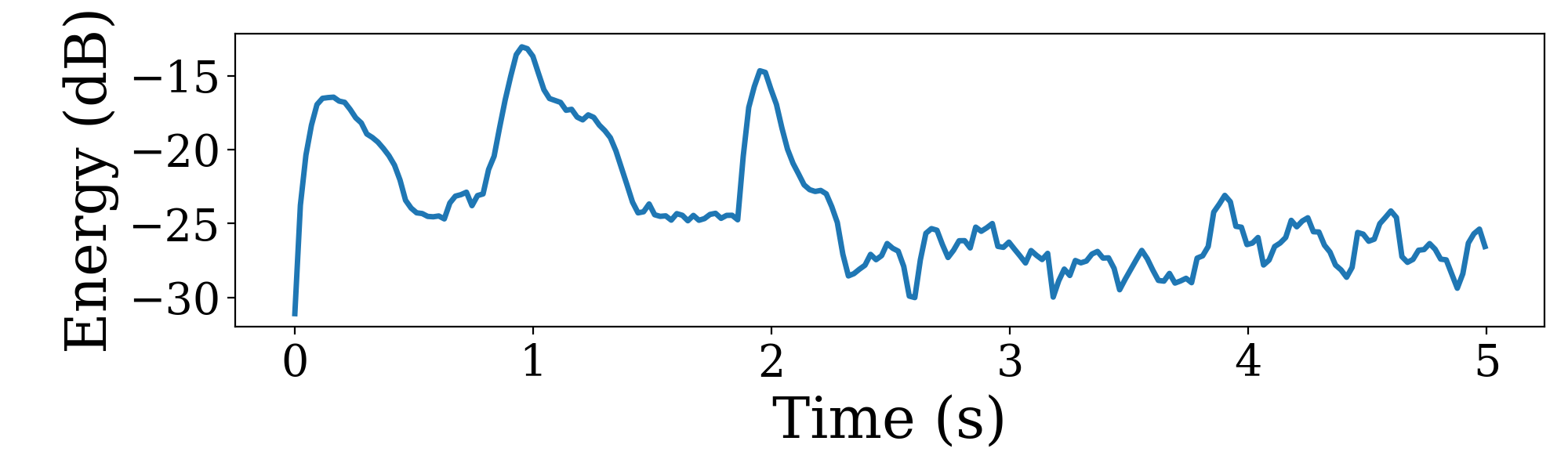}}
\newcommand{\reflaugh}{\includegraphics[width=\imgww,height=\imgww,keepaspectratio]{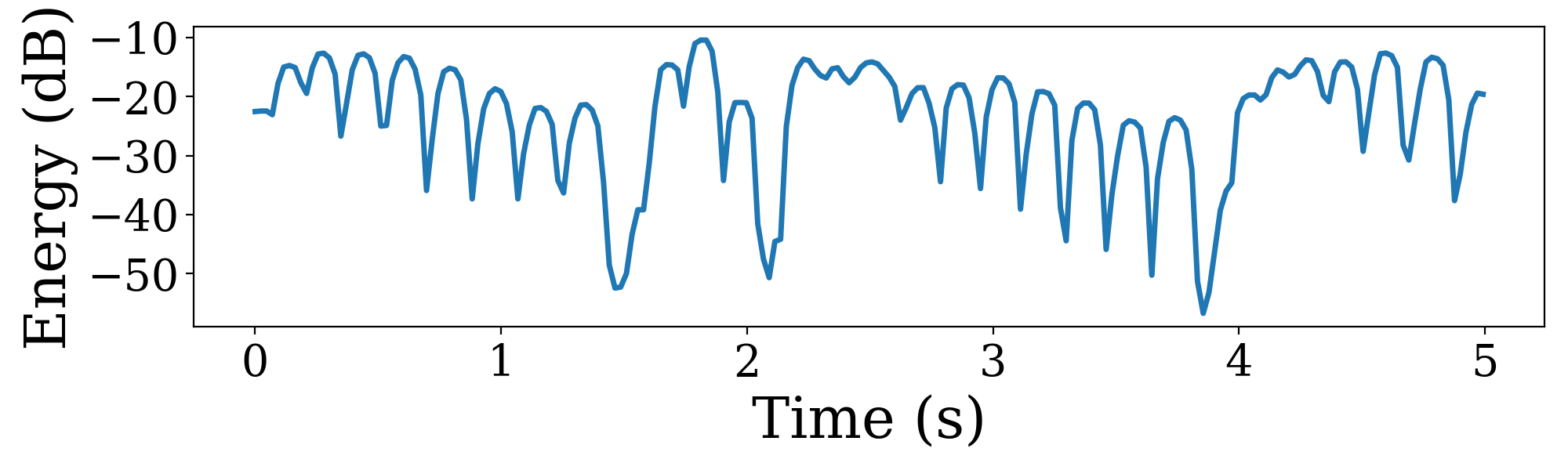}}
\newcommand{\refhandsaw}{\includegraphics[width=\imgww,height=\imgww,keepaspectratio]{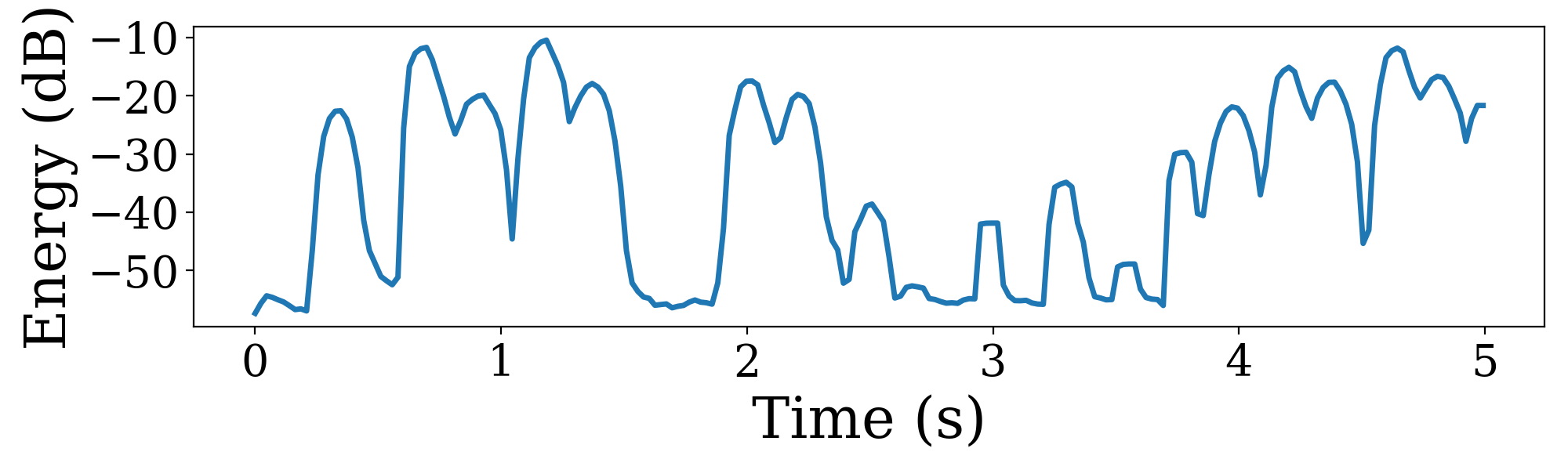}}
\newcommand{\refbirds}{\includegraphics[width=\imgww,height=\imgww,keepaspectratio]{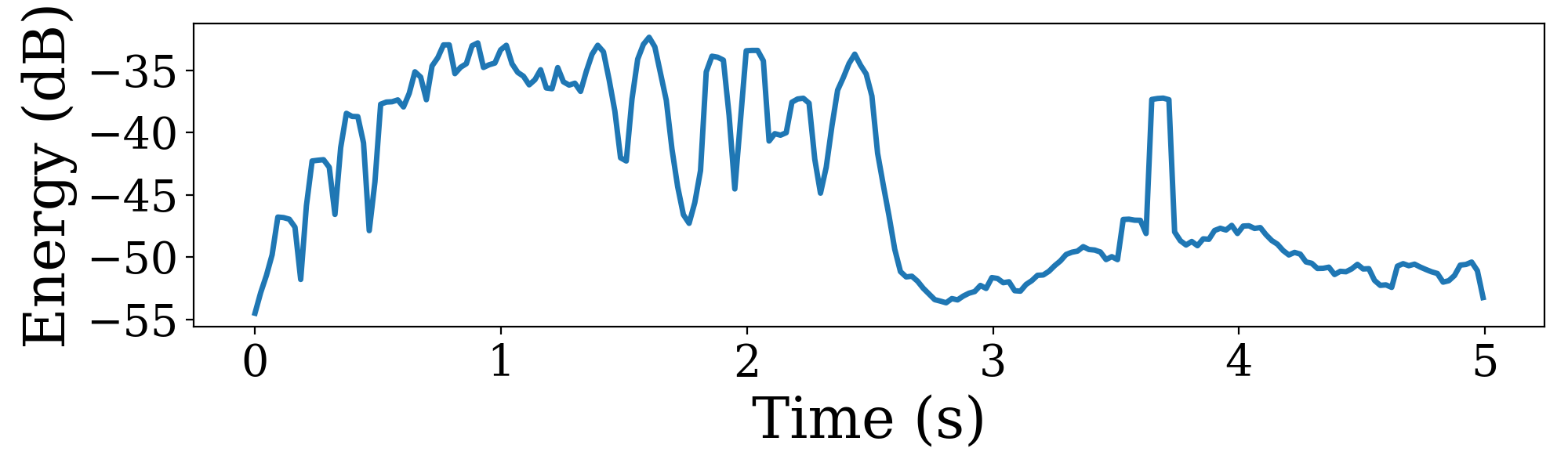}}

% ---- ThinkSound ----
\newcommand{\TScrow}{\includegraphics[width=\imgww,height=\imgww,keepaspectratio]{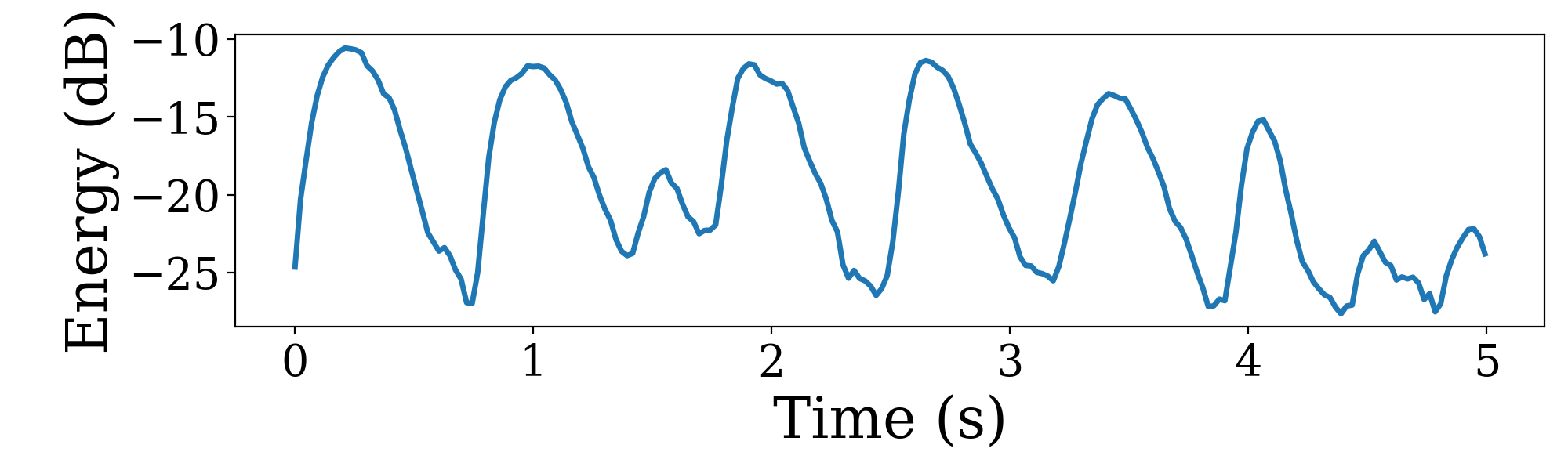}}
\newcommand{\TSlaugh}{\includegraphics[width=\imgww,height=\imgww,keepaspectratio]{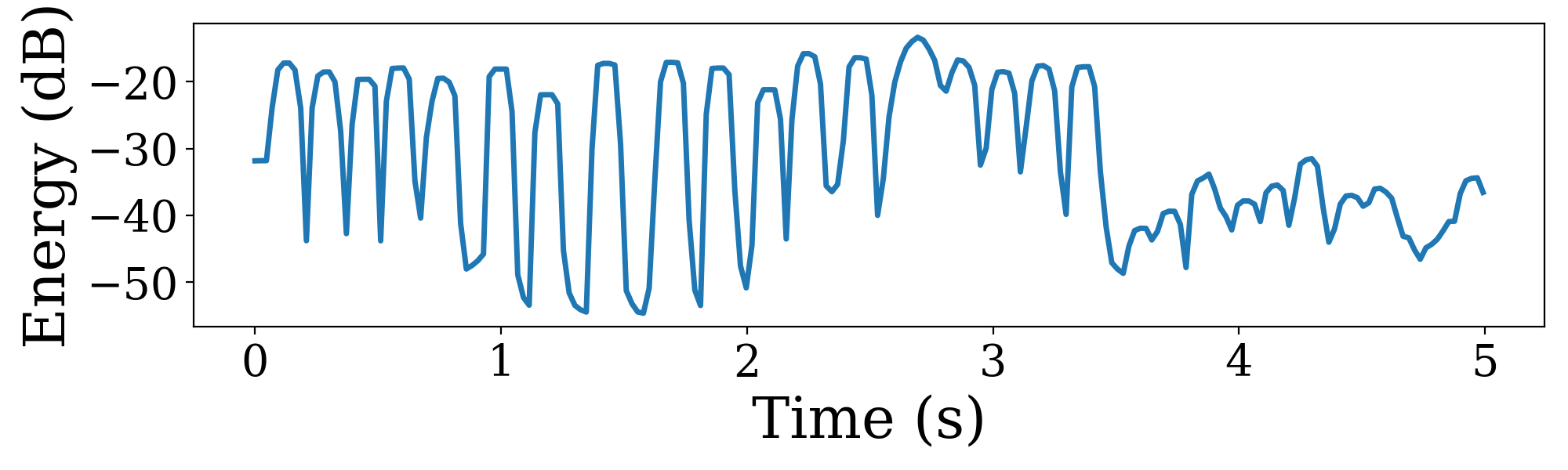}}
\newcommand{\TShandsaw}{\includegraphics[width=\imgww,height=\imgww,keepaspectratio]{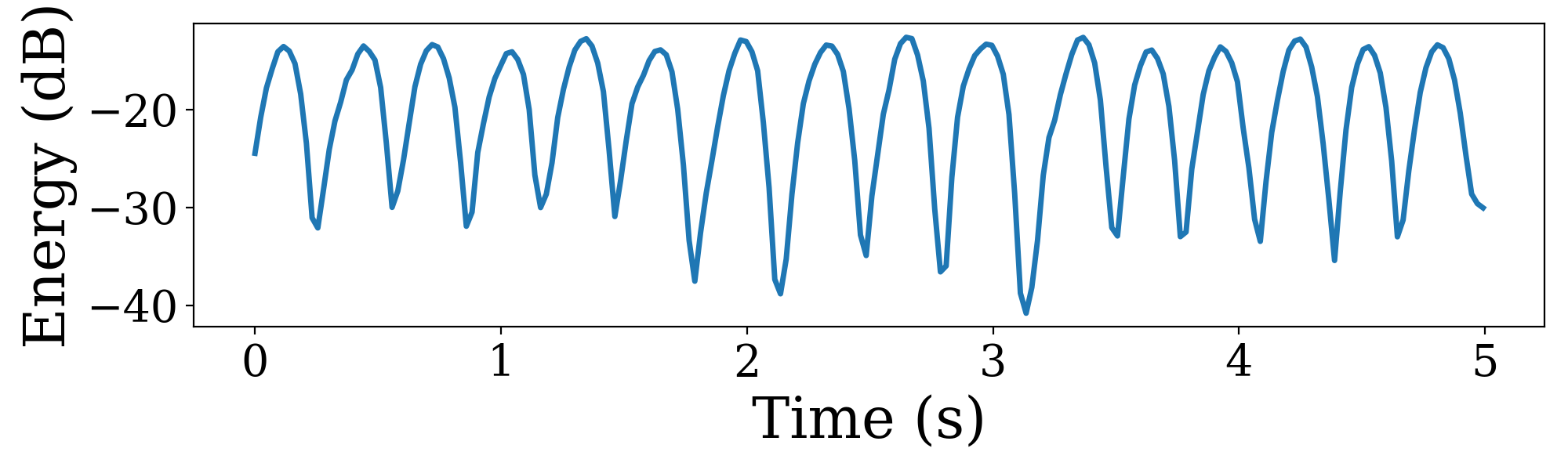}}
\newcommand{\TSbirds}{\includegraphics[width=\imgww,height=\imgww,keepaspectratio]{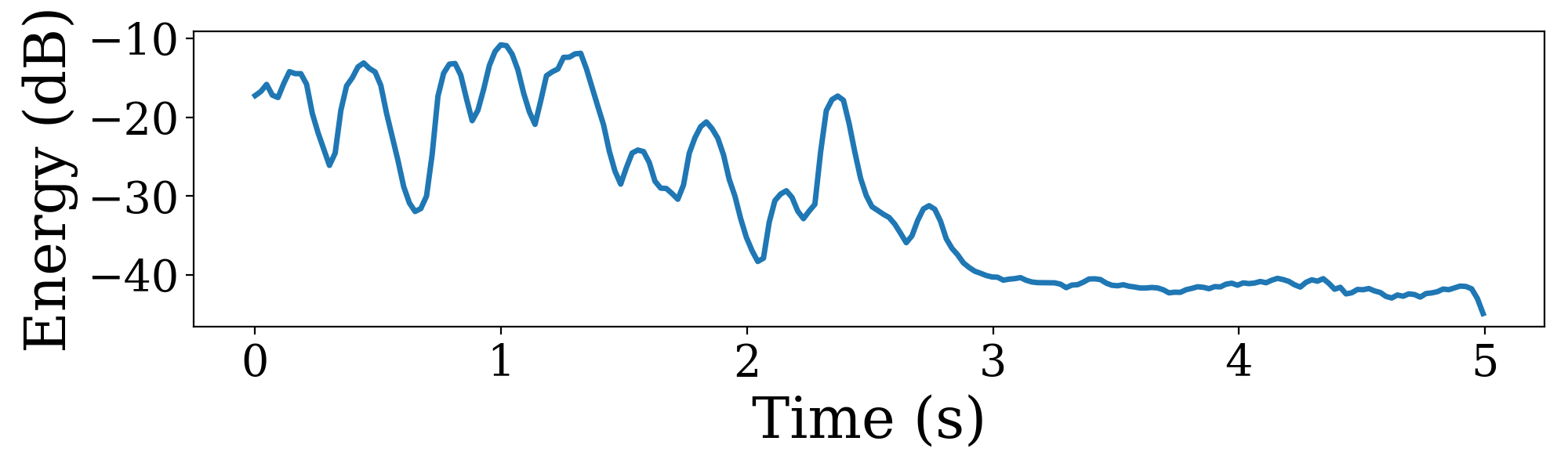}}

% ---- AudioLDM ----
\newcommand{\ALcrow}{\includegraphics[width=\imgww,height=\imgww,keepaspectratio]{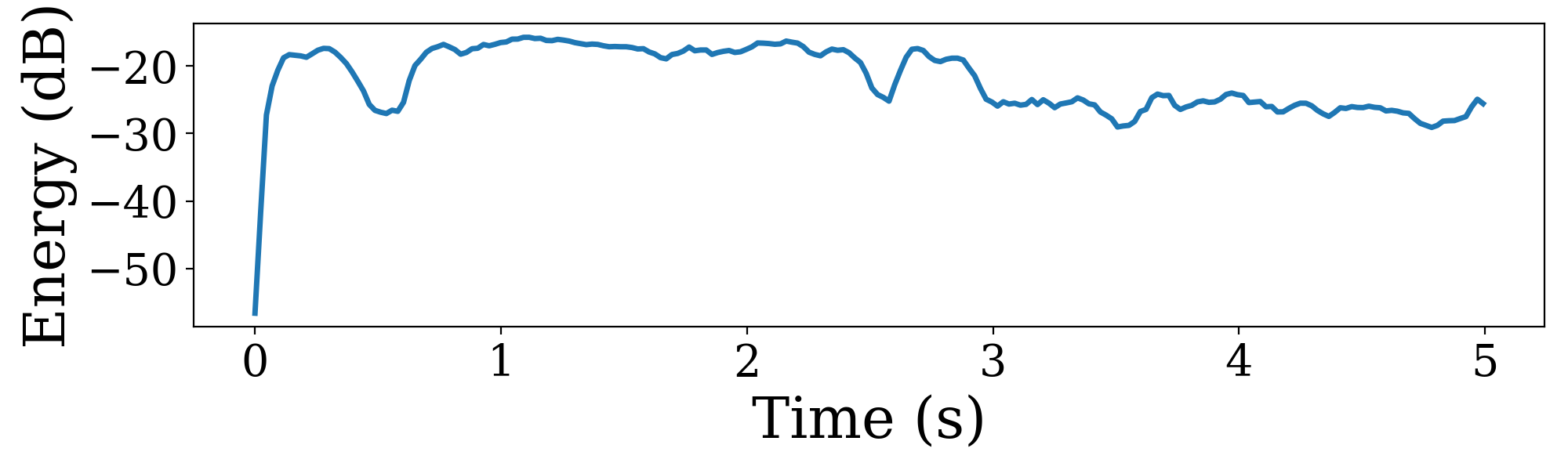}}
\newcommand{\ALlaugh}{\includegraphics[width=\imgww,height=\imgww,keepaspectratio]{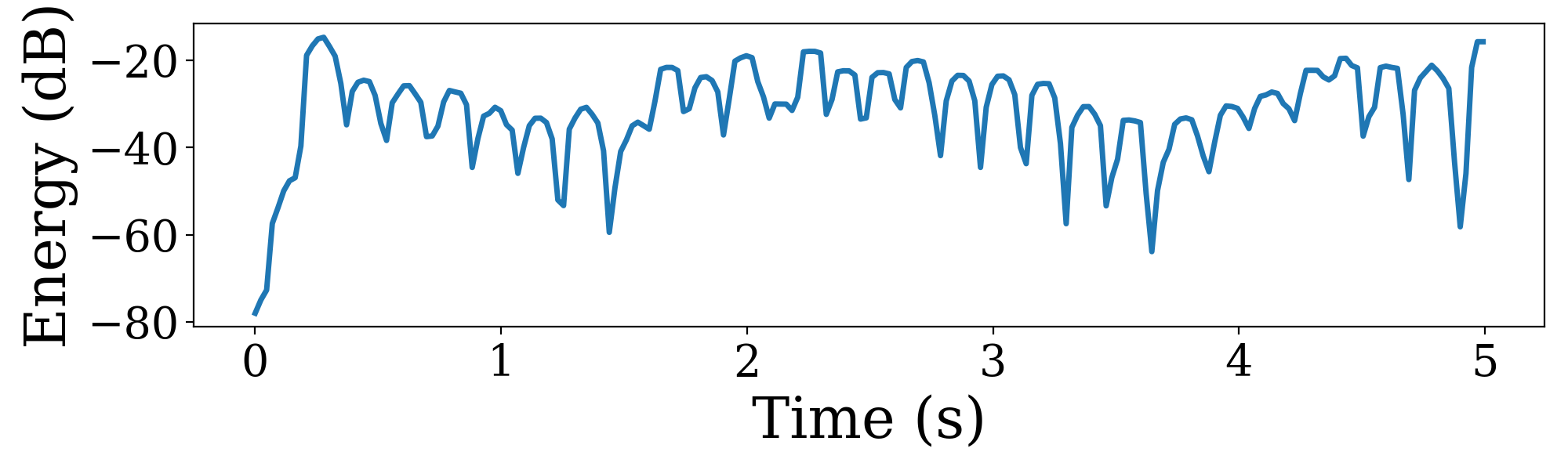}}
\newcommand{\ALhandsaw}{\includegraphics[width=\imgww,height=\imgww,keepaspectratio]{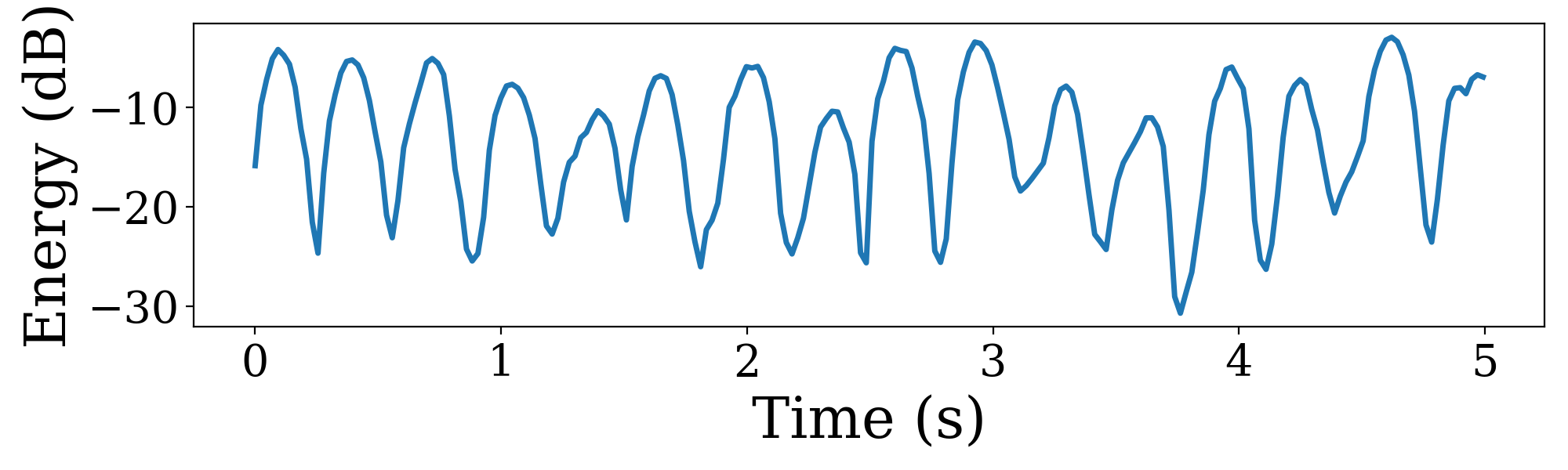}}
\newcommand{\ALbirds}{\includegraphics[width=\imgww,height=\imgww,keepaspectratio]{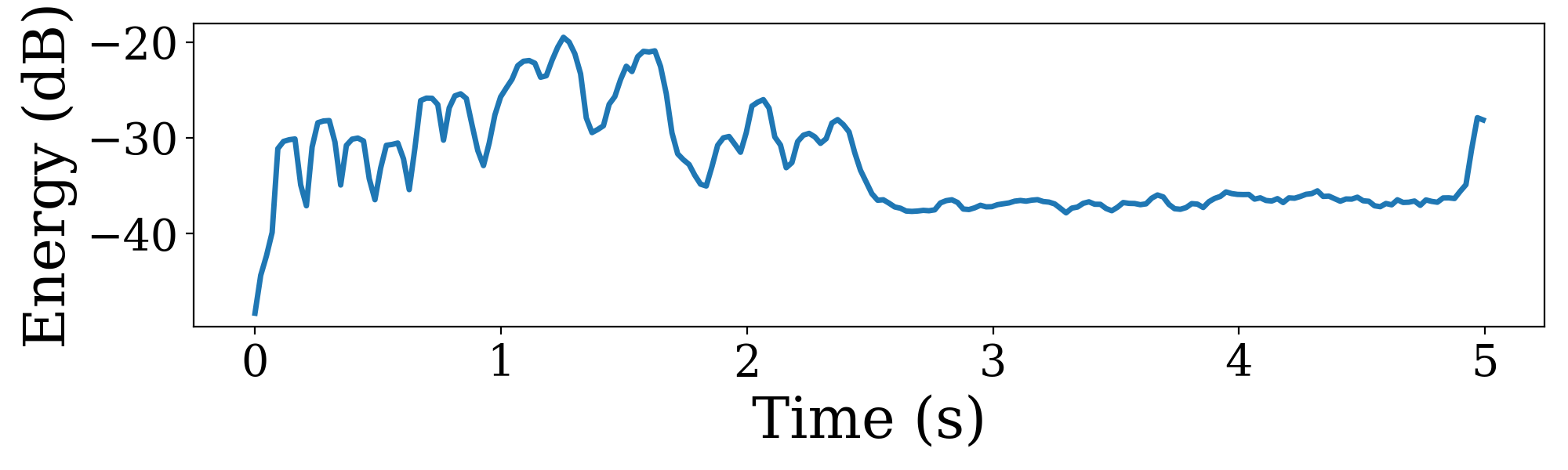}}

% ---- T-Foley ----
\newcommand{\TFcrow}{\includegraphics[width=\imgww,height=\imgww,keepaspectratio]{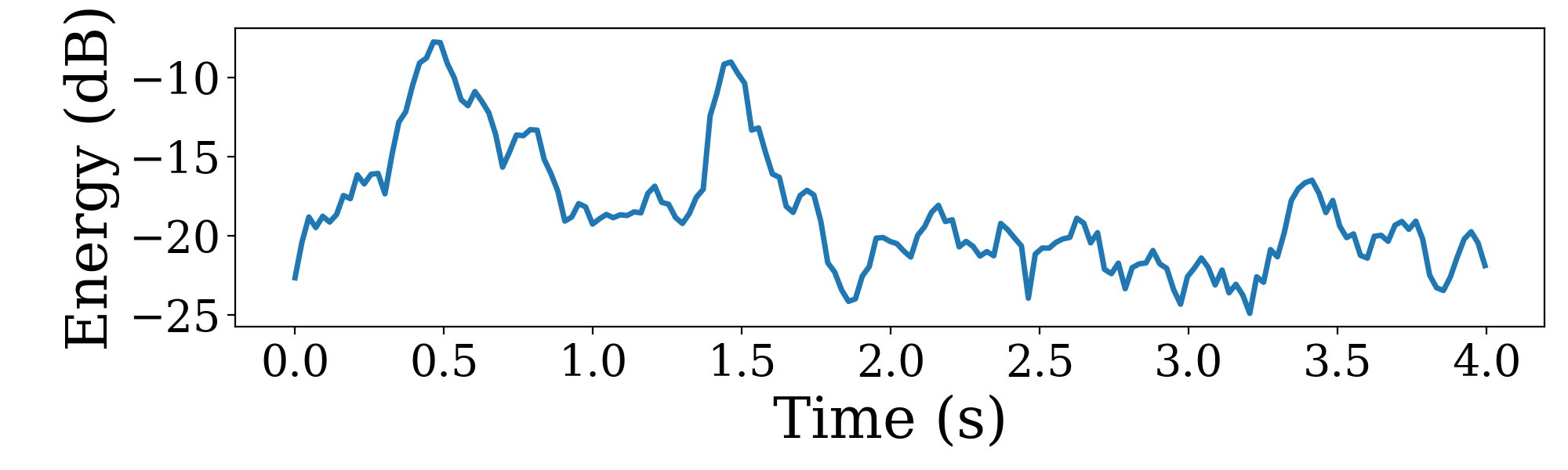}}
\newcommand{\TFlaugh}{\includegraphics[width=\imgww,height=\imgww,keepaspectratio]{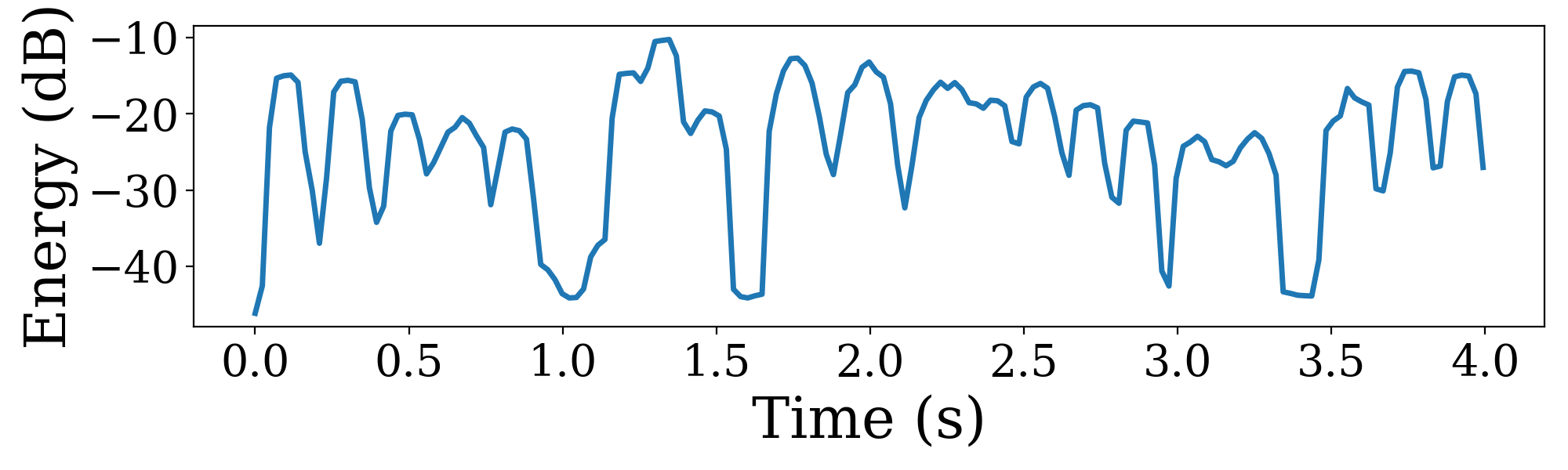}}
\newcommand{\TFhandsaw}{\includegraphics[width=\imgww,height=\imgww,keepaspectratio]{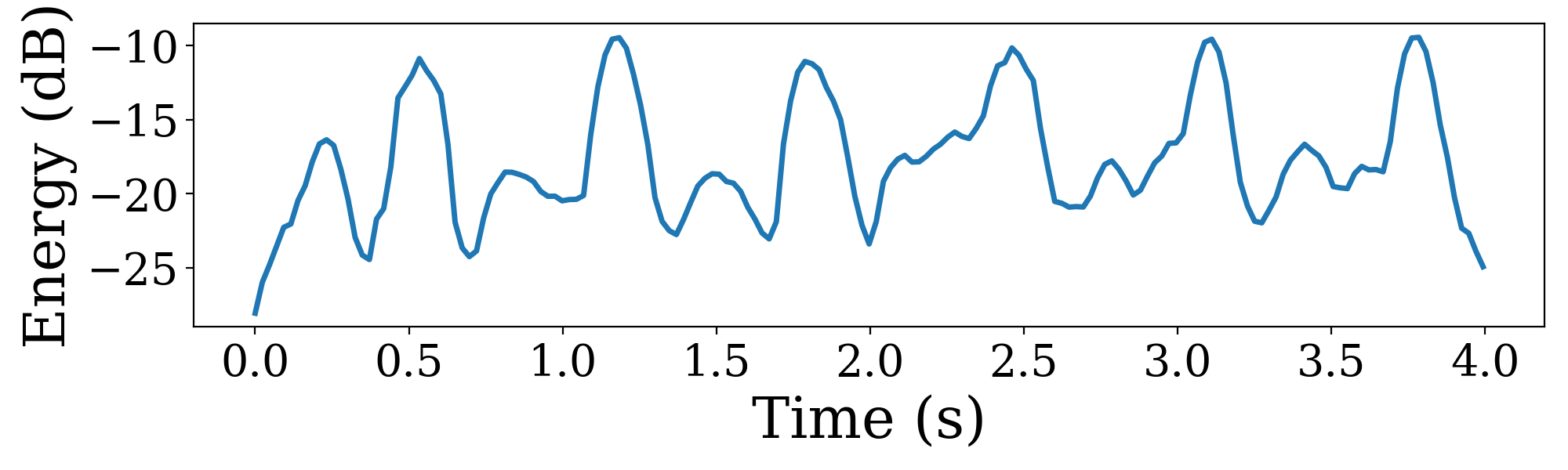}}
\newcommand{\TFbirds}{\includegraphics[width=\imgww,height=\imgww,keepaspectratio]{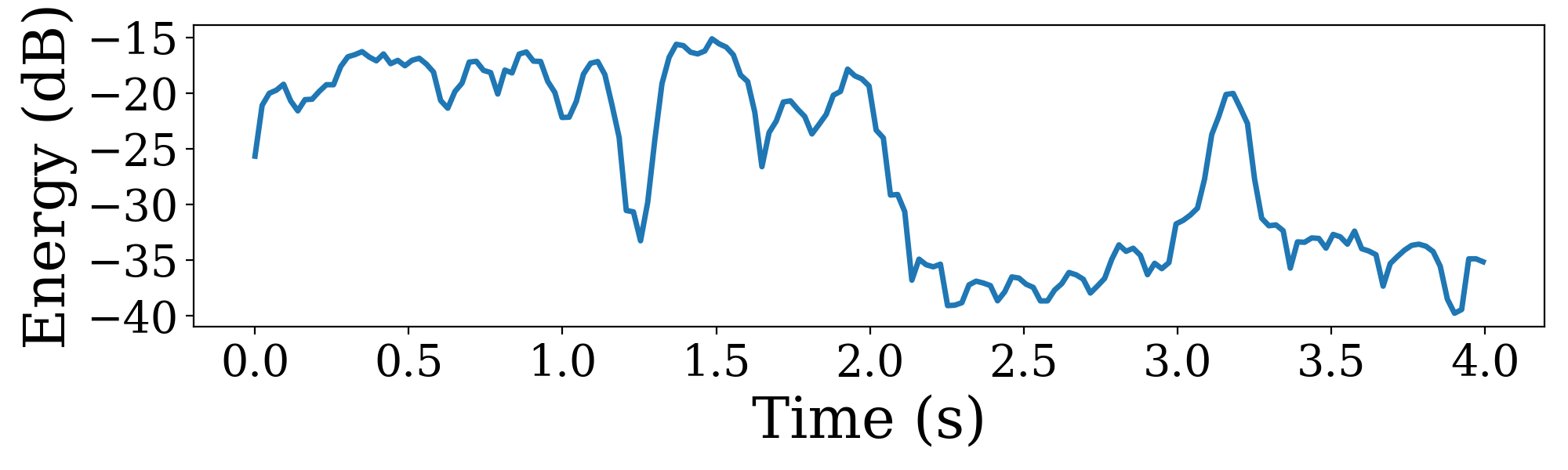}}

% ---- AudioX ----
\newcommand{\AXcrow}{\includegraphics[width=\imgww,height=\imgww,keepaspectratio]{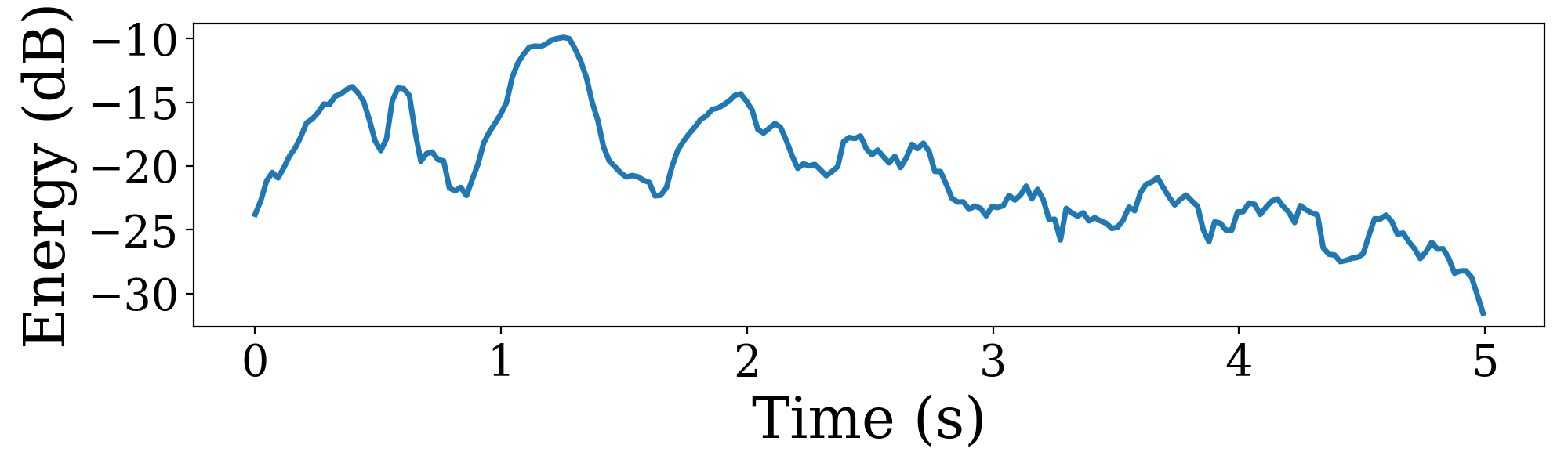}}
\newcommand{\AXlaugh}{\includegraphics[width=\imgww,height=\imgww,keepaspectratio]{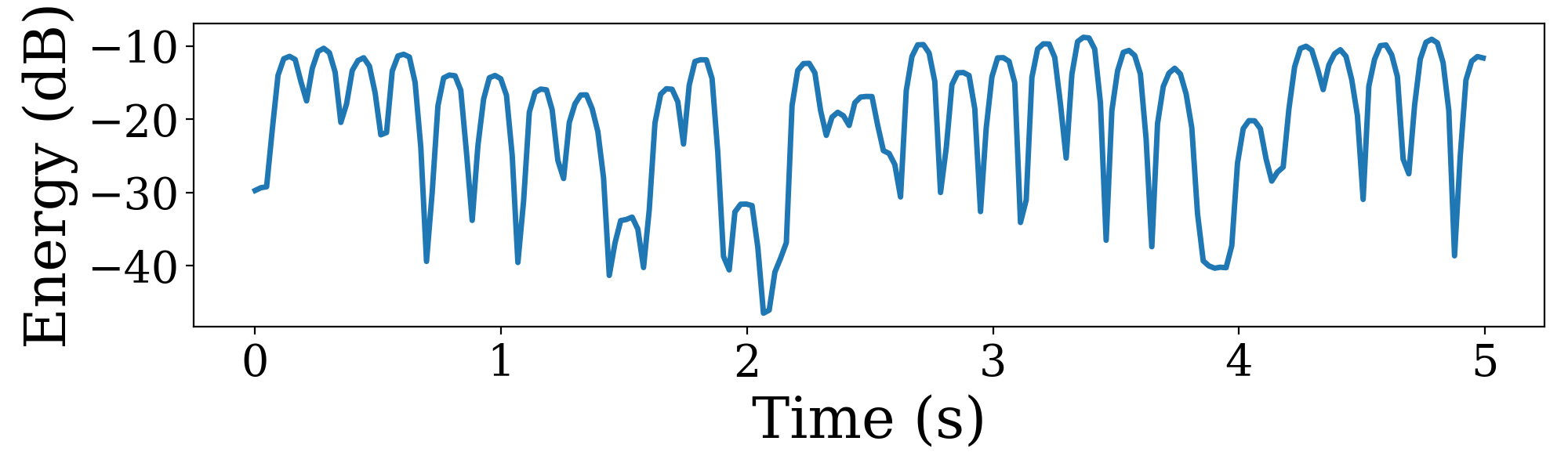}}
\newcommand{\AXhandsaw}{\includegraphics[width=\imgww,height=\imgww,keepaspectratio]{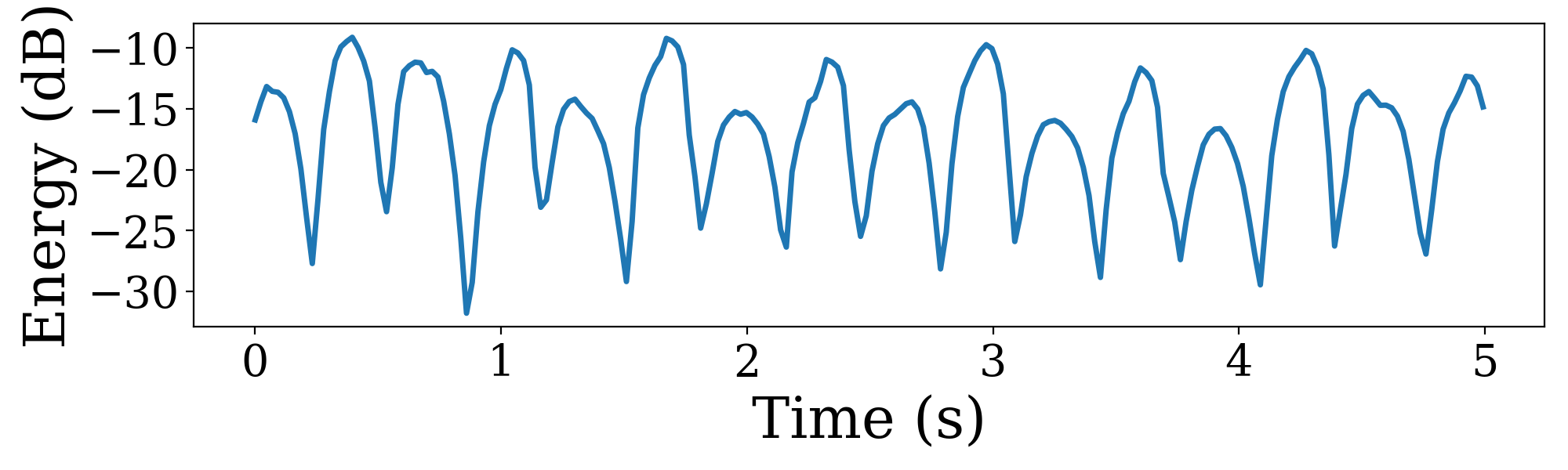}}
\newcommand{\AXbirds}{\includegraphics[width=\imgww,height=\imgww,keepaspectratio]{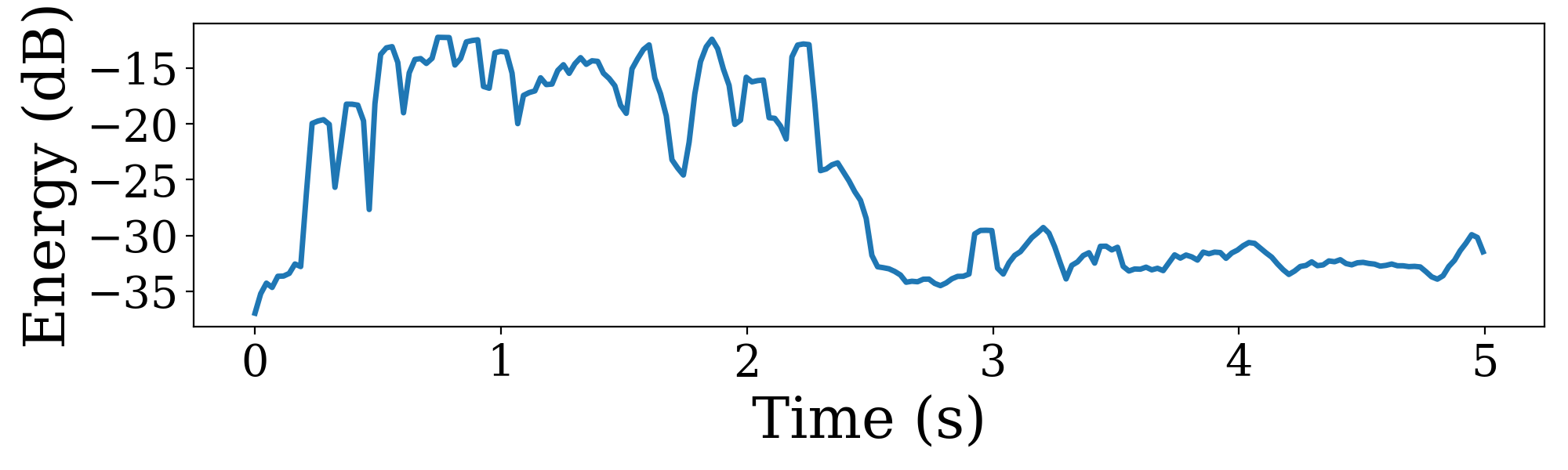}}

% ---- A2SB ----
\newcommand{\ASBcrow}{\includegraphics[width=\imgww,height=\imgww,keepaspectratio]{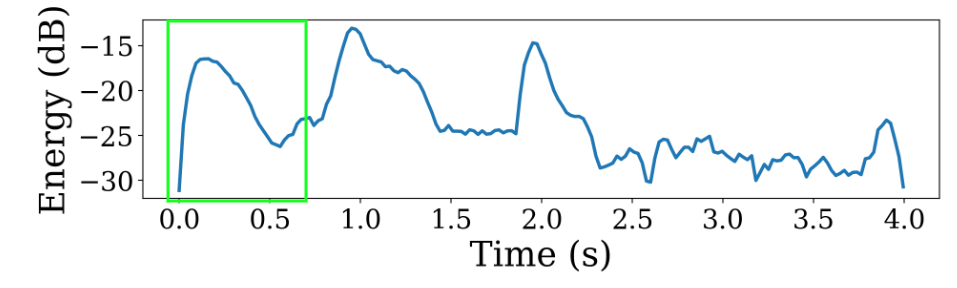}}
\newcommand{\ASBlaugh}{\includegraphics[width=\imgww,height=\imgww,keepaspectratio]{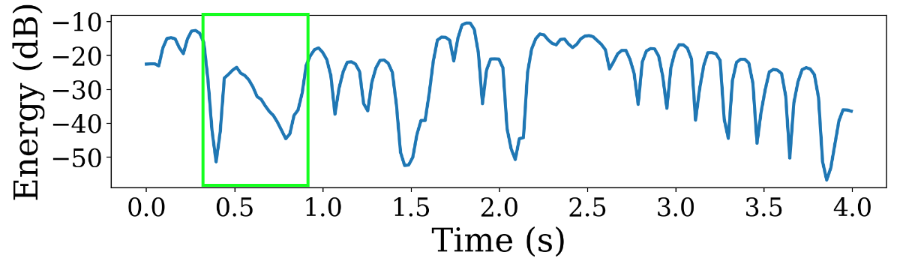}}
\newcommand{\ASBhandsaw}{\includegraphics[width=\imgww,height=\imgww,keepaspectratio]{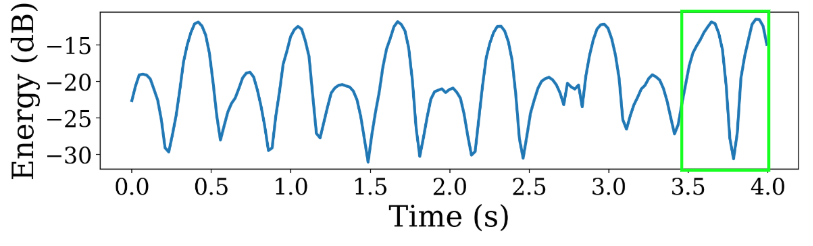}}
\newcommand{\ASBbirds}{\includegraphics[width=\imgww,height=\imgww,keepaspectratio]{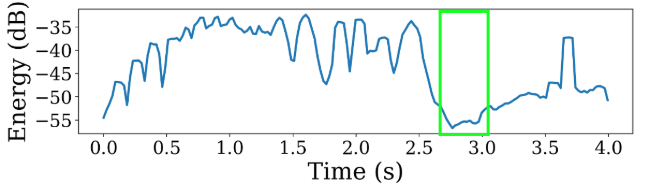}}

\begingroup
\setlength{\arrayrulewidth}{0.7pt}
\arrayrulecolor{black}

\begin{tabular}{@{}>{\raggedright\arraybackslash}m{\labelww}ccc@{}}
&
\shortstack[c]{{\footnotesize\bfseries Animals}\\{\footnotesize (crow)}} &
\shortstack[c]{\footnotesize\bfseries Natural Sounds\\\footnotesize (chirping birds)} &
\shortstack[c]{{\footnotesize\bfseries Exterior Sounds}\\{\footnotesize (handsaw)}} \\[2pt]

\cellcolor{blue!5}\textbf{Reference}
& \cellcolor{blue!5}\outlineboxxxxx{\refcrow}
& \cellcolor{blue!5}\outlineboxxxxx{\reflaugh}
& \cellcolor{blue!5}\outlineboxxxxx{\refhandsaw} \\[2pt]

\arrayrulecolor{black!90}\cmidrule(lr){1-4}

\methodcell{AudioLDM}{ICML, 2023}
& \ALcrow & \ALbirds & \ALhandsaw \\
\arrayrulecolor{black!20}\cmidrule(lr){1-4}

\methodcell{T-FOLEY*}{ICASSP, 2024}
& \TFcrow & \TFbirds & \TFhandsaw \\
\arrayrulecolor{black!20}\cmidrule(lr){1-4}

\methodcell{ThinkSound}{NeurIPS, 2025}
& \TScrow & \TSbirds & \TShandsaw \\
\arrayrulecolor{black!20}\cmidrule(lr){1-4}

\methodcell{A$^2$SB*}{ArXiv, 2025}
& \ASBcrow & \ASBbirds & \ASBhandsaw \\
\arrayrulecolor{black!20}\cmidrule(lr){1-4}

\methodcell{AudioX}{ICLR, 2026}
& \AXcrow & \AXbirds & \AXhandsaw \\
\arrayrulecolor{black!90}\cmidrule(lr){1-4}

\multicolumn{4}{p{\textwidth}}{\footnotesize
$^*$ Generated clip duration is 4 seconds, following the model restrictions.}
\end{tabular}
\endgroup

\caption{Additional qualitative comparison of ATA variations using energy curves for the same three ESC-50 classes as in Fig.~\ref{fig:Audio_melspec_a2a}. The figure visualizes how each method follows or deviates from the reference energy profile over time, thereby complementing the mel-spectrogram analysis with a temporal view of alignment and variation. T-Foley and \textcolor{orange}{A$^2$SB} generate 4-second clips following their original settings. For \textcolor{orange}{A$^2$SB}, the \outlineboxxxx{masked and regenerated regions} are outlined because only a localized segment of the reference is synthesized.}

\label{fig:Audio_energycurve_a2a}
\end{figure*}

%% file: DAFx26_tmpl.bib
@InProceedings{kim2019audiocaps,
    Author = {Kim, Chris Dongjoo  and
      Kim, Byeongchang  and
      Lee, Hyunmin  and
      Kim, Gunhee},
    Title = {AudioCaps: Generating Captions for Audios in the Wild},
    Booktitle = {NAACL-HLT},
    Pages = {119--132},
    Year = {2019}
}

@inproceedings{chen2020vggsound,
  author    = {Chen, Honglie and Xie, Weidi and Vedaldi, Andrea and Zisserman, Andrew},
  title     = {{VGGSound}: A Large-Scale Audio-Visual Dataset},
  booktitle = {ICASSP},
  year      = {2020},
  doi       = {10.1109/ICASSP40776.2020.9053174}
}

@inproceedings{liu2023audioldm,
  title     = {{AudioLDM}: Text-to-Audio Generation with Latent Diffusion Models},
  author    = {Liu, Haohe and Chen, Zehua and Yuan, Yi and Mei, Xinhao and Liu, Xubo and Mandic, Danilo and Wang, Wenwu and Plumbley, Mark D},
  booktitle = {ICML},
  year={2023}
}

@inproceedings{luo2023difffoley,
  title     = {Diff-Foley: Synchronized Video-to-Audio Synthesis with Latent Diffusion Models},
  author    = {Simian Luo and Chuanhao Yan and Chenxu Hu and Hang Zhao},
booktitle={NeurIPS},
year={2023}
}

@inproceedings{chung2024tfoley,
  title     = {{T-FOLEY}: A Controllable Waveform-Domain Diffusion Model for Temporal-Event-Guided Foley Sound Synthesis},
  author    = {Yoonjin Chung and Junwon Lee and Juhan Nam},
  booktitle = {ICASSP},
  year      = {2024},
  doi       = {10.1109/ICASSP48485.2024.10447380}
}

@inproceedings{smartdj2025,
  title     = {{SmartDJ}: Declarative Audio Editing with Audio Language Model},
  author    = {Zitong Lan and Yiduo Hao and Mingmin Zhao},
  booktitle = {ICLR},
  year      = {2026}
}

@article{liang2025audiomorphix,
  title={AudioMorphix: Training-Free Audio Editing with Diffusion Probabilistic Models},
  author={Jinhua Liang and Yuanzhe Chen and Yi Yuan and Dongya Jia and Xiaobin Zhuang and Zhuo Chen and Yuping Wang and Yuxuan Wang},
  journal={arXiv},
  year={2025}
  
}

@article{zhu2023edmsound,
    Author = {Ge Zhu and Yutong Wen and Marc-André Carbonneau and Zhiyao Duan},
    Title = {EDMSound: Spectrogram Based Diffusion Models for Efficient and High-Quality Audio Synthesis},
    Journal = {NeurIPS workshop Machine Learning for audio},
    Year = {2023}
}

@article{stableaudioopen2024,
    Author = {Zach Evans and Julian D. Parker and CJ Carr and Zack Zukowski and Josiah Taylor and Jordi Pons},
    Title = {Stable Audio Open},
    Journal = {ICASSP},
    Year = {2025},
    doi = {10.1109/ICASSP49660.2025.10888461}
}

@article{liu2023audioldm2,
  title   = {{AudioLDM} 2: Learning Holistic Audio Generation with Self-Supervised Pretraining},
  author  = {Haohe Liu and Yi Yuan and Xubo Liu and Xinhao Mei and Qiuqiang Kong and Qiao Tian and Yuping Wang and Wenwu Wang and Yuxuan Wang and Mark D. Plumbley},
  journal = {ACM TASLP},
  year    = {2024},
  doi     = {10.1109/TASLP.2024.3399607}
}

@inproceedings{kreuk2022audiogen,
  title     = {{AudioGen}: Textually Guided Audio Generation},
  author    = {Felix Kreuk and Gabriel Synnaeve and Adam Polyak and Uriel Singer and Alexandre Défossez and Jade Copet and Devi Parikh and Yaniv Taigman and Yossi Adi},
  booktitle = {ICLR},
  year      = {2023}
}

@inproceedings{majumder2024tango2,
  title     = {Tango 2: Aligning Diffusion-Based Text-to-Audio Generations through Direct Preference Optimization},
  author    = {Navonil Majumder and Chia-Yu Hung and Deepanway Ghosal and Wei-Ning Hsu and Rada Mihalcea and Soujanya Poria},
  booktitle = {ACM ICM},
  year      = {2024},
  doi       = {10.1145/3664647.3681688}
}

@inproceedings{cheng2024mmaudio,
  title     = {{MMAudio}: Taming Multimodal Joint Training for High-Quality Video-to-Audio Synthesis},
  author    = {Ho Kei Cheng and Masato Ishii and Akio Hayakawa and Takashi Shibuya and Alexander Schwing and Yuki Mitsufuji},
  booktitle = {CVPR},
  year      = {2025}
}

@article{shi2025samaudio,
    Author = {Bowen Shi and Andros Tjandra and John Hoffman and Helin Wang and Yi-Chiao Wu and Luya Gao and Julius Richter and Matt Le and Apoorv Vyas and Sanyuan Chen and Christoph Feichtenhofer and Piotr Dollár and Wei-Ning Hsu and Ann Lee},
    Title = {SAM Audio: Segment Anything in Audio},
    Journal = {arXiv},
    Year = {2025},
    doi = {10.48550/arXiv.2512.18099}
}

@article{zhao2025uniform,
    Author = {L. Zhao and others},
    Title = {UniForm: A Unified Multi-Task Diffusion Transformer for Audio-Video Generation},
    Journal = {arXiv},
    Year = {2025},
    doi = {10.48550/arXiv.2502.03897}
}

@inproceedings{ppae2024,
  title     = {Prompt-guided Precise Audio Editing with Diffusion Models},
  author    = {Lei Zhao and Linfeng Feng and Dongxu Ge and Rujin Chen and Fangqiu Yi and Chi Zhang and Xiao-Lei Zhang and Xuelong Li},
  booktitle = {PMLR},
  year      = {2024},
  publisher = {PMLR}
}

@inproceedings{audit2023,
  title     = {{AUDIT}: Audio Editing by Following Instructions with Latent Diffusion Models},
  author    = {Yuancheng Wang and Zeqian Ju and Xu Tan and Lei He and Zhizheng Wu and Jiang Bian and Sheng Zhao},
  booktitle = {NeurIPS},
  year      = {2023},
}

@inproceedings{audiox2025,
  title     = {{AudioX}: Diffusion Transformer for Anything-to-Audio Generation},
  author    = {Zeyue Tian and Zhaoyang Liu and Yizhu Jin and Ruibin Yuan and Liumeng Xue and Xu Tan and Qifeng Chen and Wei Xue and Yike Guo},
  booktitle = {ICLR},
  year      = {2026}
}

@inproceedings{liu2025thinksound,
  title     = {ThinkSound: Chain-of-Thought Reasoning in Multimodal Large Language Models for Audio Generation and Editing},
  author    = {Huadai Liu and Kaicheng Luo and Jialei Wang and Wen Wang and Qian Chen and Zhou Zhao and Wei Xue},
  booktitle = {NeurIPS},
  year      = {2025}
}

@article{kong2025a2sb,
    Author = {Zhifeng Kong and Kevin J Shih and Weili Nie and Arash Vahdat and Sang-gil Lee and Joao Felipe Santos and Ante Jukic and Rafael Valle and Bryan Catanzaro},
    Title = {A2SB: Audio-to-Audio Schrodinger Bridges},
    Journal = {arXiv},
    Year = {2025}
}

@electronic{liu2023audioldm_eval,
    Author = {H. Liu and Yixiao Zhang and Rodrigo Mira and Yongyi Zang and Ikko Eltociear Ashimine},
    Title = {{audioldm\_eval}: Audio Generation Evaluation},
    note = {GitHub repository},
    Year = {2023}
}

@inproceedings{imagebind2023,
  title     = {{ImageBind}: One Embedding Space to Bind Them All},
  author    = {Rohit Girdhar and Alaaeldin El-Nouby and Zhuang Liu and Mannat Singh and Kalyan Vasudev Alwala and Armand Joulin and Ishan Misra},
  booktitle = {CVPR},
  year      = {2023},
}

@InProceedings{piczak2015esc50,
    Author = {K. J. Piczak},
    Title = {{ESC}: Dataset for Environmental Sound Classification},
    Booktitle = {ACM MM},
    Year = {2015},
    doi = {10.1145/2733373.2806390}
}

@inproceedings{floresgarcia2024sketch2sound,
  title     = {Sketch2Sound: Controllable Audio Generation via Time-Varying Signals and Sonic Imitations},
  author    = {Hugo Flores García and Oriol Nieto and Justin Salamon and Bryan Pardo and Prem Seetharaman},
  booktitle = {ICASSP},
  year      = {2025}
}

@inproceedings{Chen2025CVPR,
  author    = {Ziyang Chen and Prem Seetharaman and Bryan Russell and Oriol Nieto and David Bourgin and Andrew Owens and Justin Salamon},
  title     = {Video-Guided Foley Sound Generation with Multimodal Controls},
  booktitle = {CVPR},
  year      = {2025},
  doi       = {10.1109/CVPR52734.2025.01749},
}

@misc{loopcopilot2023,
  title         = {Loop Copilot: Conducting AI Ensembles for Music Generation and Iterative Editing},
  author        = {Yixiao Zhang and Akira Maezawa and Gus Xia and Kazuhiko Yamamoto and Simon Dixon},
  year          = {2023},
}

@article{jia2024audioeditor,
  title         = {AudioEditor: A Training-Free Diffusion-Based Audio Editing Framework},
  author        = {Yuhang Jia and Yang Chen and Jinghua Zhao and Shiwan Zhao and Wenjia Zeng and Yong Chen and Yong Qin},
  year          = {2024},
  journal       = {arXiv preprint arXiv:2409.12466},
}

@misc{yang2024uniaudio,
  title         = {UniAudio: An Audio Foundation Model Toward Universal Audio Generation},
  author        = {Dongchao Yang and Jinchuan Tian and Xu Tan and Rongjie Huang and Songxiang Liu and Xuankai Chang and Jiatong Shi and Sheng Zhao and Jiang Bian and Zhou Zhao and Xixin Wu and Helen Meng},
  booktitle = {ICML},
  year          = {2024}
}

@article{yang2024audiobox,
  title        = {AudioBox TTA-RAG: Retrieval-Augmented Generation for Zero-Shot and Few-Shot Text-to-Audio},
  author       = {Mu Yang and Bowen Shi and Matthew Le and Wei-Ning Hsu and Andros Tjandra},
  journal      = {Arxiv},
  year         = {2025}
}

@article{yuan2025dreamaudio,
  title        = {DreamAudio: Few-Shot Customized Text-to-Audio Generation via Natural Language Supervision},
  author       = {Yi Yuan and Xubo Liu and Haohe Liu and Xiyuan Kang and Zhuo Chen and Yuxuan Wang and Mark D. Plumbley and Wenwu Wang},
  journal      = {arXiv},
  year         = {2025}
}

@article{polyak2024moviegen,
  title   = {Movie Gen: A Cast of Media Foundation Models},
  author  = {The Movie Gen team @Meta},
  journal = {CVPR},
  year    = {2024}
}

@inproceedings{fang2026acfoley,
  title={AC-Foley: Reference-Audio-Guided Video-to-Audio Synthesis with Acoustic Transfer},
  author={Pengjun Fang and Yingqing He and Yazhou Xing and Qifeng Chen and Ser-Nam Lim and Harry Yang},
  booktitle={ICLR},
  year={2026}
}

@article{xie2024audiotime,
  title   = {AudioTime: A Temporally-aligned Audio-text Benchmark Dataset},
  author  = {Zeyu Xie and Xuenan Xu and Zhizheng Wu and Mengyue Wu},
  journal = {ICASSP},
  year    = {2025}
}

@article{wang2025ttabench,
  title   = {TTA-Bench: A Comprehensive Benchmark for Evaluating Text-to-Audio Models},
  author  = {Hui Wang and Cheng Liu and Junyang Chen and Haoze Liu and Yuhang Jia and Shiwan Zhao and Jiaming Zhou and Haoqin Sun and Hui Bu and Yong Qin},
  journal = {AAAI},
  year    = {2026}
}

@misc{niu2024soundmorpher,
  title         = {{SoundMorpher}: Perceptually-Uniform Sound Morphing with Diffusion Model},
  author        = {Xinlei Niu and Jing Zhang and Charles Patrick Martin},
  year          = {2024},
  archivePrefix = {arXiv}
}

@article{liu2023ddspsfx,
  title={DDSP-SFX: Acoustically-guided sound effects generation with differentiable digital signal processing},
  author={Yunyi Liu and Craig Jin and David Gunawan},
  journal={DAFx},
  year={2023}
}

@article{okamoto2022onoma,
  title={Onoma-to-wave: Environmental sound synthesis from onomatopoeic words},
  author={Yuki Okamoto and Keisuke Imoto and Shinnosuke Takamichi and Ryosuke Yamanishi and Takahiro Fukumori and Yoichi Yamashita},
  journal={APSIPA Transactions},
  year={2021}
}

@INPROCEEDINGS{clap,
  author={Benjamin Elizalde, Soham Deshmukh, Mahmoud Al Ismail and Huaming Wang},
  booktitle={ICASSP}, 
  title={CLAP Learning Audio Concepts from Natural Language Supervision}, 
  year={2023},
  doi={10.1109/ICASSP49357.2023.10095889}}

@inproceedings{xing2024seeing,
  title     = {Seeing and Hearing: Open-domain Visual-Audio Generation with Diffusion Latent Aligners},
  author    = {Yazhou Xing and Yingqing He and Zeyue Tian and Xintao Wang and Qifeng Chen},
  booktitle = {CVPR},
  year      = {2024},
  doi       = {10.1109/CVPR52733.2024.00683}
}
